\documentclass[11pt,a4paper]{article}
\usepackage{jheppub} 
\usepackage{graphicx}
\usepackage{bm}

\usepackage[skip=3pt]{caption}
\usepackage{framed}
\usepackage{xcolor}
\usepackage{rotating}
\usepackage{empheq}
\usepackage{afterpage}

\usepackage[utf8]{inputenc}

\usepackage{tabularray}

\usepackage[skip=3pt]{subcaption}
\usepackage{float}
\usepackage{enumerate}

\newcommand{\lbar}{\lower0.2ex\hbox{$\mathchar'26$}\mkern-10mu \lambda}

\def\Im{\mathrm{Im}}

\def\mrmd{{\mathrm{d}}}

\def\th{{\mathrm{th}}}

\def\nd{{\mathrm{nd}}}
\def\exp{\mathrm{exp}}

\def\a{\alpha}

\def\d{\delta}
\def\D{\Delta}

\def\ve{\varepsilon}
\def\f{\phi}
\def\F{\Phi}

\def\G{\Gamma}

\def\k{\kappa}
\def\l{\lambda}

\def\n{\nu}
\def\o{\omega}
\def\O{\Omega}

\def\s{\sigma}

\def\t{\theta}

\def\nd{{\mathrm{nd}}}

\def\mcalB{{\mathcal{B}}}
\def\mcalC{\mathcal{C}}
\def\mcalD{{\mathcal{D}}}

\def\mcalF{\mathcal{F}}
\def\mcalG{\mathcal{G}}

\def\mcalS{\mathcal{S}}
\def\mcalV{{\mathcal{V}}}

\def\mcalN{\mathcal{N}}

\def\mcalG{\mathcal{G}}

\def\mcalK{\mathcal{K}}

\def\mfrS{\mathfrak{S}}

\def\mrmD{{\mathrm{D}}}

\def\mbbZ{\mathbb{Z}}

\def\msfw{\mathsf{w}}

\def\*{\star}

\def\ts{\tilde s }

\def\tA{\tilde A}
\def\tB{\tilde B}

\def\tS{{\tilde S}}

\def\tV{\tilde V}
\def\ttO{\tilde \O}
\def\ttP{\tilde \Psi}
\def\tPi{\tilde \Pi}

\def\ta{\tilde a}

\def\tu{\tilde u}

\def\ttau{\tilde \tau}

\def\mrmA{\mathrm{A}}

\def\mrmG{\mathrm{G}}
\def\mrmB{\mathrm{B}}

\def\mrmI{\mathrm{I}}
\def\mrmL{\mathrm{L}}

\def\mrmP{\mathrm{P}}

\def\mrmT{\mathrm{T}}

\def\mrmW{\mathrm{W}}

\def\mrmi{\mathrm{i}}
\def\mrmii{\mathrm{ii}}

\def\med{\mathrm{med}}

\def\Res{\mathrm{Res}}

\def\<{\langle}
\def\>{\rangle} 
\def\dee{\partial}

\def\aD{a^\mrmD}

\def\Im{\mathrm{Im}}

\usepackage{slashed}

\definecolor{darkbrown}{rgb}{0.4, 0.26, 0.13}
\definecolor{paleblue}{rgb}{0.69, 0.93, 0.93}
\definecolor{lightskyblue}{rgb}{0.53, 0.81, 0.98}
\definecolor{skyblue}{rgb}{0.53, 0.81, 0.92}
\definecolor{darkred}{rgb}{0.55, 0.0, 0.0}
\definecolor{darkblue}{rgb}{0.0, 0.0, 0.55}
\definecolor{darkpastelgreen}{rgb}{0.01, 0.75, 0.24}
\definecolor{lightgreen}{rgb}{0.56, 0.93, 0.56}

\usepackage{babel}
\usepackage{beramono}
\usepackage[T1]{fontenc}

\definecolor{identifiercolor}{rgb}{.4,.6,.56}
\definecolor{stringcolor}{gray}{0.5}
\definecolor{inactivecolor}{rgb}{0.15,0.15,0.5}

\usepackage{listings}
\title{\boldmath Exact WKB in all sectors I: Potentials with degenerate saddles}

\author[1,2]{Tatsuhiro Misumi,}
\emailAdd{misumi@phys.kindai.ac.jp}
\affiliation[1]{Department of Physics, Kindai University, Osaka 577-8502, Japan}
\affiliation[2]{Research and Education Center for Natural Sciences, Keio University, Kanagawa 223-8521, Japan}

\author[3]{Cihan Pazarbaşı}
\emailAdd{cihan.pazarbasi@gmail.com}
\affiliation[3]{Okinawa Institute of Science and Technology, Okinawa 904-0495, Japan}

\abstract{
	We explore the exact-WKB (EWKB) method through the analysis of Airy and Weber types, with an emphasis on the exact quantization of locally harmonic potentials in multiple sectors. The core innovation of our work lies in introducing a novel complexification approach to the energy parameter $u$, distinct from the common complexification of the (semi-classical) expansion parameter used in Borel summability. This new technique allows for continuous analytical continuation across different sectors of a potential while maintaining the exact quantization condition, even before median summation. By redefining the $A$-cycle above the potential barrier top, we ensure the quantization condition remains real and, by use of the Stokes automorphism and the median resummation, show that the resurgence structure is preserved across transitions between sectors. Furthermore, we discuss the Weber-type exact-WKB method, offering exact estimates for quantum actions around all types of saddle points, generalizing previous results. Through the analysis of these quantum actions, we reveal the presence of $S$-duality, facilitating the exchange between perturbative and non-perturbative behaviors, and we conjecture the mapping of the P-NP relations between dual theories. Our study encompasses periodic and symmetric double-well potentials, demonstrating that the exact-WKB method captures intricate structures in quantum systems in all sectors, including multi-instanton contributions and the resurgence of quantum actions.
} 

\begin{document}
	
	\maketitle
	\flushbottom	

	\section{Introduction}\label{Section: Introduction}

	The exact-WKB (Wentzel-Kramers-Brillouin) method \cite{BPV, Voros1983,Silverstone, AKT1, AKT2, Takei1, DDP2, DP1, Takei2, Kawai1, Takei3, AKT3, AKT4, Schafke1, Iwaki1} stands as one of the most powerful tools in the study of quantum mechanical systems. 
	With being combined with Borel resummation, the exact-WKB (EWKB) method offers significant improvements over semi-classical approximations, particularly in quantum systems where both perturbative and non-perturbative effects play a role. 
	In recent decades, the EWKB method has seen significant advancements, fueled by contributions from both physicists and mathematicians. 
	These developments have led to a deeper understanding of the connection between the semi-classical analysis and the exact quantization conditions \cite{Sueishi:2020rug,Sueishi:2021xti,Kamata:2021jrs}, revealing intricate structures such as Stokes phenomena and resurgence theory in quantum mechanical systems \cite{Alvarez1,Alvarez2, Alvarez3,Alvarez4,Jentschura:2004jg, Zinn-Justin:2004vcw, Zinn-Justin:2004qzw,Jentschura:2010zza,Jentschura:2011zza, Basar:2013eka,Dunne:2013ada,Dunne:2014bca,Basar:2015xna,Misumi:2015dua,Behtash:2015loa,Gahramanov:2015yxk,Kozcaz:2016wvy,Dunne:2016jsr,Dunne:2016qix,Fujimori:2016ljw,Basar:2017hpr,Fujimori:2017oab,Sueishi:2019xcj,Cavusoglu:2023bai,Cavusoglu:2024usn,Ture:2024nbi}. 
	The method has been instrumental in exploring non-perturbative effects involving multi-instanton contributions which appear in analysis of various quantum mechanical and field theoretical systems \cite{Zinn-Justin:1981qzi,Zinn-Justin:1983yiy,Unsal:2007vu,Unsal:2007jx,Shifman:2008ja,Poppitz:2009uq,Anber:2011de,Poppitz:2012sw, Misumi:2014raa,Behtash:2018voa,Fujimori:2019skd,Misumi:2019upg,Fujimori:2020zka,Unsal:2020yeh,Pazarbasi:2021ifb,Pazarbasi:2021fey}.
	It also has deep connections with 
	4d  ${\cal N}=2$ gauge theories \cite{Nekrasov:2009rc,Mironov:2009uv,Kashani-Poor:2015pca,Kashani-Poor:2016edc,Ashok:2016yxz,Yan:2020kkb}, 
	wall-crossing phenomena \cite{Gaiotto:2012rg,Allegretti:2020dyt}, 
	ODE/IM correspondence \cite{Dorey:2001uw, Dorey:2007zx,Ito:2018eon,Ito:2019jio,Imaizumi:2020fxf}, 
	TBA equations \cite{Emery:2020qqu,Ito:2024nlt}
	topological string theory \cite{Grassi:2014cla,Grassi:2014zfa,Codesido:2017dns,Codesido:2017jwp,Hollands:2019wbr,Ashok:2019gee,Coman:2020qgf,Iwaki:2023cek} and 
	other subjects \cite{Kuwagaki:2020pry,Taya:2020dco,Enomoto:2020xlf,Enomoto:2022mti,Duan:2018dvj,Imaizumi:2022dgj,vanSpaendonck:2022kit,Kamata:2023opn,Kamata:2024tyb,Bucciotti:2023trp,vanSpaendonck:2024rin}.

	The construction of the EWKB method starts with the complexification of the coordinate space and then uses the classical energy conservation relation $p_0^2 = 2(E-V(x))$ to turn the $x \in \mathbb C$ plane into a Stokes graph.   
	One determines the exact quantization condition by the Stokes graph, which remains invariant under variations of the parameters and energy as long as one does not encounter a topology change in Stokes graph. 
	On the other hand, the WKB wave function is obtained recursively from the non-linear Riccati equation, from which one can obtain the expressions for perturbative/non-perturbative cycles (Voros multipliers).  
	By taking the Borel resummation of the WKB wave function and investigating the Stokes graph, the method captures all perturbative, non-perturbative phenomena for all energy levels,  large-order/low order resurgence, low-order-low-order constructive form of the resurgence and fairly non-trivial aspects of the spectrum at once.

	In this paper, we build upon these foundations by introducing a new approach to the EWKB method with a focus on locally harmonic potentials. Specifically, we propose a novel complexification of the energy parameter $u$, as opposed to the more traditional complexification of the expansion parameter, i.e. the Planck constant $\hbar$ or some coupling parameter $g$, that has been used to track Borel summability in resurgence analysis. 
	
	Our approach allows for the continuous analytical continuation of quantum systems across different sectors\footnote{Note that with the word \textit{``sector''}, we refer to the distinct spectral regions of a given quantum system. In particular, in this paper, we consider two different sectors corresponding to the below and above barrier top regions of the classical potential.}, providing a new perspective on how exact quantization conditions evolve across sector boundaries. A similar analysis was previously performed in \cite{Basar:2015xna} by Başar and Dunne for Mathieu equation (periodic potential) in view of the analytic continuation of elliptic integrals. In this way, they showed the smooth transition of the quantized spectrum through the barrier top. The novel analytic continuation we present here introduces an equivalent analytic continuation to EWKB framework and provides a generalization to all locally harmonic potentials.

	An important property of the EWKB method is that it always fully encodes the physical quantities with their resurgence structures. Therefore, by construction our approach links the resurgence structures of different sectors. To understand this relationship better, we show that the median summations of exact quantization conditions have the same form in all sectors with an appropriate redefinition of $A$-cycle, in a way that compatible with the analytic continuation of the associated quantum action. This manifestly displays the preservation of resurgence structures in all sectors as well as the smooth transition of the spectrum across the barrier top.

	To further develop the EWKB method, we also discuss the Weber-type approach. While the Airy-type approach to EWKB has been extensively studied for its qualitative insights, the Weber-type approach offers a more quantitative framework, particularly when dealing with quantum actions around saddle points. In this work, we generalize the Weber-type EWKB method to handle all types of saddle points, i.e.~both minima and maxima of a given classical potential. Such a generalization was previously discussed in \cite{DDP2,DP1} from a mathematical point of view where authors provided double-well potential as an explicit example. In our discussion, we follow a different path, which is based on \cite{AKT1,AKT4}, in the construction of the Weber-type EWKB. Moreover, we provide exact expressions of quantum actions,  labeled as $a(u,g)$ and $\aD(u,g)$, for generic potentials, highlighting the local and non-local components of these actions as well as their meaning in the physical literature.
	
	In addition to that providing exact estimates for generic potentials, our analysis reveals an important symmetry, i.e.~$S$-duality, that governs the exchange between perturbative and non-perturbative behaviors around minima and maxima.~$S$-duality is a central concept in many areas of theoretical physics, as it relates weak and strong coupling regimes in quantum systems.
	In \cite{Codesido:2017dns,Codesido:2017jwp}, this exchange between minima and maxima was discussed in the context of the relationship between holomorphic anomaly equations and all-order WKB expansions of one dimensional quantum systems. Here, we show that it is a natural by product of the EWKB analysis and becomes manifest when the Weber-type approach is utilized. In addition to that the nature of EWKB analysis indicates that the $S$-duality, which is originally defined at the leading order (classical level), could relate the resurgence structures of dual theories. In this sense, we also conjecture a generic $S$-transformation of P-NP relations between dual theories.
	
	
	The setup we introduce is applicable to generic one-dimensional quantum systems. In this paper, we mainly focus on the subclass of potentials with degenerate saddle points, i.e. Chebyshev potentials \cite{Basar:2017hpr}. An important property of these potentials is that their WKB actions are proportional to each other. This helps us to investigate the transition across the barrier tops for this subclass of potentials in general. This provides a valuable intuition for our EWKB analysis.

	Finally, we discuss two of the Chebyshev potentials, periodic and symmetric double-well potentials in details.~They are prototypical systems for studying tunneling phenomenon as well as multi-instanton contributions and resurgence theory \cite{Sueishi:2020rug,Sueishi:2021xti,Alvarez3,Zinn-Justin:2004qzw,Zinn-Justin:2004vcw,Dunne:2013ada,Dunne:2014bca,Misumi:2015dua,Basar:2015xna}.~For our purposes, they serve as rich examples for understanding the evolution of quantum actions across different sectors while preserving the entire resurgence structure in the presence of multiple degenerate saddle points. 
	
	
	
	Our work not only refines the exact-WKB method but also extends its applicability to a wider class of quantum systems. By introducing the complexification of the energy parameter $u$ and expanding the Weber-type method, we are able to provide both qualitative and quantitative insights into the behavior of quantum systems with degenerate minima and maxima. The introduction of $S$-duality and the preservation of the resurgence structure across potential barriers further deepen our understanding of the exact-WKB method's role in quantum mechanics.


	
	\section{Generalization of the EWKB method}\label{Section: Setup_AiryWeber}
	In this section, we discuss the general formulation of  both Airy and Weber types of EWKB analyses. The setup of both approaches are well known. Therefore, the main purpose of this section is emphasizing the changes we made to serve our purpose of exact quantization at all sectors. However, we also introduce both setups briefly for clarity and completeness of the discussion. 
	
	As we mentioned before, although the EWKB method is applicable to general ordinary differential equations, in this paper we only focus on $2^\nd$ order Schrödinger type equation:
	\begin{equation}
		\left(-g^2 \frac{\mrmd^2}{\mrmd x^2} + P(x,u,g) \right) \f(x) = 0 \, ,\label{SchrodingerEquation_Generic}
	\end{equation}
	where $P(x,u,g)$ is a polynomial of $g$:
	\begin{equation}
		P(x,u,g) = \sum_{n} P_n(x,u) g^n \, .
	\end{equation}
	Moreover, we only consider the polynomial up to order $O(g)$, i.e. $P_{n \geq 2} = 0$. Specifically, for the Airy-type analysis, we set
	\begin{equation}
		P_0(x,u) = 2\left(V(x) - u\right)\, , \quad P_{1}(x,u) = 0\,, \label{CurveTerms_Airy}
	\end{equation}
	and for the Weber-type analysis, we set
	\begin{equation}
		P_0(x,u) = 2\left(V(x) - u_0\right)\, , \quad P_1(x,u) = 2\tu\, . \label{CurveTerms_Weber}
	\end{equation}
	Finally, we assume the classical potential possess only locally harmonic saddle points on the real line, i.e. $V(x) \simeq \pm \frac{1}{2}\left(x - x_0\right)^2 + \dots$ .
	
	Note that despite the apparent differences in $P(x,u,g)$, for locally harmonic potentials, both setups are in fact equivalent to each other when the quantization of \eqref{SchrodingerEquation_Generic} is considered. The only difference in two cases is simple rescaling of the parameter $u$ as $u \rightarrow u_0 + g \tu$, which makes $\tu$ a dimensionless parameter, and prepares the setup for Weber-type analysis. We clarify the last point in Section \ref{Section: WeberTypeEWKB}. 
	
	The relationship between the Airy-type and Weber-type setups enables us to use either one in the quantization of \eqref{SchrodingerEquation_Generic} and they provide equivalent but complementary information about the quantum system. On the one side, the Airy-type approach helps us to construct exact quantization conditions in terms of the quantum actions of the curve $P(x,u,g)$, which was discussed in detail in \cite{Sueishi:2020rug,Sueishi:2021xti,Kamata:2021jrs}. For our purposes, the Airy-type approach is also crucial to explicitly understand the continuous transition between different sectors of a given classical potential $V(x)$. 
	The Weber-type EWKB, on the other hand, is useful in a more quantitative manner as it provides explicit expressions for the quantum actions. For our purposes, the Weber-type approach helps us to obtain the explicit expressions for the quantum actions in all sectors and establish the relationship between the actions in different sectors which indicates an $S$-duality from the EWKB point of view.
	
	We finally note that despite the Weber-type EWKB's quantitative power, in computations of the non-perturbative actions, we realize that Airy-type integrals becomes more convenient. So, when we discuss the details of the non-perturbative actions, which we obtain via the Weber-type approach, we return to the Airy-type integrals. This is allowed by virtue of the relationship between the two approaches which is carried to the WKB expansions \cite{DP1,Sueishi:2021xti}.

	\subsection{Airy-type approach}\label{Section: AiryTypeEWKB}

	\paragraph{\underline{Overview}:}
	We first start with the Airy-type EWKB method. In the rest of the paper, we identify the curve of the Airy-type approach as
	\begin{equation}\label{Airy_Curve_General}
		P_\mrmA(x,u) = 2\left(V(x) - u\right)\,,
	\end{equation}
	for a given classical potential $V(x)$ and express the Schr\"odinger equation as
	\begin{equation}\label{SchrodingerEquation_Airy}
		\left(-g^2 \frac{\mrmd^2}{\mrmd x^2}  +  P_\mrmA(x,g,u)\right) \psi(x) = 0,
	\end{equation}
	where we picked $\psi$ to identify the wave function in the Airy-type EWKB.
	
	The Airy-type approach is very well studied for the below-barrier-top sectors of $V(x)$ \cite{DDP2, Sueishi:2020rug,Sueishi:2021xti,Kamata:2021jrs} and it was shown that it is in complete agreement with the other methods.
	In this subsection, we present a novel approach to uncover the transition to the above-barrier-top sectors. In order to be precise in  our discussion, we only focus on the inverted harmonic oscillator and postpone the \textit{exact} quantization of more generic potential to Sections \ref{Section: PeriodicPotential} and \ref{Section: DoubleWell}. Note that despite the simplicity of the inverted harmonic oscillator, since we are interested in potentials with locally harmonic saddles, it provides a great approximation for the regions around these saddles. Therefore, our discussion in this section generalizes for generic locally harmonic potentials in a straightforward manner.
	
	Since the setup of the Airy-type EWKB is well-known and thoroughly discussed in the literature, we do not present it in details and instead, emphasize differences we introduce here. In order to understand the transition between different sectors, we need to modify the setup slightly by introducing a complexification to the spectral parameter $u$. Note that this complexification is analogous to $u \rightarrow u \pm i\ve$ which we discuss in Section \ref{Section: ChebyshevTransitionDuality} and helps us to avoid the singularities associated to the top of a barrier\footnote{Another singular point in Airy-type approach corresponds to the bottom of a well. Once $u$ reaches a singular point, the geometry of Stokes diagrams changes discontinously. The new type of Stokes diagrams subject of the Weber-type EWKB which we discuss in Section \ref{Section: WeberTypeEWKB}.} while transitioning from one sector to another one.

	\paragraph{\underline{Setup}:}
	
	Let us first quickly review the setup of Airy-type EWKB and set the stage before discussing the modifications we make. We start with the WKB ansatz:
	\begin{equation}\label{AiryAnstatz}
		\psi^\pm(x) = \exp\left\{\pm \int^x  \mrmd x'\, s(x',g)\right\}\, ,
	\end{equation}
	where 
	\begin{equation}
		s(x) = \sum_{m=-1}^{\infty} s_m(x)\,g^m .
	\end{equation} 
	Inserting the ansatz in  \eqref{SchrodingerEquation_Airy}, we get the Riccati equation
	\begin{equation}
		s^2(x,g) + g s'(x,g) = P_\mrmA(x,u).
	\end{equation}
	Then, solving it order by order in $g$, we observe that
	\begin{equation}
		\hat s = -\frac{1}{2}\frac{\mrmd}{\mrmd x}\, \log \ts,
	\end{equation}
	where we defined
	\begin{equation}
		\hat s = \sum_{m=0}^{\infty} s_{2m}\,g^{2m}\, , \qquad \ts = \sum_{m=0}^\infty s_{2m-1}\, g^{2m-1}\, , \label{WKB_Expansions_Airy}
	\end{equation}
	and consequently, $\psi$ can be expressed as
	\begin{equation}\label{AiryAnsatz_OddTerms}
		\psi^\pm(x) = \frac{1}{\sqrt{\ts(x,g)}} \, \exp\left\{\pm \int^x \mrmd x' \, \ts(x,g) \right\}. \\
	\end{equation}
	For small $g$, \eqref{AiryAnsatz_OddTerms} can be expanded as
	\begin{equation}
		\psi^\pm(x) = \frac{e^{\pm \s_\mrmA(x)}}{\sqrt{\s_\mrmA(x)}}\, \sum_{m=0}^\infty \psi_m^\pm\, g^m\, , \label{FormalSeries_Airy}
	\end{equation}
	where 
	\begin{equation}\label{LeadingOrder_Airy}
		\s_\mrmA(x,u) = \frac{1}{g}\int_{x_0}^{x} \mrmd x'\, \sqrt{P_\mrmA(x,u)}.
	\end{equation}
	Note that in \eqref{LeadingOrder_Airy}, we set the lower boundary of the integral to a (simple) turning point of the classical potential $V(x)$, i.e $P_\mrmA(x_0,u) = 0$, which is also known as a \textit{normalization} point. 
	
	The power of the EWKB analysis stems from the incorporation of the resurgence theory (more specifically Borel summation) at the differential equation level. It is known that the series expansion in \eqref{FormalSeries_Airy} is an asymptotic one \cite{AKT1} and could be handled by Borel summation. We write the Borel summed $\psi$ by the following Laplace integral:
	\begin{equation}
		\psi^\pm(x) = \int_{\mp\s_\mrmA}^{\infty + \Im\left[\mp \s_\mrmA\right]} \frac{\mrmd s\, e^{-s}}{\sqrt{\s_\mrmA}}\; \mcalB\left[\psi^\pm\right](s\pm \s_\mrmA),
	\end{equation} 
	which indicates that when the integral contour hits a singularity of $\mcalB\left[\psi^\pm\right]$, the solution $\psi^\pm$ is discontinuous. On the complex $x$-plane, these discontinuities correspond to curves, called \textit{Stokes lines}, which are given by the following condition:
	\begin{equation}\label{StokesLineCondition_Airy}
		\Im \, \s_\mrmA  = 0 .  
	\end{equation} 
	
\begin{figure}[t]
		\centering
		\begin{subfigure}[h]{0.6\textwidth}
			\vspace{0pt}
			\caption{\underline{Stokes Lines and branch cut.}}	\label{Figure: Monodromy}
			\includegraphics[width=\textwidth]{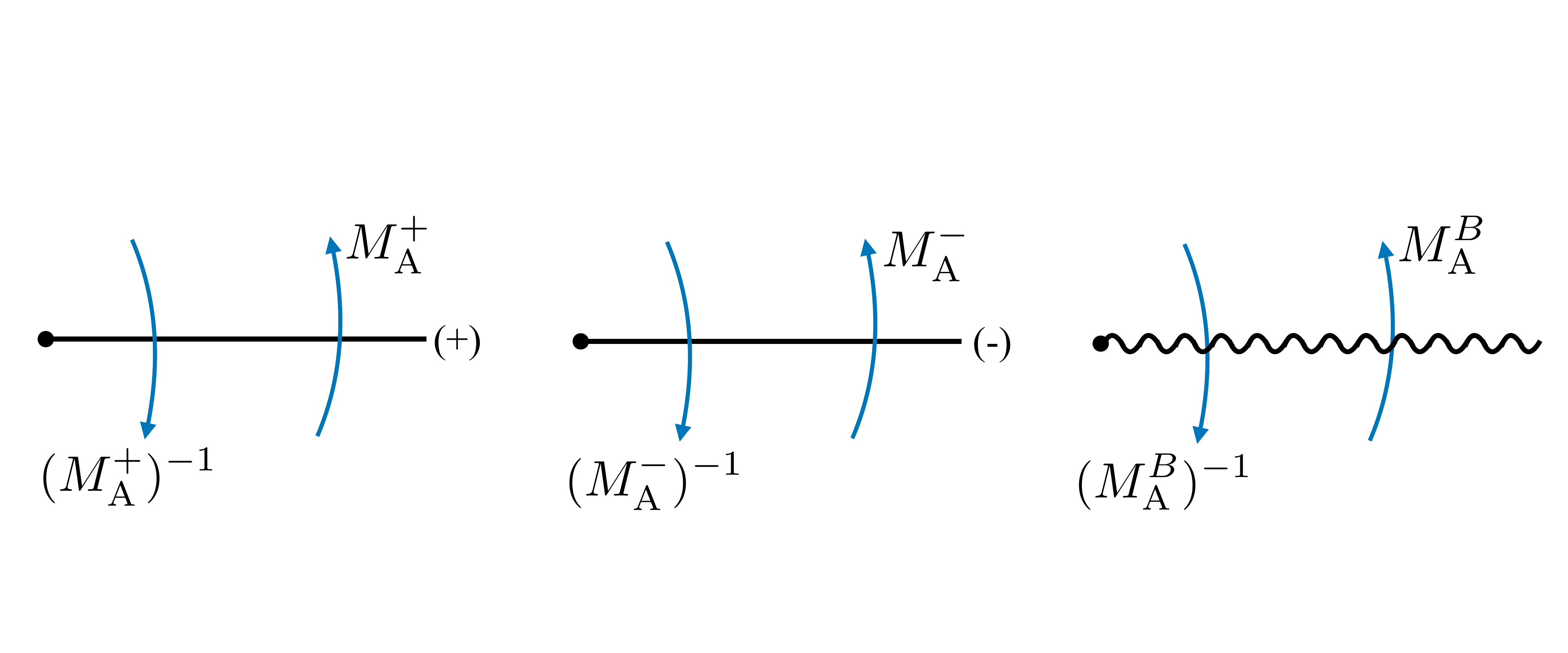}
		\end{subfigure}
		~ \hfill 
		\begin{subfigure}[h]{0.3\textwidth}
			\caption{\underline{A sample diagram.}}	\label{Figure: SampleDiagram}
			\vspace{10pt}
			\includegraphics[width=\textwidth]{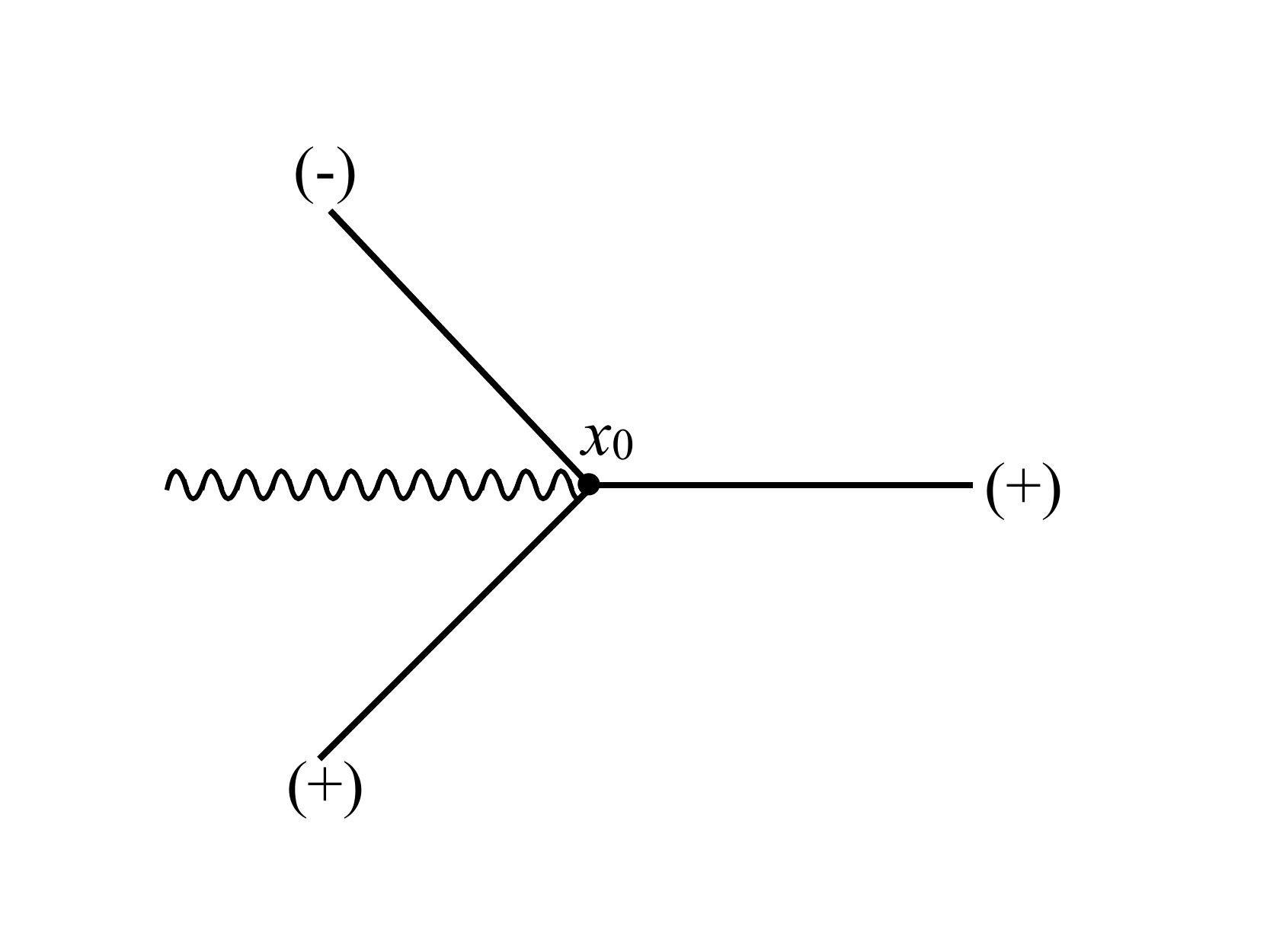}	
		\end{subfigure}
		\caption{Airy-type Stokes lines (solid lines), branch cut (wavy lines) and a sample diagram for the Airy-type approach. The monodromy matrices in \eqref{Monodromy_Airy} represent the discontinuities on the Stokes lines and branch cuts as passing in counter-clockwise direction.} \label{Figure: StokesLines_Diagram}
	\end{figure}

	On a Stokes line, one of the solutions, i.e. $\psi^+$ or $\psi^-$, is discontinuous. The set of solutions on both sides of a Stokes line is connected as
	\begin{equation}
		\begin{pmatrix}
			\psi^+_{\mrmI\mrmI} \\
			\psi^-_{\mrmI\mrmI}
		\end{pmatrix} =  \mrmT	\begin{pmatrix}
			\psi^+_{\mrmI} \\
			\psi^-_{\mrmI}
		\end{pmatrix}\, ,
	\end{equation}
	where $\mrmT$ is one of the following monodromy matrices depending on the cases described in Fig.~\ref{Figure: Monodromy}:
	\begin{equation} \label{Monodromy_Airy}
		M_\mrmA^+ = \begin{pmatrix}
			1  & i \\
			0 & 1
		\end{pmatrix} ,  \quad 
		M_\mrmA^- = \begin{pmatrix}
			1  & 0 \\
			i & 1
		\end{pmatrix} , \quad 	
		M_\mrmA^\mrmB = \begin{pmatrix}
			0  & -i \\
			-i & 0
		\end{pmatrix} \, .
	\end{equation}
	Note that last matrix is due to the square root branch cut of the term $\sqrt{P_\mrmA}$ and it governs the exchange of the solutions $\psi^\pm$. While constructing the Stokes diagram of a given curve $P_\mrmA$, the branch cuts are put by a chosen convention. See Fig.~\ref{Figure: SampleDiagram} for an example. We discuss the equivalence of different branch cut choices in Appendix~\ref{Section: BranchCut_Appendix}.
	
	Finally, we introduce a matrix (also called Voros matrix) for changing the normalization point when there are more than one turning point
	\begin{equation}\label{Airy_NormalizationChange}
		N_\mrmA^{a,b} = \begin{pmatrix}
			e^{\mcalV_{a,b}}  & 0 \\
			0 & e^{-\mcalV_{a,b}}
		\end{pmatrix} \, ,
	\end{equation}
	where $\mcalV$ is given by the integrals between the turning points $a$ and $b$: 
	\begin{equation}
		\mcalV_{a,b} = \int_{a}^{b} \mrmd x\, \ts(x,g)\,, \label{Airy_HalfCycle}
	\end{equation}
	which corresponds to the integral over half  cycles which are associated to the quantum actions. 

	\paragraph{\underline{Analytic continuation of Stokes diagrams}:}
	When a potential $V(x)$ has two or more turning points, it is possible that a Stokes line emerging from a turning point ends at another one, which indicates the degeneracy of Stokes lines\footnote{Note that a degenerate Stokes diagram is sometimes called as merging turning points in the literature \cite{AKT4,Takei3,Sueishi:2021xti}. In this paper, we stick with the term ``degenerate'' as merging turning points also means their coalescence at the saddle points of a given potential, which is the subject of the Weber-type EWKB approach.}. The degeneracy signals a need of analytic continuation\footnote{Analytic continuations which break the degeneracy are related to each other via so-called DDP formula \cite{DDP2,DP1} which is deeply related to the Borel summation process \cite{Kamata:2023opn}. We utilize it in Sections \ref{Section: PeriodicPotential} and \ref{Section: DoubleWell} to obtain the median exact quantization conditions which reveals manifestly real spectra.} 
	which deforms the Stokes diagram so that two turning points would not be connected by a Stokes line. In the literature, in general, the analytic continuation of the Stokes diagrams are handled by a deformation of the parameter $g$. For example, if the condition \eqref{StokesLineCondition_Airy} is satisfied when $\arg g =0 $, then introducing a small imaginary part to $g$ would deform the Stokes diagrams and break the degeneracy. 
	
	\begin{figure}[t]
		\centering
		\includegraphics[width=0.85\textwidth]{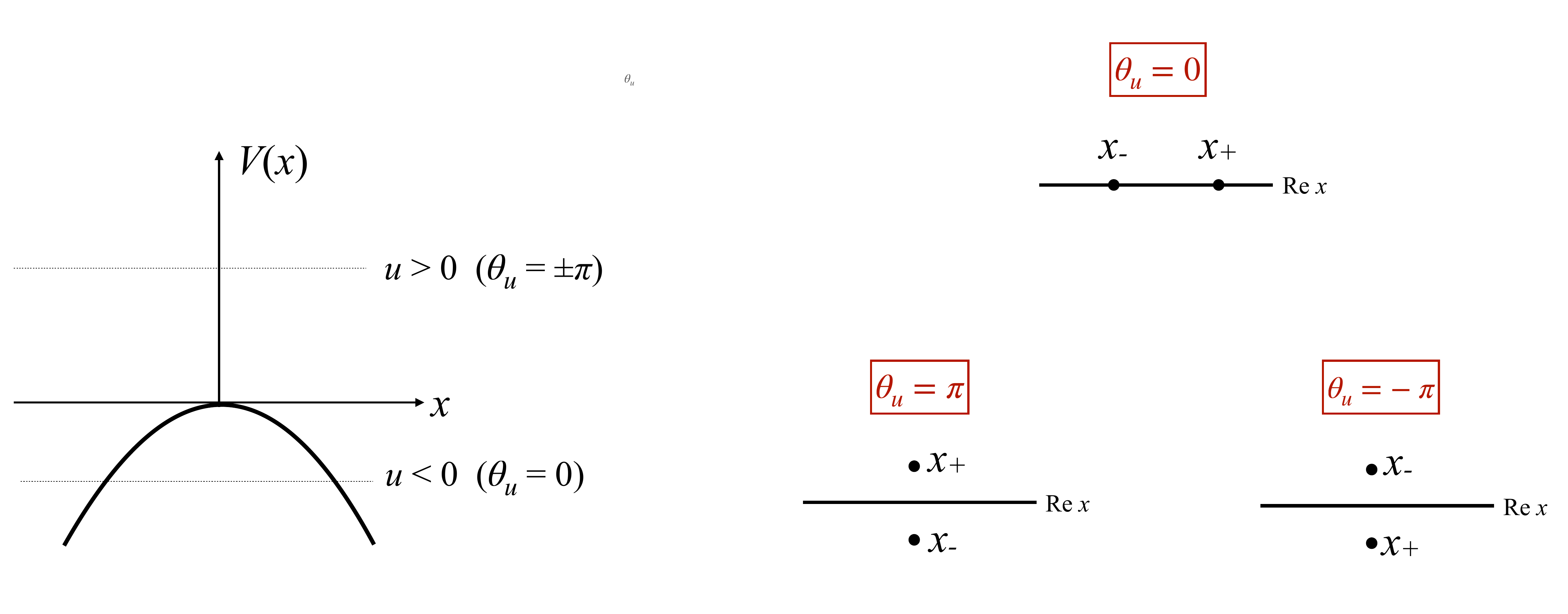}
		\caption{Inverted harmonic oscillator and associated turning points for $u<0$ (equivalently $\t_u=0$)~and $u>0$~(equivalently $\t_u=\pm\pi$).~Note that for $\t_u=\pm \pi$, the locations of the turning points $x_\pm$ reversed, which indicates two different analytic continuations.~(See Fig.~\ref{Figure: IHO_Transition} for its effect on the Stokes geometry.)~This is also the case for generic locally harmonic potentials and we encounter with it when we discuss periodic and symmetric double-well potentials in Sections \ref{Section: PeriodicPotential} and \ref{Section: DoubleWell}.} \label{Figure: TurningPoints_IHO_BelowAbove}
	\end{figure}
	
	For our purposes, on the other hand, we analytically continue the parameter $u$ by introducing a phase as $u = |u|e^{i\t_u}$. This is equivalent to introducing an imaginary part to $g$. In order to see this, let us investigate $\s_\mrmA$ for an inverted harmonic oscillator. Without losing any generality we set the origin at $x=0$ and $u=0$. Then, introducing a phase for the parameter $u$, we write $P_\mrmA$ as
	\begin{equation} 
		P_\mrmA = 2\left(-x^2 + |u| e^{i \t_u}\right)\,, \label{Airy_Curve_IHO}
	\end{equation}
	which admits (complex) turning points, i.e. $x_\pm = \pm e^{\frac{i \t_u}{2}} \sqrt{|u|}$ as plotted for specific values $\t_u=0, \pm \pi$ in Fig.~\ref{Figure: TurningPoints_IHO_BelowAbove}. 
	
\begin{figure}[t]
		\centering
		\includegraphics[width=0.64\textwidth]{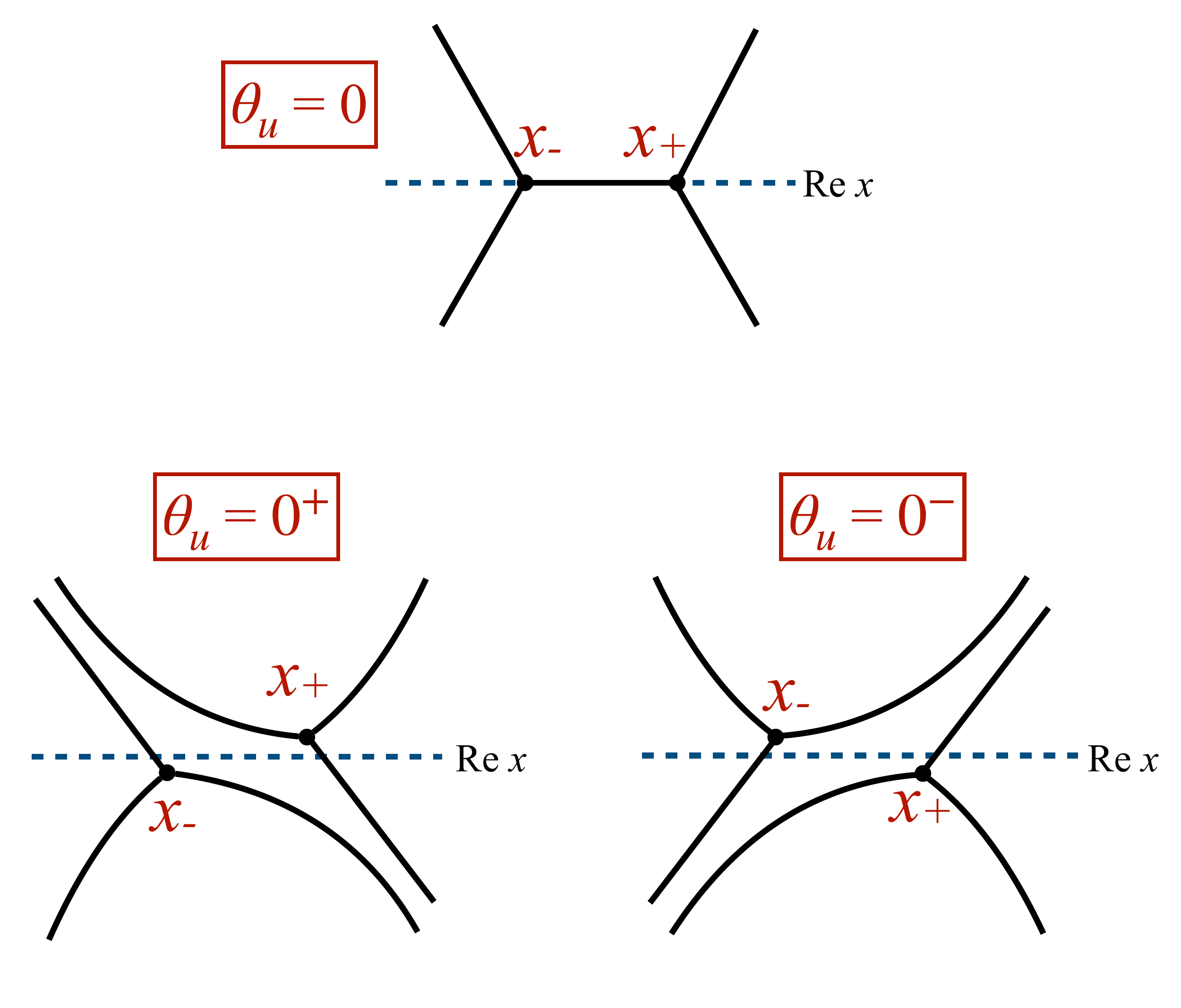}
		\caption{Degeneracy breaking of a Stokes diagram due to the phase of the parameter $u$.} \label{Figure: DegeneracyBreaking}
	\end{figure}

	Around turning points $x=x_\pm$, $\s^\pm_\mrmA$ are approximated as
	\begin{align}
		\s^\pm_\mrmA & \simeq \mp \frac{4}{3} \frac{\left|u \right|^{1/4}}{|g|}\frac{e^{\frac{i \t_u}{4}}}{e^{i \t_g}} \left[\mp\left(x - x_\pm\right) \right]^{3/2} + O\left(\left(x - x_\pm\right)^{5/2}\right)\, , \label{Airy_AroundTurningPoint}
	\end{align}
	where we introduced a phase for $g$. Equation \eqref{Airy_AroundTurningPoint} manifestly shows that phases $\t_u$ and $\t_g$ deform the Stokes diagrams in opposite directions and both could be used to break the degeneracy of Stokes diagrams. For example around $\t_u = 0$, the deformation and degeneracy breaking of the Stokes diagram for the curve \eqref{Airy_Curve_IHO} is depicted in Fig.~\ref{Figure: DegeneracyBreaking}. Note that the geometries of the diagrams for $\t_u = 0^\pm$ are the same as the ones for $\t_g = 0^\mp$. 
	
	An important difference between changing the phases $\t_g$ and $\t_u$ is that the latter also rotates the turning point. This is crucial for our purposes: As we describe in Fig.~\ref{Figure: TurningPoints_IHO_BelowAbove}, phases $\t_u=0$ and $\t_u = \pm \pi$ correspond to the sectors $u<0$ and $u>0$ for the inverted oscillator. We utilize this property to \textit{continuously} connect these two sectors and it is crucial for our discussions in Sections \ref{Section: PeriodicPotential} and \ref{Section: DoubleWell}.
	
\begin{figure}[t]
		\centering
		\begin{subfigure}[h]{0.48\textwidth}
			\caption{\underline{Transition for $x=x+$.}}
			\vspace{2pt}
			\includegraphics[width=\textwidth]{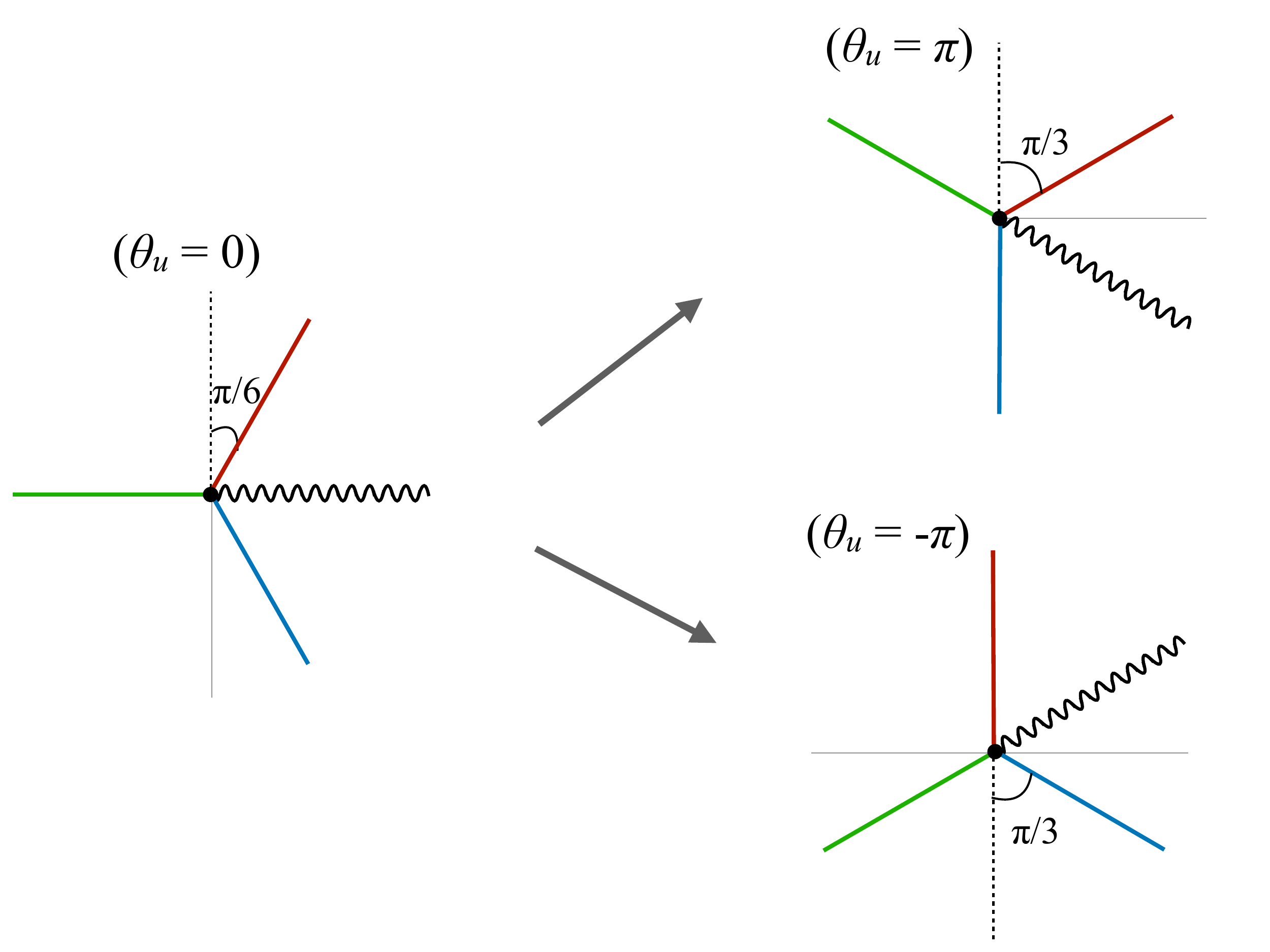}
			\label{Figure: StokesDiagramRotation1}
		\end{subfigure}
		\hfill 
		\begin{subfigure}[h]{0.48\textwidth}
			\caption{\underline{Transition for $x=x_-$.}}
			\vspace{2pt}
			\includegraphics[width=\textwidth]{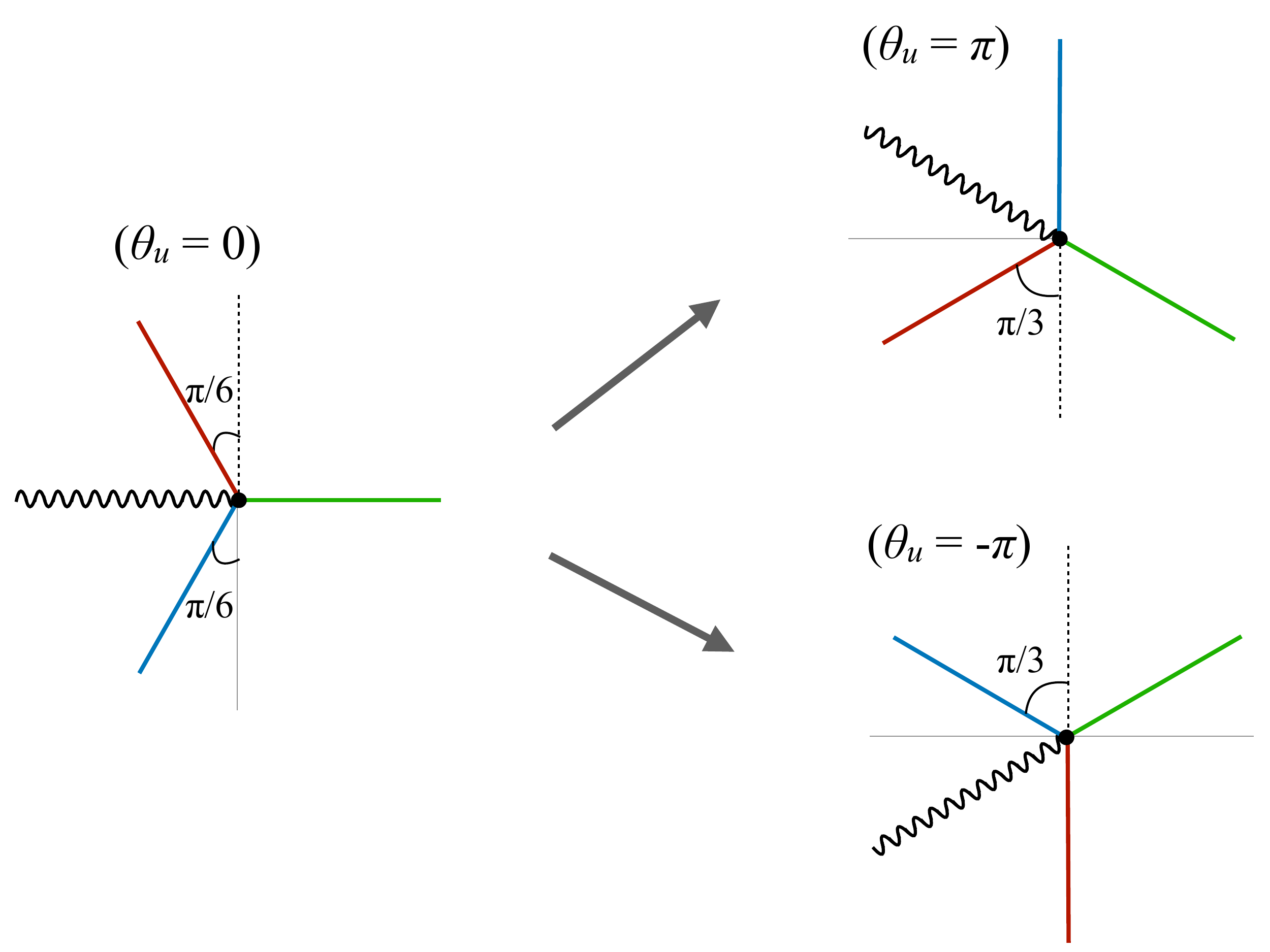}
			\label{Figure: StokesDiagramRotation2}
		\end{subfigure}
		\caption{The transition of Stokes lines associated with the turning points $x=x_\pm$ by the phase rotation $\t_u = 0 \rightarrow \pm \pi$.} \label{Figure: StokesDiagramRotation_All}
	\end{figure}

	Let us first investigate the effect of altering $\t_u$ on the Stokes diagrams emerging from the turning points $x_\pm$: We define the angles, i.e. $\t_\pm$, between an individual Stokes line in each case and positive real axis as
	$x - x_\pm = e^{i \t_\pm} \left|x- x_\pm \right|$.  
	\begin{itemize}
		\item \textbf{\underline{Around $\bm {x_+}$}}: From \eqref{Airy_AroundTurningPoint}, we get the Stokes lines along the directions of
		\begin{equation}
			\t_+ = \frac{2(n-\frac{1}{2})\pi}{3} - \frac{\t_u}{6}, \quad n \in \mbbZ \, , \label{Airy_StokesLine_Generic+}
		\end{equation}
		which shows that when the phase of $u$ changes by $\t_u$, a Stokes line normalized at $x=x_+$ rotates by $\frac{\t_u}{6}$. Then, entire Stokes diagram rotates by $\mp \frac{\pi}{6} $ when $\t_u = 0 \rightarrow \pm \pi$ as described in Fig.~\ref{Figure: StokesDiagramRotation1} where we picked $n=0,1,2$.	
		\item \textbf{\underline{Around $\bm{x_-}$}:} A similar analysis shows us that the Stokes lines emerging from $x_-$ lies along the directions 
		\begin{equation}
			\t_- = \frac{2n \pi}{3} - \frac{\t_u}{6}, \quad n \in \mbbZ\, , \label{Airy_StokesLine_Generic-}
		\end{equation}
		and when $\t_u = 0 \rightarrow \pm \pi$, the entire Stokes diagram rotates as in Fig.~\ref{Figure: StokesDiagramRotation2} where we picked $n=-1,0,1$. 
	\end{itemize}

	Note that, our choices of the values of $n$ for the Stokes diagrams in Fig.~\ref{Figure: StokesDiagramRotation_All} determine which solution is discontinuous on the associated Stokes line. We point out this with the $(\pm)$ signs on the diagrams. Moreover, we placed the branch cuts along the directions corresponding to $\t_+ = 0$ and $\t_- = \pi$ for the diagrams associated to the turning points at $x=x_+$ and $x=x_-$ respectively. Other branch cut conventions are equivalent to our choice with appropriate changes in the signs of individual Stokes lines, i.e. picking different values of $n$ in \eqref{Airy_StokesLine_Generic+} and $\eqref{Airy_StokesLine_Generic-}$. In the rest of this paper, unless otherwise stated, we continue using the conventions in Fig.~\ref{Figure: StokesDiagramRotation_All}. 
	
\begin{figure}[t]
		\centering
		\includegraphics[width=0.9\textwidth]{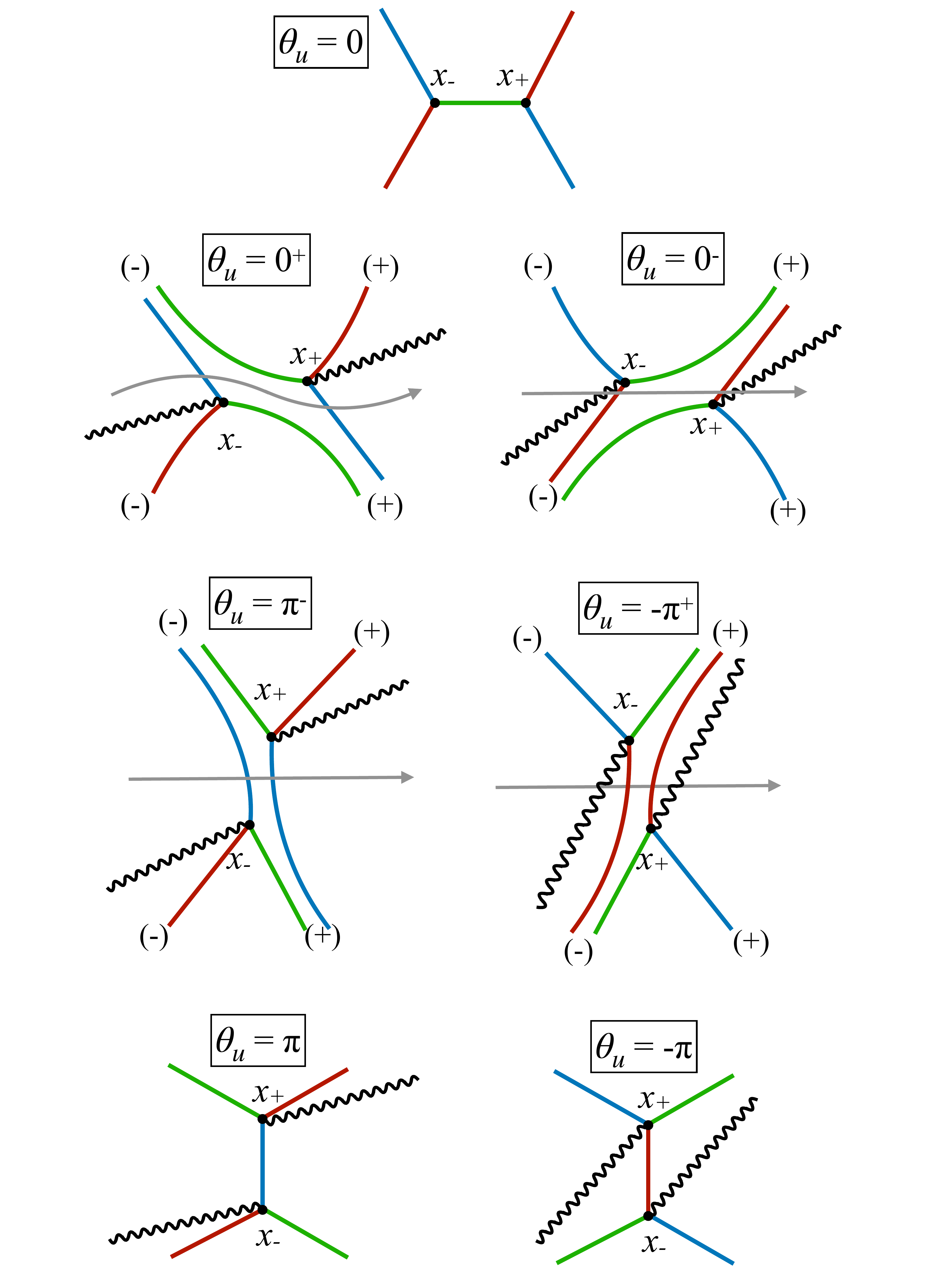}
		\vspace{10pt}
		\caption{The analytic continuations of the Stokes diagram while the phase changing as $\t_u= 0 \rightarrow \pm\pi$. Note that all the diagrams are originated from the same degenerate diagram at $\t_u=0$ and two possible analytic continuations lead to different branch cut conventions at $\t_u = \pm \pi$. Throughout the transitions, except $\t_u=0$ and $\t_u = \pm \pi$, there is no Stokes lines degeneration which implies the deformation is a continuous one.} \label{Figure: IHO_Transition}
	\end{figure}

	Finally, we present the transition of the Stokes diagram of Fig.~\ref{Figure: TurningPoints_IHO_BelowAbove} from the sector $u<0$ to $u>0$. Two possible analytic continuations corresponding to the transitions $\t_u= 0^+ \rightarrow \pi^-$ and $\t_u = 0^- \rightarrow -\pi^+$ are depicted in Fig.~\ref{Figure: IHO_Transition}, which clearly shows the continuous transition between the two sectors since no Stokes line or branch cut become degenerate throughout the rotation. Note that although the Stokes lines are again degenerated at $\t_u = \pm \pi$ in the same way, the difference in the branch cuts' orientation indicates that two Stokes diagrams are not identical\footnote{This becomes an important aspect of the analytic continuation we introduce in this section, when we discuss more general potentials in Sections~\ref{Section: PeriodicPotential} and \ref{Section: DoubleWell}. We also elaborate the relationship between different branch cut conventions in Appendix~\ref{Section: BranchCut_Appendix}.}.

	The continuous connection between $u<0$ and $u>0$ sectors shows itself also in the transition matrix connecting the same two points in complex $x$-plane. For example, for all cases of $\t_u \in \left(0,\pi\right)$, the transition matrix connecting $x = - \infty$ to $x = +\infty$ is written as
	\begin{equation}
		T_{\t_u\in \left(0,\pi\right)} = M_+ N_{12}M^{-1}_-\, .  \label{Airy_TransitionMatrix_IHO+}
	\end{equation}
	For $\t_u \in \left(-\pi, 0\right)$, there is again only one transition matrix connecting $x=-\infty$ to $x = +\infty$:
	\begin{equation}
		T_{\t_u \in \left(-\pi, 0\right)} = M^{-1}_\mrmB M^{-1}_+ N_{12}M_-M_\mrmB\, . \label{Airy_TransitionMatrix_IHO-}
	\end{equation} 
	Note that from the geometric point of view, the continuous connection can be observed from the fact that the gray lines in Fig.~\ref{Figure: IHO_Transition} pass through the same Stokes lines for all cases.

	\paragraph{\underline{Remark}:}These analytic continuation arguments can easily be generalized for the transition across the barrier top of any locally harmonic potentials. To see this, let us consider the Airy-type curve of around a harmonic barrier top at $x = x_0$ associated to the energy level $u = u_\mrmT$:
	\begin{equation}\label{Airy_Curve_BarrierTop}
		P_\mrmA = 2 \left(V(x) - u_\mrmT + |\d| e^{i\t_\d}\right) \simeq 2\left(-\frac{1}{2}(x-x_0)^2 + |\d|e^{i\t_\d}\right)\, ,
	\end{equation}
	where we used the expansion of the potential at $x=x_0$. Note that \eqref{Airy_Curve_BarrierTop} is nothing but \eqref{Airy_Curve_IHO} with the saddle point at $x=x_0$. Therefore, the quantization of all potentials with locally harmonic degenerate saddle points is encoded by a single transition matrix. As a result, for each analytic continuation across the barrier top, there is a single quantization condition for all sectors.
	Before discussing the details of the transition quantization conditions and the relationship between different analytic continuations, we first present the Weber type EWKB  which provides us a complementary quantitative picture in the next subsection.

	\subsection{Weber-type approach} \label{Section: WeberTypeEWKB}
	
	\paragraph{\underline{Overview}:}
	In this paper, we use the Weber-type approach as a complementary one to the Airy-type EWKB. Despite the equivalence of both approaches, their Stokes geometries are quite different. As we stated before, the main difference between the two approaches is the rescaling in the function $P(x,g,u)$. Due to this rescaling, the classical part $P_0$ in \eqref{CurveTerms_Weber} and the associated turning points, which are given by $P_0(x,u_0) = 0$, are independent of the parameter $u$. In the Weber-type EWKB, we also set $u_0$ such that $P_0 = 0$ is satisfied by the saddle points of the classical potential $V(x)$. As a result of this, the Stokes lines in the Weber-type EWKB originate from the saddle points which are \textit{double} turning points rather than simple ones which we encountered in the Airy-type EWKB.  
	
	Another important consequence of the rescaling in the Weber-type setup is that it prevents us from observing the continuous change between different sectors as in the Airy-type EWKB. On the other hand, it provides a very powerful quantitative tool to uncover the explicit expressions of the perturbative and non-perturbative parts WKB cycles, which can be very cumbersome to obtain via the Airy-type approach. The quantitative power of the Weber-type approach arises from the locally harmonic behaviour of the saddle points. By the means of a set of equations \cite{AKT1, Sueishi:2021xti}, this property enables a construction of a mapping from \textit{local} coordinates where the harmonic oscillator is formulated to \textit{global} coordinates where locally harmonic potentials are formulated.   
	
	So far, in the literature, the analysis of one dimensional quantum mechanics in view of the Weber-type EWKB is well-discussed with a focus on the region below the barrier top \cite{Sueishi:2021xti,Kamata:2021jrs,Kamata:2023opn,Kamata:2024tyb}. In these cases, the information about the WKB cycles arises from the minima of the potential: The perturbative part is given by a residue integration around the saddle point. The non-perturbative part, on the other hand, is more intricate; but it is possible to obtain a generic expression, 
	which can be separated into two parts: One originates from the locally harmonic saddle point and carries the \textit{local} information. The other one arises from the connection of successive turning points and can be identified as a \textit{non-local} expression. 
	
	The analysis of barrier top region with the Weber-type EWKB, on the other hand, is much less discussed in the literature. In highly sophisticated mathematical papers \cite{DDP2,DP1}, Delabaere, Dillinger and Pham investigated the Weber-type EWKB around all type of locally harmonic saddles including the barrier tops. In this subsection, we approach the problem from a different point of view\footnote{This difference was also emphasized in \cite{DP1} where the authors call their method uses the transformations of resurgent functions while the constructions in \cite{AKT1,AKT4} are based on micro-differential analysis which we utilize in this paper.} based on \cite{AKT1,AKT4} (see also \cite{Kawai1}), which was applied to the quantum mechanical problems in \cite{Sueishi:2021xti,Kamata:2021jrs}. We also provide a construction from a physicist point of view with a connection to the relevant physics literature.

	In the following, we follow the construction in \cite{Sueishi:2021xti} and generalize it to all types of locally harmonic saddles, i.e.~both minimum and maximum on the real line, of a given 1D potential. Our main purpose is obtaining the generic expressions for perturbative and non-perturbative WKB actions around both the bottom of wells and the top of barriers. Previously, this was done for non-perturbative cycles connecting two degenerate minima in \cite{Sueishi:2021xti}, where the non-perturbative action was expressed as\footnote{The notation will be clarified later in this section.}
	\begin{equation}\label{NP_cycle_Weber_2SaddleConnecting}
		e^{\frac{2\pi i}{g} \aD(\tu,g)} = \mcalK(\tu,g) \, \frac{2\pi }{\left( \G \left(\frac{1}{2} + \mcalF(\tu,g)\right) \right)^2}\, g^{-2 \mcalF(\tu,g)}. 
	\end{equation}
	Note that in \cite{DP1}, the application of Weber-type EWKB to the maxima of double-well potential was discussed where the authors computed $\mcalF$ and $\mcalK$ order by order in $g$ for the double-well potential and inferred the prefactor $\mcalK$ from these expansions. 
	
	In our discussion, on the other hand, we provide generic expressions of the $\mcalK$ factor in \eqref{NP_cycle_Weber_2SaddleConnecting} under general assumptions about the turning points that the Weber-type WKB cycles connect. Note that for the saddles at the bottom of the wells, it  was first obtained by Zinn-Justin and Jentschura (ZJJ) in relation with the multi-instanton (both bions and bounces) configurations \cite{Zinn-Justin:2004vcw, Zinn-Justin:2004qzw,Jentschura:2010zza}. In this section, we recover them by the Weber-type EWKB and generalize them to all saddles of a given potential. With this generalization, we observe the exchange of the perturbative and non-perturbative cycles around the maxima of the potential in comparison to its minima. This manifestly points out an $S$-duality between minima and maxima of a classical potential from the EWKB point of view.
	

	Finally, note that in the following discussion, we also handle the Weber-type EWKB analysis in a slightly different way than \cite{Sueishi:2021xti}. Specifically, we keep the normalization of the formal WKB solutions at the turning points rather than setting it to infinity as in \cite{Sueishi:2021xti}. We observe that our setup simplifies the analysis of the non-perturbative cycles and differentiates the local and non-local parts in a more explicit manner.

	\paragraph{\underline{Local vs global}:}
	As the name suggests, the Weber-type EWKB is originated from the Weber equation. For our purposes, it represents the \textit{local} behaviour around saddle points and its EWKB analysis is linked to the EWKB analysis of the locally harmonic potentials via \textit{local-global} relations. 
	
	Throughout this section, we express the corresponding Schrödinger equations in \textit{local} and \textit{global} coordinates as in the following:
	\begin{itemize}
		\item \underline{Local coordinates}:
		\begin{equation}\label{WeberEquation}
			\left(-g^2 \frac{\mrmd^2 }{\mrmd y^2}  + Q(y,\tu)\right) \Psi_\mrmL(y) = 0,
		\end{equation}
		where 
		\[Q(y,\tu) = \frac{\o^2 y^2}{4} - g \tu \,, \] 
		and $\o$ is a general complex parameter. Note that in this paper, we are interested in the special cases $\o^2 =\pm 1$, which correspond to harmonic and inverse harmonic oscillators. However, for the sake of the generality of our discussion, we keep $\o$ as an arbitrary parameter until we need the specific information about harmonic minima or maxima. 
		
		\item \underline{Global Coordinates}:
		
		\begin{equation}\label{SchrodingerEquation_WeberGlobal}
			\left(-g^2 \frac{\mrmd^2}{\mrmd x^2}  + P_\mrmW(x,g,\tu)\right) \Psi_\mrmG(x) = 0\, , 
		\end{equation}
		where
		\begin{equation}\label{Weber_Curve_General}
			P_\mrmW(x,g,\tu) = 2\left(\tV(x)  + g \tu\right)\, .
		\end{equation}
		Note that  $\tV(x) = V(x) - u_0$ is the rescaled classical potential, i.e. $O(g^0)$ term in \eqref{CurveTerms_Weber}. Moreover, we assume around a saddle point $x=x_0$, $V(x)$ behaves as
		\begin{equation}\label{LocallyHarmonicPotential}
			V(x) \simeq \frac{\o^2}{2} (x - x_0)^2 + O((x - x_0)^3) \dots\,, \; 
		\end{equation} 
		and we set $u_0$ such that $V(x_0) = 0$.
	\end{itemize}

	At the formal level, the Weber-type analysis follows the same route with the Airy-type EWKB: We start with the formal WKB ansatz: 
	\begin{equation}\label{WeberAnstatz}
		\Psi^\pm_\mrmL(y) = \exp\left\{\int^y \mrmd y'\, \O^\pm(y',g)\right\}\, , \qquad 	\Psi^\pm_\mrmG(x) = \exp\left\{\int^x \mrmd x'\, S^\pm(x',g)\right\}\,,
	\end{equation}
	and via the Riccati equation, we get
	\begin{align}\label{WeberAnsatz_Decomposed}
		\Psi_\mrmL^\pm(y)
		&= \exp\left\{\mp \frac{1}{2}\int_{0}^{y}\mrmd \log \ttO(y',g)  \right\}\, \ttP_\mrmL(y)\, ,  \\
		\Psi_\mrmG^\pm(x) &= \exp\left\{\mp \frac{1}{2}\int_{x_0}^{x}\mrmd \log \tS(y(x'),g)  \right\} \, \ttP^\pm_\mrmG(x)\, , \label{WeberAnsatz_Decomposed_Global}
	\end{align}
	where we defined
	\begin{equation}\label{WeberAnsatz_OddTermPart}
		\ttP_\mrmL(y) =  \exp\left\{\pm \int_0^y\mrmd y'\, \ttO(y,g) \right\} ,\qquad  	\ttP^\pm_\mrmG(x) = \exp\left\{\pm \int_{x_0}^y\mrmd x'\, \tS(x,g) \right\} \, .
	\end{equation}
	The integrands $\ttO$ and $\tS$ are formal power series in $g$:
	\begin{equation}
		\ttO(y,g) = \sum_{n=-1}^{\infty} \ttO_n(y) g^n , \qquad  \tS(x,g) = \sum_{n=-1}^{\infty} \tS_n(x) g^n\, , \label{WKB_Expansions_Weber}
	\end{equation}
	where the first two terms are given as 
	\begin{align}
		\ttO_{-1}(y) = \frac{\o y}{2}&, \qquad \ttO_0(y) = -\frac{\tu}{\o y}\, ,\label{FirstTerms_Weber}\\
		\tS_{-1}(x) = \sqrt{2\tV(x)}&, \qquad \tS_0(x) = -\frac{\tu}{\sqrt{2\tV(x)}} \, . \label{FirstTerms_Global}
	\end{align}
	Higher order terms can be found recursively if needed. Note that for a given potential $V(x)$, the expansions of $\tS$ in the Weber type EWKB and $\ts$ in the Airy type EWKB are equal to each other as $\tS$ can be obtained by expanding $\ts$ for around $u=u_0$. 
	
	Two Schr\"odinger equations and their WKB solutions are related to each others via the following set of equations \cite{Sueishi:2021xti}:
	\begin{align}
		P(x,g) &= \left(\frac{\dee y(x,g)}{\dee x}\right)^2 Q(x,g) - \frac{g^2}{2}\left\{y(x,g);x\right\}\, \label{LocalToGlobal_1}, \\
		S^{(\pm)}(x,g) &= \frac{\dee y(x,g)}{\dee x}\, \O^{(\pm)}(y(x,g),g) - \frac{1}{2}\frac{y''(x,g)}{y'(x,g)}\, \label{LocalToGlobal_2}, \\ 
		\tS (x,g) &= \frac{\dee y}{\dee x} \, \ttO(y(x,g),g)\, \label{LocalToGlobal_3},
	\end{align}
	where 
	\begin{equation}
		\{y;x\} = \frac{y'''(x,g)}{y'(x,g)} - \frac{3}{2}\left(\frac{y''(x,g)}{y(x,g)}\right)^2 \,,
	\end{equation}
	is the Schwarzian derivative and the local coordinate $y$ is written as
	\begin{equation}
		y(x,g) = \sum_{m=0}^{\infty} y_m(x) \, g^m. \label{Expansion_y}
	\end{equation}
	
	Equation \eqref{LocalToGlobal_2}, on the other hand, induces an explicit relationship between the solutions in local and global coordinates
	\begin{equation}
		\Psi_\mrmG^\pm(x) = \frac{ \Psi_\mrmL^\pm(y(x,g))}{\sqrt{y'(x,g)}}\, . \label{GlobalSolution_mapping}
	\end{equation}
	Note that this is an exact map of the solution from local coordinate $y$ to global coordinate $x$. The same expression appears in the \textit{uniform WKB approach} \cite{Dunne:2014bca} where $\Psi_\mrmL(y) = D_\n(y)$ with $D_\n$ being parabolic cylinder equation. Thus, the local-global transition of Weber-type approach is equivalent to the uniform WKB approach. 
	
	Finally, following the arguments in \cite{Kawai1}, we can show that the Stokes discontinuities of local and global solutions are the same. Therefore, it is enough to obtain the monodromy matrices in the local coordinates and the same matrices applies in the global coordinates provided by appropriate coordinate transformations. This equivalence is the source of the ``universal'' behavior of the Stokes discontinuities for generic potentials in global coordinates as noticed before in \cite{DDP2,DP1,Kawai1,Bucciotti:2023trp}. Note that as we will see later, the Stokes discontinuities determine the local structure of non-perturbative cycles which also contains a non-local part. Despite this additional part, we observe that non-perturbative cycles also admit  a ``universal'' structure.
	
	\paragraph{\underline{Analysis of the Weber equation}:}
	
	Let us first focus on the local part and perform a EWKB analysis on the Weber equation. In the following, we briefly review the construction of the Stokes geometry for the Weber equation with generic $\o$. This was discussed in detail in \cite{Sueishi:2021xti,Kamata:2021jrs,Bucciotti:2023trp} for $\o^2=1$. Here, we mainly emphasize the generalization to generic $\o$.

	The formal series solution to \eqref{WeberEquation} is written as
	\begin{equation}
		\Psi_\mrmL^\pm(y) = e^{\pm \frac{\o y^2}{4 g}}\, y^{\mp\frac{\tu}{\o}} \sum_{m=0}^\infty \psi_m^{(\pm)}\, y^{-2n-\frac{1}{2}}\,g^n,
	\end{equation}
	which is known to be factorially divergent. Its Borel summation is given by the following Laplace integral:
	\begin{equation}\label{BorelSummation_Weber_Integral}
		\Psi_\mrmL^\pm(y) = \int_{\mp \s_\mrmW}^\infty \frac{\mrmd s\, e^{-\frac{s}{g}}}{g}\, \frac{2\, y^{-3/2}}{\Gamma\left(\frac{1\pm\tu}{2}\right)}\left(\pm \frac{2g }{\o}\right)^{\frac{1\mp\tu}{2}} \mcalB\left[\Psi^\pm_\mrmL\right](s),
	\end{equation}
	where 
	\begin{equation}
		\s_\mrmW = \int_{0}^y \mrmd y'\, \ttO_{-1}(y') = \frac{\o y^2}{2}\,,
	\end{equation}
	and
	\begin{equation}\label{BorelTransform_Weber}
		\mcalB\left[\Psi^\pm_\mrmL\right] = t_\pm^{-\frac{1}{2}\pm\frac{\tu}{2}}\, _{2}F_{1}\left(\frac{1}{4}\pm\frac{\tu}{2}, \frac{3}{4}\pm\frac{\tu}{2},\frac{1\pm\tu}{2}; t_\pm\right).
	\end{equation}
	Note that we also define a shorthand notation
	\begin{equation}\label{parametrizeArgument_hypergeometric}
		t_\pm = \pm \left(\frac{2 s}{\o y^2} \pm \frac{1}{2}\right).
	\end{equation}
	The integral in \eqref{BorelSummation_Weber_Integral} is well defined as long as the integration contour does not hit the branch point of the hypergeometric function at $t_\pm =1$. On the other hand, the direction that the contour hits the branch cut indicates a discontinuity of the function $\Psi_\mrmL(x)$ from which we construct the monodromy matrices. 
	
	Since the branch point $t_\pm =1$ corresponds to $s=\pm \s_\mrmW$ in $s$-plane, the singular directions of the Borel integral in \eqref{BorelSummation_Weber_Integral} is controlled by $\arg (\o y^2)$. For generic $\o$ and $y$, it is straightforward to see that these singular directions, i.e. Stokes lines, are given by the condition
	\begin{equation}\label{StokesLineCondition_Weber}
		\Im \s_\mrmW = 0,
	\end{equation}
	which, together with \eqref{BorelSummation_Weber_Integral}, implies that
	\begin{itemize}
		\item $\Psi^+_\mrmL$ is not Borel summable when $\arg(\o y^2) = 0,\; 2\pi$,
		\item $\Psi^-_\mrmL$ is not summable when $\arg(\o y^2) = \pm \pi$.
	\end{itemize}
	
	Let us first consider the case $\arg (\o y^2) = 0$ and suppose that for a given $\o$, $\arg y = \t_y$ is the direction of the Stokes line in the $y$-plane. Then, following the arguments in \cite{Sueishi:2021xti}, we get the discontinuity of $\Psi^+_\mrmL$ as
	\begin{equation}\label{BorelDiscontinuity_Weber}
		\D\Psi_{\mrmL,\t_y}^+ (y) = \frac{i \sqrt{2\pi}\left(\frac{g}{\o}\right)^{-\frac{\tu}{\o}}}{\Gamma\left(\frac{1}{2} + \frac{\tu}{\o}\right)} \int_{ \s_\mrmW}^\infty \frac{\mrmd s\, e^{-\frac{s}{g}}}{g}\, \frac{2\, |y|^{-3/2}\, e^{\frac{-3i\t_y}{2}}}{\Gamma\left(\frac{1}{2}-\frac{\tu}{2\o}\right)}\left(- \frac{2g }{\o}\right)^{\frac{1}{2}+\frac{\tu}{2\o}} \mcalB\left[\Psi^-_\mrmL\right](s),
	\end{equation}
	where the integral is $\Psi_\mrmL^-$ in \eqref{BorelSummation_Weber_Integral} for $y = |y|\, e^{i\t_y}$. Then, the discontinuity of $\Psi^+_{\mrmL,\t_y}$ is written as
	\begin{align}
		\D\Psi^+_{\mrmL,\t_y}  =  \frac{i \sqrt{2\pi}\left(\frac{g}{\o}\right)^{-\frac{\tu}{\o}}}{\Gamma\left(\frac{1}{2} + \frac{\tu}{\o}\right)} \, \Psi^-_{\mrmL,\t_y}\;, \qquad \arg(\o y^2) = 0 \,.\label{StokesJump_R+}
	\end{align}

	Other Stokes lines, $\arg(\o y^2) =  \pm \pi$ and $\arg(\o y^2) = 2\pi$, lie on the directions $\t_y  \pm \frac{\pi}{2}$ and $\t_y+\pi$, respectively. The discontinuities on these directions are written as
	\begin{align}
		\D\Psi^+_{\mrmL,\t_y+\pi} &=  e^{-2\pi i \frac{\tu}{\o}} \frac{i\sqrt{2\pi} \left(\frac{g}{\o}\right)^{-\frac{\tu}{\o}}}{\Gamma\left(\frac{1}{2} + \frac{\tu}{\o}\right)} \, \Psi^-_{\mrmL,\t_y+\pi}\; , &&\quad \arg(\o y^2) = 2\pi\; ,  \label{StokesJump_R-}\\
		\D\Psi^-_{\mrmL,\t_y+\frac{\pi}{2}} &=  e^{+\pi i \frac{\tu}{\o}} \frac{i \sqrt{2\pi}\left(\frac{g}{\o}\right)^{\frac{\tu}{\o}}}{\Gamma\left(\frac{1}{2} - \frac{\tu}{\o}\right)} \, \Psi^+_{\mrmL,\t_y+\frac{\pi}{2}}\; , && \quad \arg(\o y^2) = \pi\; , \label{StokesJump_I+}\\
		\D\Psi^-_{\mrmL,\t_y-\frac{\pi}{2}} &=  e^{-\pi i \frac{\tu}{\o}} \frac{i \sqrt{2\pi}\left(\frac{g}{\o}\right)^{\frac{\tu}{\o}}}{\Gamma\left(\frac{1}{2} - \frac{\tu}{\o}\right)} \, \Psi^-_{\mrmL,\t_y-\frac{\pi}{2}}\; ,&&\quad \arg(\o y^2) = -\pi\;. \label{StokesJump_I-} 
	\end{align}
	
	Note that the prefactors on RHS of \eqref{StokesJump_R+}-\eqref{StokesJump_I-} have the same structure up to some signs and a phase which is governed by $\arg (\o y^2)$. Setting $\a = \arg(\o y^2)$, we can write the discontinuities in a compact form as
	\begin{equation}
		\D\Psi^+_{\mrmL,\t} = e^{-\frac{i\a \tu}{\o}}\, \F ^{+}_{\o}(\tu,g)\, \Psi^-_{\mrmL,\t}\,, \qquad  \qquad \D\Psi^-_{\mrmL,\t} = e^{\frac{i \a \tu}{\o}}\, \F ^{-}_{\o}(\tu,g)\, \Psi^+_{\mrmL,\t} , \label{StokesJumps_Generic}
	\end{equation}
	where 
	\begin{equation}\label{UniversalDisc_Weber}
		\F ^{\pm}_\o(\tu,g)  =  \frac{i \sqrt{2\pi}\left(\frac{g}{\o}\right)^{\mp \frac{\tu}{\o}}}{\Gamma\left(\frac{1}{2} \pm \frac{\tu}{\o}\right)}.
	\end{equation}
	The generic structure of \eqref{StokesJumps_Generic} appears in the analysis of any locally harmonic potential. Only difference occurs in the parameter $\tu$ which we explain later.
	
	As we mentioned, in this paper, we are interested in $\o^2=+1$ and $\o^2 = -1$ cases which are associated with the locally harmonic wells and barriers in the global coordinates. Therefore, we observe that the generic structure of \eqref{UniversalDisc_Weber} for wells and barriers differs only with several imaginary factors. Note that in each case, the double-valued nature of the square root operator leaves two choices for $\o$ in \eqref{StokesJumps_Generic} and \eqref{UniversalDisc_Weber}. We observe that while the Stokes diagrams have the same geometry, i.e. Stokes lines lie along the same directions on $y$-plane for each choice, the discontinuous solutions on the Stokes line lying on the same direction are not identical.

\begin{figure}[H]
		\centering
		\begin{subfigure}{0.75\textwidth}
			\caption{\underline{Diagrams when $\arg \o =0 $ and $\arg \o = \pi$}}
			\vspace{2pt}
			\includegraphics[width=\textwidth]{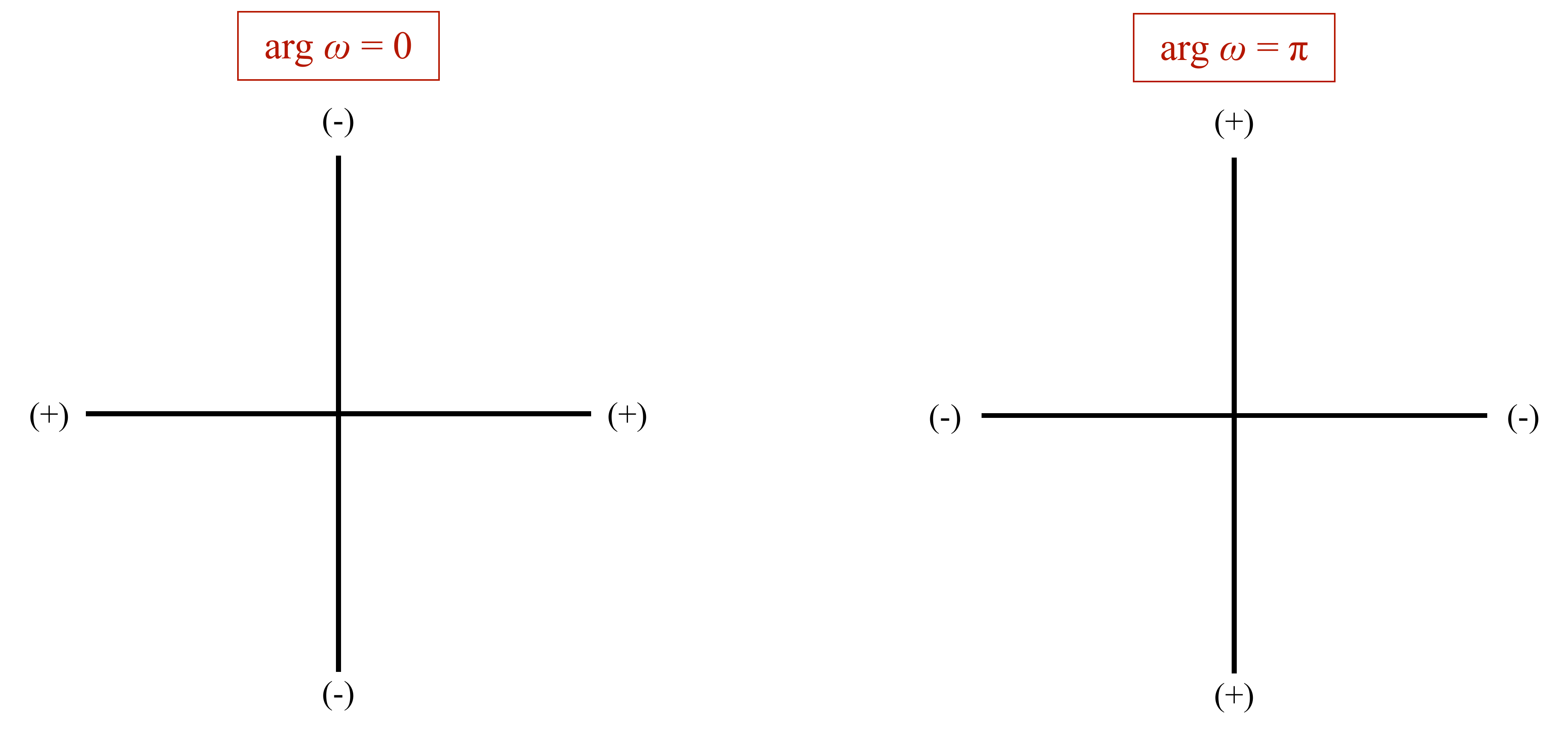}
			\label{Figure: WeberDiagrams_Wells}
		\end{subfigure}
		~ 
		\begin{subfigure}{0.75\textwidth}
			\caption{\underline{Diagrams when $\arg \o = \pm \frac{\pi}{2}$}}
			\vspace{2pt}
			\includegraphics[width=\textwidth]{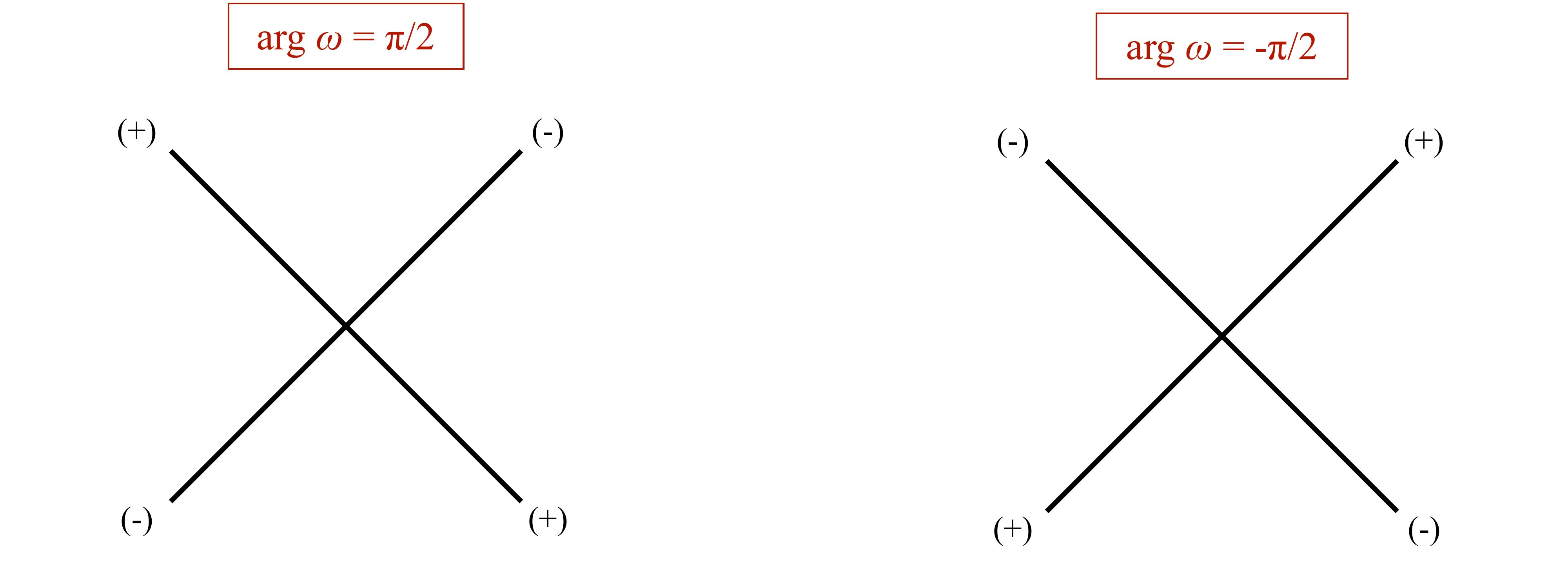}
			\label{Figure: WeberDiagrams_Barrier}
		\end{subfigure}
		\vspace{-5pt}
		\caption{Stokes Diagrams of the Weber equation for $\o^2  = \pm 1$} \label{Figure: WeberDiagrams_All}
	\end{figure}

	To elaborate this fact, let us consider the cases $\o^2=+1$ and $\o^2=-1$ separately. $\o^2=+1$ is associated to $\arg \o =0$ and $\arg \o =\pi$, while $\o^2 = -1$ is associated to $\arg \o = \pm \frac{\pi}{2}$.The associated Stokes diagrams are plotted in Fig.~\ref{Figure: WeberDiagrams_All} and the discontinuities on the Stokes lines are listed in Tables \ref{Table: StokesJumps_LocalHarmonic_Weber} and \ref{Table: StokesJumps_LocalInverse_Weber}.

	In addition to the new results in Table \ref{Table: StokesJumps_LocalInverse_Weber}, we present the results for $\arg \o =0$ in Table~\ref{Table: StokesJumps_LocalHarmonic_Weber}, which recover the discontinuities in \cite{Sueishi:2021xti, Bucciotti:2023trp}. For $\arg \o = \pi$, on the other hand, the discontinuities slightly differ from \cite{Sueishi:2021xti}, where the associated Stokes diagram was considered without the change in $\arg \o$ being taken into account. These differences, however, are merely related to our convention on Stokes diagrams and it doesn't affect the final result when Weber-type EWKB is discussed in global coordinates.

	\vspace{-0cm}
	\begin{table}[H]
		\caption{Stokes Jumps for $\o^2 =1$ in local coordinates. (See Fig.~\ref{Figure: WeberDiagrams_Wells}.)} \label{Table: StokesJumps_LocalHarmonic_Weber}
		\vspace{-8pt}
		\begin{center} 
			\begin{tblr}{|c | c || c |c|}	
				\hline
				\SetCell[c=2]{c}{$\arg \o =0$ } &  & \SetCell[c=2]{c} $\arg \o =\pi$ \\ 
				\hline 
				$\t_y$ = 0 	& $\D\Psi^+_\mrmL = \F _{\o=1}^+ \, \Psi^-_\mrmL$  & $\t_y=0$ & $\D\Psi^-_\mrmL = e^{i \pi \tu}\, \F ^-_{\o=-1}\, \Psi^+_\mrmL$ \\
				\hline
				$\t_y$ = $\frac{\pi}{2}$ 	& $\D\Psi^-_\mrmL = e^{i \pi \tu}\; \F _{\o=1}^- \, \Psi^+_\mrmL$ & $\t_y=\frac{\pi}{2}$ & $\D\Psi^+_\mrmL = e^{-2\pi i \tu}\, \F ^+_{\o=-1}\, \Psi^-_\mrmL$ \\
				\hline
				$\t_y$ = $\pi$ 	& $\D\Psi^+_\mrmL = e^{-2i\pi \tu} \;\F _{\o=1}^+\, \Psi^-_\mrmL$  & $\t_y = -\pi$&  $\D\Psi^-_\mrmL = e^{-i \pi \tu}\, \F^-_{\o=-1}\, \Psi^+_\mrmL$  \\
				\hline 
				$\t_y$ = $-\frac{\pi}{2}$ 	& $\D\Psi^-_\mrmL = e^{-i \pi  \tu}\; \F _{\o=1}^+\, \Psi^+_\mrmL$ & $\t_y=-\frac{\pi}{2}$&  $\D\Psi^+_\mrmL =  \F ^+_{\o=-1}\, \Psi^-_\mrmL$  \\
				\hline
			\end{tblr}
		\end{center}	
	\end{table}
	\vspace{-0.7cm}
	\begin{table}[H]
		\caption{Stokes Jumps for $\o^2 = -1$ in local coordinates. (See Fig.~\ref{Figure: WeberDiagrams_Barrier}.)} \label{Table: StokesJumps_LocalInverse_Weber}
		\vspace{-10pt}
		\begin{center} 
			\begin{tblr}{|c | c || c |c|}
				\hline
				\SetCell[c=2]{c}{$\arg \o =\frac{\pi}{2}$ } &  & \SetCell[c=2]{c} $\arg \o =-\frac{\pi}{2}$\\ 	
				\hline 
				$\t_y$ = -$\frac{\pi}{4}$ 	& $\D\Psi^+_\mrmL = \F _{\o=i}^+ \, \Psi^-_\mrmL$  & $\t_y=-\frac{\pi}{4}$ & $\D\Psi^-_\mrmL = e^{-i \pi (i \tu)}\, \F ^-_{\o=-i}\, \Psi^+_\mrmL$ \\
				\hline
				$\t_y$ = $\frac{\pi}{4}$ 	& $\D\Psi^-_\mrmL = e^{i \pi (- i \tu)}\; \F _{\o=i}^- \, \Psi^+_\mrmL$ & $\t_y= \frac{\pi}{4}$ & $\D\Psi^+_\mrmL =  \F ^+_{\o=-i}\, \Psi^-_\mrmL$ \\
				\hline
				$\t_y$ = $\frac{3\pi}{4}$ 	& $\D\Psi^+_\mrmL = e^{-2i\pi (-i \tu)} \; \F _{\o=i}^+\, \Psi^-_\mrmL$  & $\t_y = \frac{3\pi}{4}$&  $\D\Psi^-_\mrmL = e^{i \pi (i\tu)}\, \F ^-_{\o=-i}\, \Psi^+_\mrmL$  \\
				\hline 
				$\t_y$ = $-\frac{3\pi}{4}$ 	& $\D\Psi^-_\mrmL = e^{-i \pi  (-i \tu)}\; \F _{\o=i}^+\, \Psi^+_\mrmL$ & $\t_y=  \frac{5\pi}{4}$&  $\D\Psi^+_\mrmL = e^{-2\pi i(i \tu)} \F^+_{\o=-i}\, \Psi^-_\mrmL$  \\
				\hline
			\end{tblr}
		\end{center}
	\end{table}
	\paragraph{\underline{Transition to global coordinates}:} As we discussed before, in the transition to the global coordinates, the universal character of the discontinuities remains intact. Only change in the function $\F_\o^\pm$ occurs in the parameter $\tu$. Following the arguments in \cite{Sueishi:2021xti}, we briefly summarize how the Stokes discontinuities change in the global coordinates: 
	
	In the local coordinates, the parameter $\tu$ does not get any quantum corrections and totally independent of $g$. In fact, it is possible to obtain $\tu$ directly from the asymptotic behaviour of $\ttO$ around the turning point by observing 
	\begin{equation}\label{Residue_Local}
		\Res \left[\ttO(y,g)\right]_{y=0} = -\frac{\tu}{\o},
	\end{equation}
	which stems from 
	\[ \ttO(y,g) = \frac{\o y}{2 g} - \frac{\tu}{\o y }+ O\left(\frac{1}{y^2}\right)\, .\]
	In the global coordinates, on the other hand, the $\tu$ is no longer a simple constant parameter. Around the vicinity of a saddle point, it can be represented by an asymptotic series in $g$, i.e. $\tu(g) = \sum \tu_n\, g^n$. This suggests that, upon the transition from the local coordinates to the global ones, the residue around a saddle point $x_0$ becomes a function of $\tu$ and $g$, which we represent by $\mcalF(\tu,g)$:
	\begin{equation}\label{Action_Perturbative_Weber}
		\frac{\mcalF(\tu,g)}{\o} = \Res\left[\tS(x,g)\right]_{x=x_0}. \\
	\end{equation}
	Note that the last expression is motivated from the following equality:
	\begin{equation}
		\Res\left[\tS (x,g)\right]_{x=x_0} = \Res\left[\ttO(y(x),g)\right]_{y(x_0)=0} \, ,
	\end{equation}
	which stems from \eqref{LocalToGlobal_3} and the expression $\frac{\dee y}{\dee x}$ being free of singularities by definition of $y(x)$, i.e. \eqref{Expansion_y}. Then, we observe that as a result of the \textit{local} $\rightarrow$ \textit{global} coordinate transition, the function $\F$ transforms as
	\begin{equation}\label{Transition_LocalTerms}
		\F_\o^\pm(\tu,g) \rightarrow \F_\o^\pm(\mcalF(\tu,g),g).
	\end{equation}
	Note that $\F_\o^\pm$ is still obtained from the discontinuities associated with a single saddle point. In this sense, it is associated to the local information in the global coordinates .
	
	Finally, given the transformation from local to global coordinates, let us explicitly express, the monodromy matrices in the global coordinates for generic $\o$ and $\t = \arg (\o y^2)$:
	\begin{equation}
		M_\mrmW^+(\o,\t) = \begin{pmatrix}
			1  & \quad  e^{-\frac{i \t \mcalF}{\o}}\F_\o^+(\mcalF,g) \\[1em]
			0 & \quad 1
		\end{pmatrix}\, ,  \qquad 	M_\mrmW^-(\o,\t) = \begin{pmatrix}
			1 & \quad 0 \\[1em]
			e^{\frac{i\t \mcalF}{\o}}	\F_\o^-(\mcalF,g) & \quad  1
		\end{pmatrix}. \label{Monodromy_Weber} 
	\end{equation}
	\vspace{0cm}

	\paragraph{\underline{Connection to the WKB cycles}:}
	
\begin{figure}[t]
		\centering
		\begin{subfigure}{0.48\textwidth}
			\caption{\underline{2 saddle points}}
			\includegraphics[width=\textwidth]{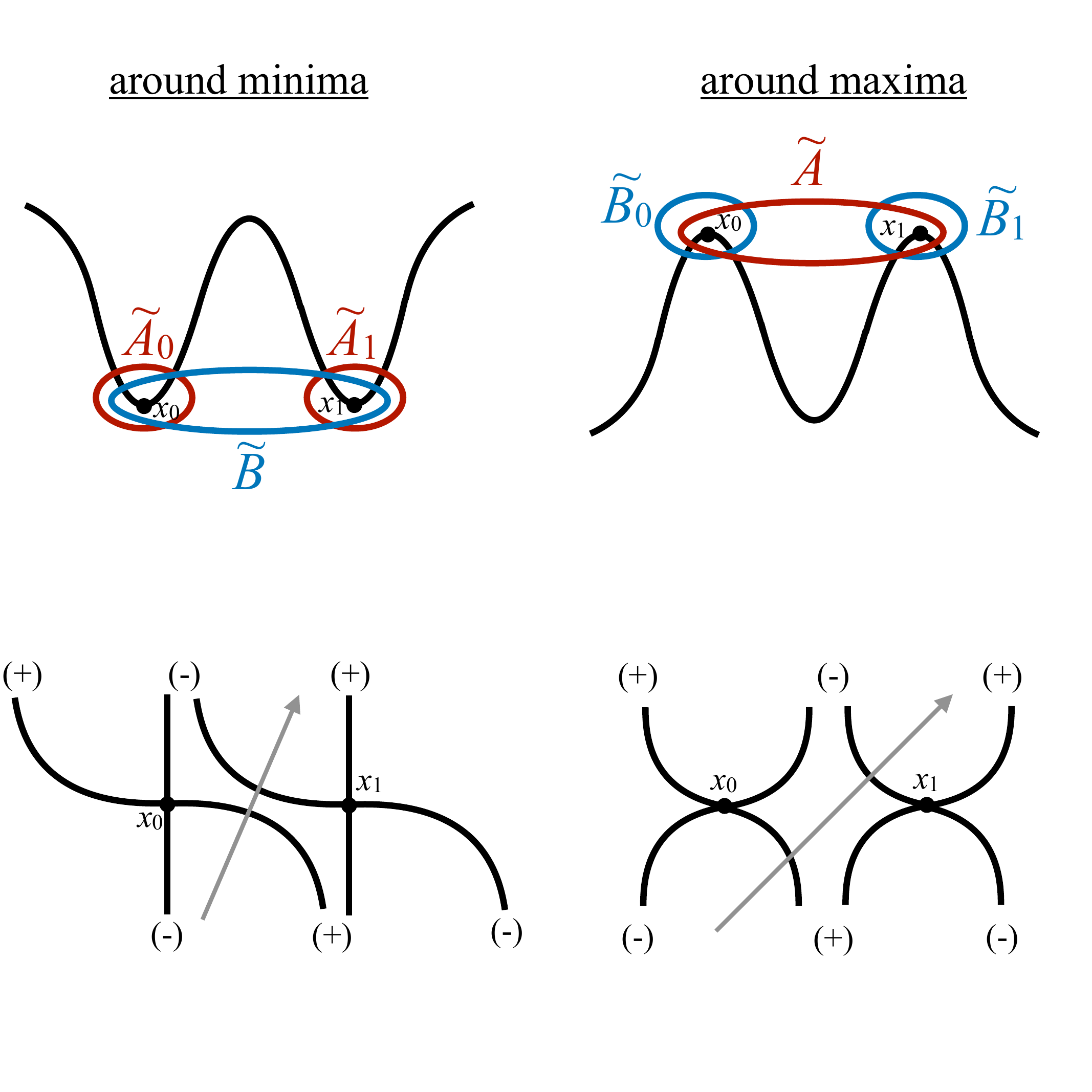}
			\label{Figure: Weber_NPcycle_DoubleSaddle}
		\end{subfigure}
		~\hfill 
		\begin{subfigure}{0.48\textwidth}
			\caption{\underline{1 saddle point and 1 simple turning point}}
			\includegraphics[width=\textwidth]{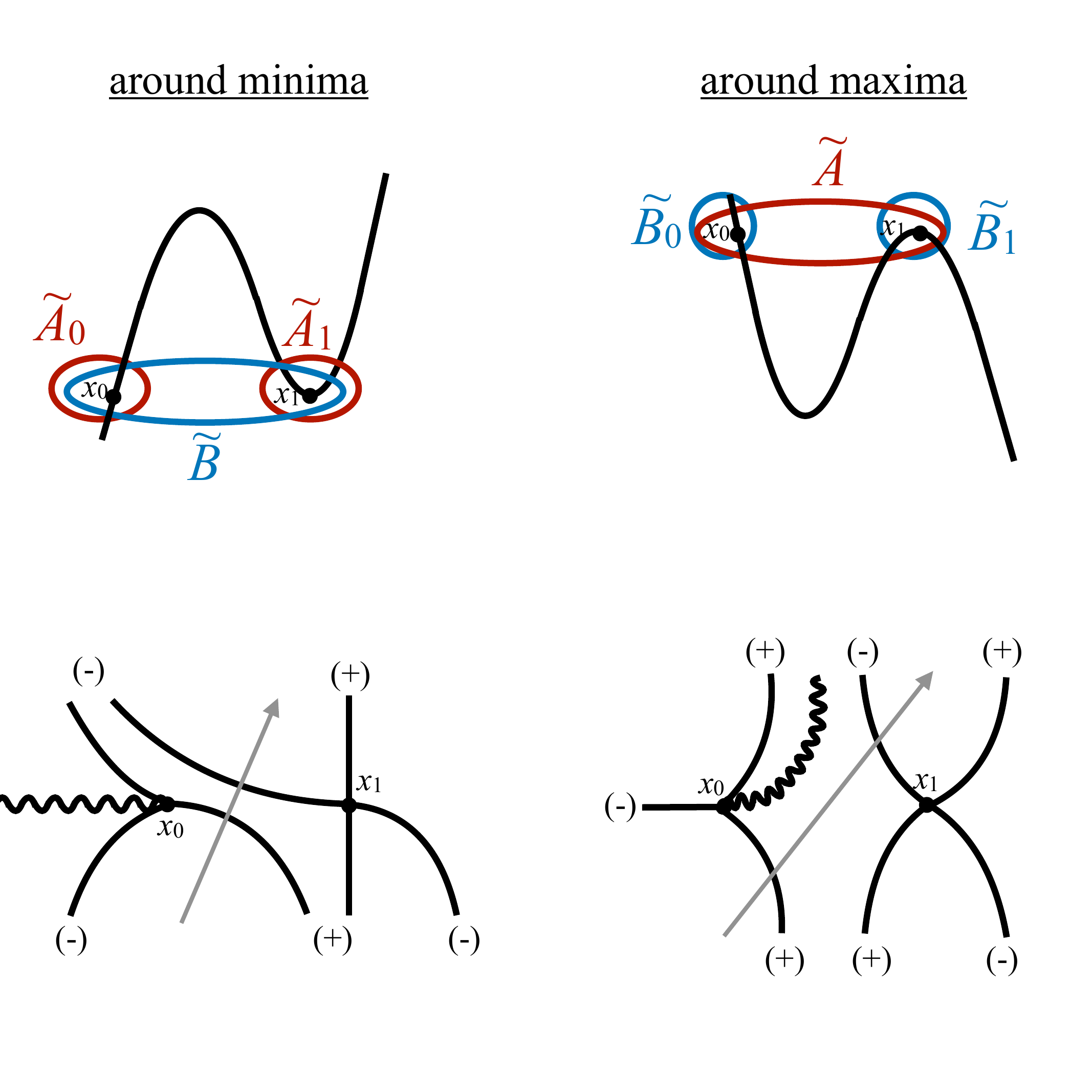}
			\label{Figure: Weber_NPcycle_SaddleTurning}
		\end{subfigure}
		\vspace*{-25pt}
		\caption{Perturbative and non-perturbative cycles in the Weber-type EWKB approach.} \label{Figure: Weber_NPcycle_All}
	\end{figure}

	Now, we can construct the contribution of WKB cycles in the Weber-type approach for possible Stokes diagrams of generic locally harmonic saddles. In Fig.~\ref{Figure: Weber_NPcycle_All}, we present all possible types of the Weber-type Stokes diagrams around the bottom of wells and the top of barriers. Again, we emphasize that the transition to the Weber-type Stokes diagrams is only related by 
	$u \rightarrow g\tu$ and 
	$V(x) \rightarrow V(x) - u_0$. As a result, the Airy-type Stokes diagrams at $\t_u = 0$ and $\t_u = \pm \pi$ in Fig.~\ref{Figure: IHO_Transition} turn into the Weber-type Stokes diagrams around minima and around maxima respectively as in Figures~\ref{Figure: WeberDiagrams_Wells} and~\ref{Figure: WeberDiagrams_Barrier}. 
	
	In the following, as in the rest of this subsection, our discussion mainly follows the prescription given in \cite{Sueishi:2021xti} with adaptations to our setup and generalizations to the cases which were not covered in \cite{Sueishi:2021xti}. We also present explicit generic expressions for the leading order contribution to the non-perturbative cycles.
	
	Let us start with perturbative cycles which encircle individual saddle points. They can be considered as ``singular'' limits of the Airy-type ones and they encode the perturbative information about the quantum system. Using \eqref{Action_Perturbative_Weber}, we get around the bottom of wells, 
	\begin{align}
		\tPi_A &= \exp\left\{\oint_{\tA_k}\, \mrmd x\, \tS(x,\tu,g) \right\} = e^{\mp 2 i \pi \mcalF^{\{k\}}(\tu,g)}\, , \label{PerturbativeCycle_Well_Weber} 
	\end{align}
	and around the top of the barriers, 
	\begin{align}
		\tPi_B &= \exp\left\{\oint_{\tB_k}\, \mrmd x\, \tS(x,\tu,g) \right\} = e^{\mp 2 i \pi \mcalF^{\{k\}}(\tu,g)}\, , \label{PerturbativeCycle_Barrier_Weber} 
	\end{align}
	where $\mp$ signs are chosen accordingly to the sign of $\o$ at the corresponding saddle point.
	
	The non-perturbative cycles, i.e. $\tB$ around the bottom of wells and $\tA$ around the top of barrier, need a more involved analysis. Following the guidelines in \cite{Sueishi:2021xti}, we obtain the non-perturbative actions from the connection of regions $\mrmI$ and $\mrmI\mrmI$ of the Stokes diagrams in Fig.~\ref{Figure: WeberDiagrams_All} using the boundary condition $\Psi^+_{\mrmG,\mrmI\mrmI} = 0$.  
	
	As we observe from Fig.~\ref{Figure: Weber_NPcycle_All}, the non-perturbative cycles in the Weber-type approach can be classified into two types:
	\begin{enumerate}[(i)]
		\item  Cycles connecting two saddle points:	
		\item  Cycles connecting a saddle point and a simple turning point:
	\end{enumerate}
	For generic $\o$, the transition matrices are written as
	\begin{equation}
		T^{(\mathrm{i})}_{\mrmI\rightarrow\mrmI\mrmI} = \left(M^{-}_\mrmW(\o,\t_1)\right)^{-1} \mcalN^{(0,1)}_\mrmW\, M^+_\mrmW(\o,\t_0) \,,\label{TransitionMatrixNP_Weber1}
	\end{equation}
	and
	\begin{equation}
		T^{(\mathrm{ii})}_{\mrmI\rightarrow\mrmI\mrmI} = \left(M^{-}_\mrmW(\o,\t_1)\right)^{-1}  \mcalN^{(0,1)}_\mrmW\, M^+_\mrmA\,, \label{TransitionMatrixNP_Weber2}
	\end{equation}
	where 
	\begin{equation}
		\mcalN_\mrmW^{(0,1)} = 	\begin{bmatrix}
			e^{-\int_{x_0}^{x_1} \mrmd x\, \tS(x,g,\tu)} & 0 \\
			0 & 	e^{\int_{x_0}^{x_1} \mrmd x\, \tS(x,g,\tu)} 
		\end{bmatrix}\,,
	\end{equation}
	is the matrix for changing the normalization point and $M^+_\mrmA$ is the monodromy matrix for Airy type diagrams originated from the simple turning point at $x=x_0$ in Fig.~\ref{Figure: Weber_NPcycle_SaddleTurning} . Then, we get the associated non-perturbative actions as
	\begin{equation}
		\tPi^\mathrm{i}_\mathrm{NP} = e^{-W_{0,1}(\tu,g)}\, e^{i\pi \mcalF_{\o_1}}\,\frac{2\pi \left(\frac{g}{\o_0}\right)^{-\mcalF_{\o_0}}\, \left(\frac{g}{\o_1}\right)^{\mcalF_{\o_1}}}{\G\left(\frac{1}{2} + \mcalF_{\o_0} \right)\,\G\left(\frac{1}{2}-\mcalF_{\o_1}\right)}\,, \label{NP_Weber_Generic01}
	\end{equation}
	and
	\begin{equation}
		\tPi^\mathrm{ii}_\mathrm{NP}  =  e^{-W_{0,1}(\tu,g)}\, e^{i\pi\mcalF_{\o_1}} \frac{\sqrt{2\pi}\, \left(\frac{g}{\o_1}\right)^{\mcalF_{\o_1}}}{\G\left(\frac{1}{2} - \mcalF_{\o_1}\right)}\,, \label{NP_Weber_Generic02}
	\end{equation}
	where we identify $\mcalF_{\o_k}$ as the perturbative expansion arising from the residue computation around the $k^\th$ saddle point.

	Note that in all cases we present in Fig.~\ref{Figure: Weber_NPcycle_All}, we assume degenerate and symmetric wells and barriers. This means that $\o_0$ and $\o_1$ have a phase difference by $\pi$, i.e. $\arg \o_0 = \arg \o_1 + \pi$. Moreover, $\mcalF_{\o_0}$ and $\mcalF_{\o_1}$ differs only in their signs as well. Then, we can simplify the expressions in \eqref{NP_Weber_Generic01} and \eqref{NP_Weber_Generic02}, and rewrite them using only $\o_0$ and $\mcalF_{\o_0}$
	\begin{align}
		\tPi^\mathrm{i}_\mathrm{NP} &= e^{-W_{0,1}(\tu,g)}\,\frac{2\pi \left(\frac{g}{\o_0}\right)^{-2\mcalF_{\o_0}}}{\left[\G\left(\frac{1}{2} + \mcalF_{\o_0} \right)\right]^2}\, , \qquad  
		\tPi^\mathrm{ii}_\mathrm{NP}  =  e^{-W_{0,1}(\tu,g)}\,  \frac{\sqrt{2\pi}\, \left(\frac{g}{\o_0}\right)^{-\mcalF_{\o_0}}}{\G\left(\frac{1}{2} + \mcalF_{\o_0}\right)}\, . \label{NP_WeberGeneric} 
	\end{align}  
	In this way, without any confusion due to the double signs arising from $\sqrt{\o^2}$, we can simply use a single value for $\o$ at all saddle points of a given potential.   
	
	Equation \eqref{NP_WeberGeneric} shows that the non-perturbative actions around the bottom of wells and the top of barriers have the same structure which indicates a ``universal'' behaviour. 
	Note that the monodromy matrices contribute locally in \eqref{NP_WeberGeneric},  while the non-local contributions is produced by $\mcalN_\mrmW^{0,1}$. The latter one is represented by $e^{W_{0,1}(\tu,g)}$ whose exponent is given by 
	\begin{equation}
		W_{0,1}(\tu,g) = 2\int_{x_0}^{x_1} \mrmd x\, \tS(x,g,\tu). \label{WeberBion_Generic}
	\end{equation}
	
	Investigating this term makes the link with the path integral formalism explicit: At the leading order, $\tS(x,g,\tu) = \sqrt{2\tV(x)}$ and its integration are associated with multi-instanton action, 
	i.e. 
	\begin{equation}
		\mcalB = \frac{2}{g} \int_{x_0}^{x_1}\mrmd x\, \sqrt{2 \tV(x)}. \label{InstantonAction}
	\end{equation}
	More specifically, if we focus on the bottom of wells, in the case of Fig.~\ref{Figure: Weber_NPcycle_DoubleSaddle}, $\mcalB$ contributes to $\Pi_\mathrm{NP}^\mrmi$ as the (neutral) \textit{bion} action which is a configuration of an instanton and an anti-instanton; 
	while, in the case of Fig.~\ref{Figure: Weber_NPcycle_SaddleTurning}, $\mcalB$ contributes to $\Pi_\mathrm{NP}^\mrmii$ as a \textit{bounce} moving on a closed path between a saddle point, i.e. $x_1$ in this case, and a simple turning point, i.e. $x_0$.  
	
\begin{figure}[t]
		\centering
		\includegraphics[width=0.9\textwidth]{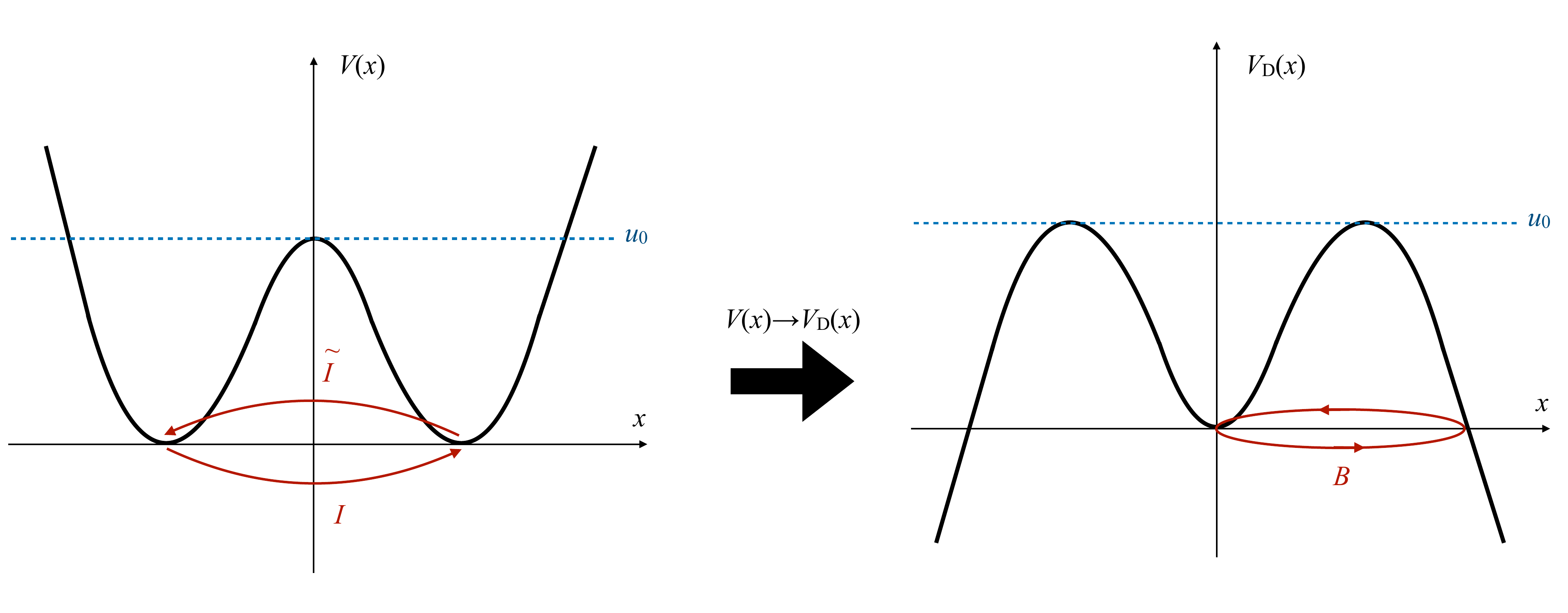}
		\caption{An example of inversion and rescaling of the potential.} \label{Figure: InvertingPotential}
	\end{figure}

	When, we focus on the top of barrier, 
	$\mcalB$ can be associated with (multi-)instanton contributions of the inverted potential. The inversion stems from the rescaling \[V(x) \rightarrow V(x) - u_0 \] which brings the top of the corresponding barrier to the level of zero energy. (See e.g. Fig.~\ref{Figure: InvertingPotential}.) Then, inverting the potential, the associated saddles become a minimum of the inverted $V(x)$ which admits zero energy classical solutions in Euclidean path integral formulation, i.e. instantons. 
	
	The computations of higher order corrections to $W_{0,1}$, on the other hand, are more involved. For example, at the next order, we encounter the following integral, when $x_0$ and $x_1$ are double saddle points as in Fig.~\ref{Figure: Weber_NPcycle_DoubleSaddle}:
	\begin{align}
		\mcalK_\mrmi &= \int_{x_0}^{x_1} \mrmd x\,\frac{\tu}{\o_0} \frac{\o_0}{ \sqrt{2\tV(x)}} \nonumber \\
		& \simeq \frac{\tu}{\o_0}\Big[ \log\mcalC^\mrmi_0  -  \log\left(x_1-x\right) \Big]_{x=x_1} - \frac{\tu}{\o_0}\Big[ - \log\mcalC^\mrmi_0  +  \log\left(x-x_0 \right) \Big]_{x=x_0}\,. \label{LogDivTerm1}
	\end{align} 
	When only $x_1$ is the double turning point as in Fig.~\ref{Figure: Weber_NPcycle_SaddleTurning}, we have
	\begin{align}
		\mcalK_\mrmii &= \int_{x_0}^{x_1} \mrmd x\,\frac{\tu}{\o_0} \frac{\o_0}{ \sqrt{2\tV(x)}} \simeq \frac{\tu}{\o_0}\Big[  \log\mcalC^\mrmii_0  - \log\left(x_1-x \right) \Big]_{x=x_1}\,  .\label{LogDivTerm2}
	\end{align}
	In both cases, $\mcalC^\mrmi_0$ and $\mcalC^\mrmii_0$ are constants specified by the potential $\tV(x)$. 
	
	Note that in \eqref{LogDivTerm1} and \eqref{LogDivTerm2}, we inserted $\o_0$ on purpose since $\frac{\tu}{\o_0}$ is the term appearing in $\mcalF_{\o_0}$. This property can be deduced from the residue computation in the local coordinates in \eqref{Residue_Local} which turns into the leading order term of $\mcalF_{\o_0}(\tu,g)$ at $g=0$. In fact, at any order contributing to \eqref{WeberBion_Generic}, we encounter such integrals which generate the corrections to $\log \mcalC^\mrmi_0$ and $\log\mcalC^\mrmii_0$ terms in \eqref{LogDivTerm1} and \eqref{LogDivTerm2} respectively. Eventually, their total contribution to $e^{-W_{0,1}}$ becomes 
	\begin{equation}
		e^{-W_{0,1}} = e^{-\mcalG(\tu,g)} e^{-2 \mcalF(\tu,g) \log\mcalC^\mrmi}\, ,  \label{NonLocalWeber1}
	\end{equation}
	and 
	\begin{equation}
		e^{-W_{0,1}} = e^{-\mcalG(\tu,g)} e^{-\mcalF(\tu,g) \log\mcalC^\mrmii}\, , \label{NonLocalWeber2}
	\end{equation}
	where $\mcalG$ is the collection of the terms which don't have logarithmic contributions. In other words, it constitutes the analytic part of the $W_{0,1}$ and is written in a series expansion in $g$:
	\begin{equation}
		\mcalG(\tu,g) = \sum_{m=0}^\infty \mcalG_m(\tu)\, g^m \, .
	\end{equation}

	Finally, in order to make the final results in \eqref{NonLocalWeber1} and \eqref{NonLocalWeber2} finite, the singular terms in \eqref{LogDivTerm1} and \eqref{LogDivTerm2}, which also occur in higher order corrections as well, should be eliminated properly.  At this point, it is more convenient to recall the equivalence between the Airy-type and Weber-type WKB or more specifically between the expansions of $\tS$ in \eqref{WKB_Expansions_Weber} and $\ts$ in \eqref{WKB_Expansions_Airy}. Using this equality, we can rewrite \eqref{WeberBion_Generic} as 
	\begin{equation}
		W_{0,1}(\tu,g) = 2 \int_{x_0}^{x_1} \mrmd x\, \ts(x,g,g \tu) \, .\label{AiryBion_Generic}
	\end{equation}
	In this form, the computations of $W_{0,1}$ can be carried out by using Mellin transformation at each order as described in \cite{DP1,Zinn-Justin:2004qzw,Kamata:2021jrs}. Then, following \cite{Zinn-Justin:2004qzw}, we introduce normalizations at the double turning points. Then, dropping $\o_0$ subscripts, we write the final generic expressions for $\Pi_\mathrm{NP}^\mrmi$ and $\Pi_\mathrm{NP}^\mrmii$ as
	\begin{align}
		\tPi^\mathrm{i}_\mathrm{NP} &= e^{-\mcalG(\tu,g) }\,\frac{2\pi \left(\frac{g}{\mcalC^{\mathrm{i}}}\right)^{-2\mcalF(\tu,g)}}{\left[\G\left(\frac{1}{2} + \mcalF(\tu,g) \right)\right]^2}\, , \qquad 
		\tPi^\mathrm{ii}_\mathrm{NP}  =  e^{-\mcalG(\tu,g)}\,  \frac{\sqrt{2\pi}\, \left(\frac{g}{\mcalC^{\mathrm{ii}}}\right)^{-\mcalF(\tu,g)}}{\G\left(\frac{1}{2} + \mcalF(\tu,g)\right)}\, , \label{NP_Explicit} 
	\end{align}
	where 
	\begin{align}
		\mcalC^\mrmi &= 2 \, \o_0  \left(x_1 - x_0\right)^2\, \exp\left\{\int_{x_0}^{x_1}\mrmd x\, \left[\frac{\o_0}{\sqrt{2 \tV(x)}} - \frac{1}{x- x_0} - \frac{1}{x_1 - x}\right] \right\} \, , \label{PrefactorLog1} \\ \nonumber \\ 
		\mcalC^\mrmii &= 2 \, \o_0 \,  \left(x_1 - x_0\right)^2 \exp\left\{2\int_{x_0}^{x_1}\mrmd x\,  \left[\frac{\o_0}{\sqrt{2 \tV(x)}} - \frac{1}{x_1 - x}\right]\right\}  \, .  \label{PrefactorLog2}
	\end{align}
	Note that we inserted the prefactor $\o_0$, which originally appears in $\frac{g}{\o_0}$ terms in \eqref{NP_WeberGeneric} in the definitions of $\mcalC^\mrmi$ and $\mcalC^\mrmii$ for convenience. 
	
	These expressions recover ZJJ's results in \cite{Zinn-Justin:2004vcw,Zinn-Justin:2004qzw} for the bottom of wells and generalizes them for all types of locally harmonic saddles. At the barrier tops, $\o_0$ is an imaginary constant and another imaginary contribution arises from $\sqrt{2 \tV(x)}$. As a result, the term $\frac{\o_0}{\sqrt{2 \tV(x)}}$ is real and only possible imaginary contribution is due to $\o_0$ term. Moreover, the log divergence(s) at double turning point(s) is(are) cancelled by the normalization terms in \eqref{PrefactorLog1} and \eqref{PrefactorLog2}.

	\subsection{Dictionary}\label{Section: AiryWeber_Dictionary}
	Let us finally discuss the connection between the Airy and Weber type EWKB in a more quantitative manner. 
	In principal, the WKB cycles in the Weber-type approach are limits of the Airy-type ones at $u=u_0$, corresponding to minima or maxima of the potential. While the limit of cycles is not directly apparent in the EWKB framework, from a geometric point of view, it is a singular limit of the smooth (hyper)-elliptic curve corresponding to either $A$ or $B$ Airy-type cycle. Moreover, the singular curve is linked to the Euclidean instanton configurations, i.e bions or bounces, as we observed in the Weber-type EWKB. An explicit construction in these lines was made in \cite{Behtash:2017rqj}, where the link between elliptic curves appearing in SUSY QM and complex-bion configuration was presented\footnote{We thank Alireza Behtash for pointing out the geometric construction linking the singular limits of the WKB cycles and instanton configurations.}. 
	
	In the EWKB framework, the connection between Airy and Weber type cycles is provided by a dictionary. Note that this was originally discussed in \cite{Sueishi:2021xti,DDP2}, but our construction covers more general cases. Moreover, stating the explicit dictionary serves as a guideline in later sections.
	
	To make a direct connection to the literature which uses the all-order WKB expansion (e.g. see \cite{Basar:2017hpr,Basar:2015xna}), let us recall the canonical definitions of the WKB action and its dual:
	\begin{equation}\label{WKB_Actions}
		a(u,g) = \frac{1}{2\pi} \oint_A \mrmd x\, p(x,g)\, , \quad a^\mrmD(u,g) = \frac{1}{2\pi} \oint_B \mrmd x\, p(x,g)\,,
	\end{equation}
	where
	\[p(x,g) = \sum_{n=0}^{\infty} p_n(x)\, g^n\, .\] 
	Note that we have the following relation:
	\begin{equation}
		p(x,g) = i g \, \ts(x,g) \, ,
	\end{equation} 
	where $\ts$ is defined in \eqref{WKB_Expansions_Airy} and it enters the Voros connection matrix in \eqref{Airy_NormalizationChange}. In that sense, we get
	\begin{equation}
		2\mcalV_A = \oint_{A}\mrmd x\, \ts(x,g) = -\frac{2\pi i}{g}a(u,g)\, , \qquad 2\mcalV_B = \oint_{B}\mrmd x\, \ts(x,g) = -\frac{2\pi i}{g}\aD(u,g)\,.
	\end{equation}
	Then, recalling the construction in  \cite{Sueishi:2021xti} and the relationship $u=g\tu$, we put forward a dictionary between Airy and Weber type cycles as follows:
	\begin{itemize}
		\item \underline{Perturbative Parts}: This is encoded by $A$-cycle when $u<u_\mrmT$ (i.e. $u_0=0$) and $B$-cycle when $u>u_\mrmT$ (i.e. $u_0=u_\mrmT$)
		\begin{align}
			\Pi_A &= e^{-\frac{2\pi i}{g} a(u, g)} = e^{-2\pi i\mcalF(\frac{u}{g},g)}\, , \qquad \quad \;\; (u_0 = 0)\, , \label{Dictionary_Perturbative_Well}\\ \nonumber \\
			\Pi_B &= e^{-\frac{2\pi i}{g} a^\mrmD(u,g)} = e^{-2\pi i \mcalF^D(i\frac{u}{g},i g)}\, , \qquad \;  (u_0=u_\mrmT) \, . \label{Dictionary_Perturbative_Barrier}
		\end{align}
		\item \underline{Non-perturbative parts}: This is encoded by $B$-cycle when $u<u_\mrmT$ (i.e. $u_0=0$) and $A$-cycle when $u>u_\mrmT$ (i.e. $u_0=u_\mrmT$)
		\begin{itemize}
			\item \underline{In the presence of degenerate saddles}: 
			\begin{align}
				\Pi_B &=  e^{-\frac{2\pi i}{g} a^\mrmD(u,g)} =  e^{-\mcalG(\frac{u}{g},g) }\,\frac{2\pi \left(\frac{g}{\mcalC_{\mathrm{i}}}\right)^{-2\mcalF(\frac{u}{g},g)}}{\left[\G\left(\frac{1}{2} + \mcalF\left(\frac{u}{g},g\right) \right)\right]^2}\, , \qquad \quad \;\; (u_0 = 0)\, , \label{Dictionary_NP_Well_Bion}\\ \nonumber \\
				\Pi_A &=  e^{-\frac{2\pi i}{g} a(u,g)} =  e^{-\mcalG^\mrmD(i\frac{u}{g},i g) }\,\frac{2\pi \left(\frac{g}{\mcalC_{\mathrm{i}}}\right)^{-2\mcalF^\mrmD(i\frac{u}{g},ig)}}{\left[\G\left(\frac{1}{2} + \mcalF^\mrmD\left(i\frac{u}{g},ig\right) \right)\right]^2} \, , \quad \,  (u_0=u_\mrmT) \, .\label{Dictionary_NP_Barrier_Bion}
			\end{align}
			\\
			\item \underline{In the presence of one saddle point and one simple turning point}:
			\begin{align}
				\Pi_B &=  e^{-\frac{2\pi i}{g} a^\mrmD(u,g)} =  e^{-\mcalG(\frac{u}{g},g) }\,\frac{\sqrt{2\pi} \left(\frac{g}{\mcalC_{\mathrm{ii}}}\right)^{-\mcalF(\frac{u}{g},g)}}{\G\left(\frac{1}{2} + \mcalF\left(\frac{u}{g},g\right) \right)}\, , \qquad \quad \;\; (u_0 = 0)\, , \label{Dictionary_NP_Well_Bounce} \\ \nonumber \\
				\Pi_A &=  e^{-\frac{2\pi i}{g} a(u,g)} =  e^{-\mcalG^\mrmD(i\frac{u}{g},i g) }\,\frac{\sqrt{2\pi} \left(\frac{g}{\mcalC_{\mathrm{ii}}}\right)^{-\mcalF^\mrmD(i\frac{u}{g},i g)}}{\G\left(\frac{1}{2} + \mcalF^\mrmD\left(i\frac{u}{g},i g\right) \right)} \, , \qquad  (u_0=u_\mrmT) \, .\label{Dictionary_NP_Barrier_Bounce}
			\end{align}
		\end{itemize}
	\end{itemize}
	where $\mcalC_\mrmi$ and $\mcalC_\mrmii$ are given by \eqref{PrefactorLog1} and \eqref{PrefactorLog2} respectively.
	
	Note that we introduce $\mcalF^\mrmD$ and $\mcalG^\mrmD$ in the expressions \eqref{Dictionary_Perturbative_Barrier}, \eqref{Dictionary_NP_Barrier_Bion} and \eqref{Dictionary_NP_Barrier_Bounce}. They refer to the \textit{$S$-dual} case characterized by $V_\mrmD = u_\mrmT - V$. The appearing imaginary factors are related to the inversion of a well or equivalently to $\o_0=i$ case. This is the situation when we compute the functions $\mcalF$ and $\mcalG$ around a barrier top which are related to the dual functions as
	\begin{equation}
		\mcalF^\mrmD\left(\tu,g\right) = \mcalF^{u\sim u_\mrmT}\left(-i\tu,-ig\right)\, , \qquad \mcalG^\mrmD\left(\tu,g\right) = \mcalG^{u\sim u_\mrmT}\left(-i\tu,-ig\right)\, . 
	\end{equation}

	
	\section{Chebyshev potentials, transition and duality}\label{Section: ChebyshevTransitionDuality}
	
	In the last section, we analyzed both Airy and Weber type EWKB methods and expressed the exponentiated actions $\Pi_A$ and $\Pi_B$ in both languages. We showed that $\Pi_A$ and $\Pi_B$ have different characters around the bottom of wells and the top of barriers which become manifest in the explicit expressions provided by the Weber-type approach. However, this doesn't yield  the overall change of the resurgence structure of a particular quantum system. In this section, focusing on genus-1 potentials of Chebyshev type, we explore generalities on how the resurgence relations for the sector $u<u_\mrmT$, which are well-known \cite{Basar:2017hpr}, are translated to the $u>u_\mrmT$ sector.

	Throughout this section, we identify the Chebyshev potentials with $V(x) = T_m^2(x)$, where $T_m(\cos x) = \cos(m x)$ is the Chebyshev polynomials of the first kind\footnote{To obtain the specific Chebyshev potentials, it is more convenient to use the following recursion relation: $T_0(x)=1,\; T_1(x)=1,\; T_{n+1}(x) = 2 x T_n(x) - T_{n-1}(x)$.}. In this way, we fix all the minima  and the maxima of the potentials to $u=0$ and $u=u_\mrmT=1$, respectively\footnote{In Sections \ref{Section: PeriodicPotential} and \ref{Section: DoubleWell}, we scale the potentials in such a way that their behaviour around a saddle point becomes $V(x)\simeq \frac{\o^2}{2}(x-x_0)^2$ which is in accordance with our assumptions in Section~\ref{Section: AiryTypeEWKB}. This does not effect the generality of the discussions in this section.}. Note that our discussion is based on the facts presented in \cite{Basar:2017hpr} and some of them were mentioned in this paper. In this section,
	\begin{enumerate}[i)]
		\item  We elaborate on in their discussion with a focus on the transition from $u<1$ sector to $u>1$ sector, our findings in this section provide guidelines for the later sections where we analyze periodic and symmetric double-well potentials.
		\item We also provide a conjecture on an $S$-duality transformation of the well-known P-NP relation \cite{Alvarez2,Alvarez3,Alvarez4,Dunne:2013ada,Dunne:2014bca} for genus-1 potentials in sector $u<1$.
	\end{enumerate} 
	\paragraph{\underline{Setup}:}A resurgence relation between perturbative and non-perturbative physics can be expressed by the WKB actions $a(u,g)$ and $a^\mrmD(u,g)$ or equivalently $\mcalF(\tu,g)$ and $\mcalG(\tu,g)$. In the EWKB framework, it is more natural to obtain a relation between $\mcalF$ and $\mcalG$: Around the bottom of wells and for genus-1 potentials\footnote{More specifically, \eqref{DunneUnsal_General} is for the cases when all the actions of the wells are proportional to each others which indicates there are only two independent WKB actions. Recently, a generalization was presented in \cite{Cavusoglu:2023bai,Cavusoglu:2024usn}. The authors showed a paramteric form of P-NP holds for the deformed quartic potential which reduces to cubic and quartic cases in extreme limits and P-NP relation takes the form of \eqref{DunneUnsal_General}.}, this relationship is well-known in various forms \cite{Alvarez2,Alvarez3,Alvarez4,Dunne:2013ada,Dunne:2014bca} and in general, it can be expressed as
	\begin{equation}\label{DunneUnsal_General}
		\mcalG(\tu,g)= - \int \frac{\mrmd g}{g}\, \left[\a_1\, \frac{S_\mrmI}{g}\frac{\dee \tu(\mcalF,g)}{\dee \mcalF} + \a_2\, \mcalF(\tu,g)\right]\, ,
	\end{equation}
	where $\a_1$ and $\a_2$ are specified by the classical potential of the quantum system. This is also equivalent to the Matone relation \cite{Matone:1995rx,Flume:2004rp} which can be obtained from \eqref{DunneUnsal_General} using the dictionary in Section \ref{Section: AiryWeber_Dictionary} \cite{Basar:2015xna,Gorsky:2014lia}. 
	
	The change upon the transition from  $u<1$ to $u>1$, on the other hand, is best observed by investigating the relationship between the WKB actions, $a(u)$ and $a^\mrmD(u)$.  For genus-1 potentials, an effective way to put the resurgence relation is using the Wronskian relation between $a(u)$ and $a^\mrmD(u)$ \cite{Basar:2017hpr}. In the following, we examine the Wronskian relation at the classical level. It is written as
	\begin{equation}\label{classicalWronskian}
		\msfw_0^\mrmD(u) \, a_0(u) - a_0^\mrmD(u) \, \msfw_0(u) = 2 i a_0^\mrmD(0)\, \msfw_0(0)\, ,
	\end{equation}
	where $\msfw_0 = \frac{\mrmd a_0}{\mrmd u}$ and $\msfw_0^\mrmD = \frac{\mrmd a_0^\mrmD}{\mrmd u}$ are classical period and its dual, respectively. 
	
	For Chebyshev potentials, in the sector $u<1$, $\msfw_0$ and $\msfw^\mrmD_0$ for the outermost well and the neighbouring barrier are \cite{Basar:2017hpr}
	\begin{align}
		\msfw_0(u)& = \frac{\sqrt{2} \pi}{m}\, \sin\left(\frac{\pi}{2m}\right)\, u \, {}_2 F_1\left(\frac{1}{2} - \frac{1}{2m},\frac{1}{2}+\frac{1}{2m},1,u\right)\, , \label{classicalPeriod_GeneralChebyshev}\\
		\msfw^\mrmD_0(u)& = i \frac{\sqrt{2} \pi}{m}\, \sin\left(\frac{\pi}{m}\right)\, (1-u) \, {}_2 F_1\left(\frac{1}{2} - \frac{1}{2m},\frac{1}{2}+\frac{1}{2m},1,1-u\right)\, .\label{classicalPeriodDual_GeneralChebyshev} 
	\end{align}
	Then, the associated classical action and their dual are written as
	\begin{align}
		a_0(u)& = \frac{\sqrt{2} \pi}{m}\, \sin\left(\frac{\pi}{2m}\right)\, u \, {}_2 F_1\left(\frac{1}{2} - \frac{1}{2m},\frac{1}{2}+\frac{1}{2m},2,u\right)\, , \label{classicalAction_GeneralChebyshev} \\
		a^\mrmD_0(u)& = -i \frac{\sqrt{2} \pi}{m}\, \sin\left(\frac{\pi}{m}\right)\, (1-u) \, {}_2 F_1\left(\frac{1}{2} - \frac{1}{2m},\frac{1}{2}+\frac{1}{2m},2,1-u\right)\, ,\label{classicalActionDual_GeneralChebyshev} 
	\end{align}
	where $m = \frac{3}{2},2,\frac{5}{2},3,\cdot\cdot\cdot$. An important relationship between the periods (actions) and their dual periods (actions) is
	\begin{equation}
		\msfw^\mrmD(u) = i \sqrt{\k_m}\, \msfw(1-u)\, , \qquad a^\mrmD(u) = -i \sqrt{\k_m}\, a(1-u)\, . 
		\label{Action_Dual_RelationGeneral}
	\end{equation}
	Finally, note that periodic potential is a special case of the Chebyshev potentials. In this case, the classical action and its dual are written as
	\begin{align}
		a_0(u) &=  \sqrt{2}\, \pi u {}_2 F_1\left(\frac{1}{2},\frac{1}{2},2,u\right)\, ,  \label{classicalAction_Periodic}\\ 
		a_0^\mrmD(u) &= -i \sqrt{2} \pi\, (1-u) {}_2  F_1\left(\frac{1}{2},\frac{1}{2},2,1-u\right)\, ,  \label{classicalActionDual_Periodic}
	\end{align}
	and the period(action) and its dual are related to each other by
	\begin{equation}
		\msfw^\mrmD_0(u) = i \, \msfw_0(1-u)\, , \qquad a^\mrmD_0(u) = -i \, a_0(1-u)\,. \label{Action_Dual_RelationPeriodic}
	\end{equation}
	which corresponds to the case of  $\k_\mrmP=1$ in comparison with \eqref{Action_Dual_RelationGeneral}.

	\paragraph{\underline{Transition}:}
	Let us start with the transition from the sector $u<1$ to the $u>1$ sector. Using the expressions in \eqref{classicalPeriod_GeneralChebyshev}-\eqref{classicalActionDual_GeneralChebyshev}, it is straightforward to capture the evolution of \eqref{classicalWronskian} during the transition. Note that for $u>0$, $a_0^\mrmD(u)(\msfw^\mrmD_0(u))$ is an analytic function while $a_0(u)(\msfw_0(u))$ has a branch-cut along the real axis for $u \geq 1$. Then, in the $u>1$ sector, an analytic continuation is needed to be imposed for $a_0(u)(\msfw_0(u))$. For this reason, we introduce $u\rightarrow u \pm i\ve$. Note that this is equivalent to the analytic continuation of the Stokes diagrams for phases $\t_u= \pm \pi^\mp$ we discuss in Section \ref{Section: AiryTypeEWKB}. 
	
	As $a_0$ and $a_0^\mrmD$ are given by the hypergeometric function, we use its analytic properties \cite[Eqs. 15.2.3, 15.8.1]{NIST:DLMF} to obtain the discontinuity on the branch cut as
	\begin{align}
		\D a_0(u) &= \lim_{\ve \rightarrow 0} \left(a_0(u +i\ve) - a_0(u-i\ve)\right) \nonumber \\
		&= - i \frac{\sqrt{2}\, \pi}{m}\, \sin\left(\frac{\pi}{m}\right)\, (1-u)\, {}_2 F_1\left(\frac{1}{2} - \frac{1}{2m},\frac{1}{2} + \frac{1}{2m},2, 1-u\right) = a_0^\mrmD(u) \, .\label{DiscontinuityGeneral_Outermost}
	\end{align}
	An important point of the discontinuity of the classical action \eqref{DiscontinuityGeneral_Outermost} is that at least for genus-1 potentials, it holds at any order in the WKB expansion. This can be seen from the general expression of the quantum corrections in Eq.~(8.2) of \cite{Basar:2017hpr}, where at each order the corrections was written in terms of $a_0$ and $a_0^\mrmD$. Then, we get the analytical continuation of $a(u,g)$ as 
	\begin{equation}\label{ActionTransition_GeneralOutermost}
		a(u<1,g) \rightarrow a(u>1,g) \pm \frac{1}{2} a^\mrmD(u>1,g)\, ,
	\end{equation}
	where $\ve\rightarrow 0$ is assumed in $u>1$ sector. This provides a guideline for the analytic continuations of $\Pi_A$ and $\Pi_B$ in specific examples of periodic and symmetric double-well potentials which we discuss in Sections \ref{Section: PeriodicPotential} and \ref{Section: DoubleWell}. Note that the actions of inner wells and barriers are known to be proportional to the outermost ones \cite{Basar:2017hpr}. 
	Therefore, a similar relationship can be obtained for them as well. 
	We don't examine these transitions in detail.
	
	Following the same arguments for the periodic potential, the discontinuity is found as
	\begin{equation}
		\D a_0(u) = - i 2\sqrt{2} \pi (1-u) {}_2 F_1\left(\frac{1}{2},\frac{1}{2},2,1-u\right) = 2a_0^\mrmD(u)\, , \label{DiscontinuityPeriodic}
	\end{equation}
	and we get the analytic continuation of $a(u,g)$ from $u<1$ to $u>1$ as
	\begin{equation}\label{ActionTransition_Periodic}
		a(u<1,g) \rightarrow a(u>1,g) \pm  a^\mrmD(u>1,g)\, ,
	\end{equation}
	where $\ve\rightarrow 0$ is assumed in $u>1$ sector. 
	
	Both \eqref{ActionTransition_GeneralOutermost} and \eqref{ActionTransition_Periodic} can be used to analyze the transition of the resurgence relations between $a(u)$ and $a^\mrmD(u)$. At the classical level, inserting these equations in \eqref{classicalWronskian} leads to the same equation. It is straightforward to generalize this observation to all orders through the quantum corrections to the Wronskian relation in \cite{Basar:2017hpr}. This explicitly shows that although $a(u)$ and $a^\mrmD(u)$ change as $u$ varies, the resurgence relation between them stays intact. In Sections \ref{Section: PeriodicPotential} and \ref{Section: DoubleWell}, we discuss how this observation realizes in the EWKB framework.

	\paragraph{\underline{Duality transformation}:}
	Now we examine the Wronskian relation under the modular $S$-transformation, which maps the original theory to its $S$-dual:
	\begin{align}
		S: \tau(u) \mapsto \tilde{\tau}(u_\mcalD) =-\frac{1}{\tau(1-u_\mcalD)} \, , \label{S-transform_ModularParameter}
	\end{align}
	where 
	\begin{equation}
		\tau(u) = \frac{\msfw_0^\mrmD(u)}{\msfw_0(u)}, \qquad \ttau(u_\mcalD) = \frac{\tilde \msfw_0^\mrmD(u_\mcalD)}{\tilde \msfw_0(u_\mcalD)}\,, \label{modularParameters}
	\end{equation}
	are the modular parameters for the original theory and its dual, respectively. 
	
	For Chebyshev potentials, the modular $S$ transformation also generates $u \rightarrow 1-u$. 
	This is nothing but an exchange of the singular points $u=0$ and $u=1$ on the moduli space and it results in the exchange of perturbative and non-perturbative WKB cycles. The latter one is manifest in the explicit expressions of $a_0(u)$ and $a_0^\mrmD(u)$ in \eqref{Dictionary_Perturbative_Well}-\eqref{Dictionary_NP_Barrier_Bounce} which we got via the Weber-type EWKB approach. In that sense, the transformation of $V\rightarrow V_\mrmD  = 1 - V $ can be seen as a consequence of the $S$-transformation. Or equivalently, we claim that the Weber-type EWKB setup makes the $S$-duality between minima and maxima of a potential manifest to all orders in WKB expansion. 
	
	A straightforward consequence of the duality is that a barrier top of the potential $V(x)$ acts as the bottom of a well for the dual potential $V_\mrmD$ and in this sense, a Weber-type EWKB analysis around the barrier top probes the dual theory. In \cite{Codesido:2017dns,Codesido:2017jwp}, this aspect of the $S$-duality was discussed  in relation with the topological string free energies. The authors showed that all order expansion of the free energies of dual theories can be transformed to each other via $S$-transformation. 
	
	As the duality acts on all order WKB expansions or equivalently, it displays itself in EWKB setup, the entire resurgence structures should have a duality transformation as well. In this regard, it is expected to obtain P-NP relation of the dual theory via the modular $S$-transformation \eqref{S-transform_ModularParameter} for genus-1 Chebyshev potentials.
	
	Let us first make an observation at the classical level: Using \eqref{Action_Dual_RelationGeneral}, we observe that the periods of the dual theory are identified as
	\begin{align}
		\tilde \msfw_0(u_\mcalD) = \sqrt{\k_m} \, \msfw_0(u_\mcalD)\, , \qquad \tilde \msfw_0^\mrmD(u_\mcalD)  = \frac{\msfw_0^\mrmD(u_\mcalD)}{\sqrt{\k_m}}\, .\label{S-duality_Periods}
	\end{align}
	Equivalently, we get the actions of the dual theory as
	\begin{equation}
		\ta_0(u_\mcalD) = \sqrt{\k_m} \, a_0(u_\mcalD)\, , \qquad \ta_0^\mrmD(u_\mcalD)  = \frac{a_0^\mrmD(u_\mcalD)}{\sqrt{\k_m}} \, .\label{S-duality_Actions} 
	\end{equation}
	Then, $S$-transformation of the classical Wronskian relation \eqref{classicalWronskian} becomes
	\begin{equation}
		2i\, \ta^\mrmD_0(0)\, \tilde \msfw_0(0) = \ta_0(u_\mcalD)\,\tilde \msfw^\mrmD_0(u_\mcalD) - \tilde \msfw_0(u_\mcalD)\, \ta_0^\mrmD(u_\mcalD)\, ,
	\end{equation}
	which is in the same form with the original theory. Note that an equivalent transformation of the P-NP relation \eqref{DunneUnsal_General} at the classical level is given by 
	\begin{align}
		\mcalF^D_0(\tu) = \sqrt{\k_m} \mcalF_0(\tu), \qquad \mcalG^\mrmD_0(\tu) = \frac{\mcalG_0}{\sqrt{\k_m}}\, , \label{Duality_WeberObjects_Classical}
	\end{align}
	as well as $S_\mrmI^\mrmD = \frac{S_\mrmI}{\sqrt{\k_m}}$, which is the equality of the first terms in the latter relation. 
	
	The relations \eqref{S-duality_Periods} and \eqref{S-duality_Actions}, however, doesn't generalize to the quantum corrections except special cases of periodic and cubic potentials for which $\k=1$. The same is true for all order expressions of $\mcalF(\tu,g)$ and $\mcalG(\tu,g)$ and the equations in \eqref{Duality_WeberObjects_Classical} don't hold at all orders. On the other hand, we notice that introducing $S$-duality transformations based on \eqref{Duality_WeberObjects_Classical} as  
	\begin{equation}
		\mcalF \rightarrow \frac{\mcalF^\mrmD}{\sqrt{\k_m}} , \qquad \mcalG \rightarrow \sqrt{\k_m}\,\mcalG^\mrmD \,,  \label{Duality_WeberObjects_AllOrder}
	\end{equation}
	maps the P-NP relation \eqref{DunneUnsal_General} for the potential $V$ to the one for the dual theory characterized by $V_\mrmD = 1 - V$ and conjecture 
	\begin{equation}
		\mcalG^\mrmD(\tu,g) = - \int \frac{\mrmd g}{g}\left[\a_1^\mrmD \, \frac{S_\mrmI^\mrmD}{g} \, \frac{\dee \tu\left(\mcalF^\mrmD,g\right)}{\dee \mcalF^\mrmD} + \a_2^\mrmD \mcalF^\mrmD \right]\,, \label{DunneUnsal_GeneralDual}
	\end{equation}
	as the P-NP relation for the dual theory with the following identifications:
	\begin{equation}
		\a_1^\mrmD = \sqrt{\k_m} \a_1\, , \qquad \a_2^\mrmD = \frac{\a_2}{\k_m}. \label{DU_parameterDuality}
	\end{equation}
	We check this conjecture against the periodic and symmetric double-well potentials in Sections~\ref{Section: PeriodicPotential} and~\ref{Section: DoubleWell}, where we also provide modifications for the region around barrier tops.

	
	\section{Periodic potential}\label{Section: PeriodicPotential}
	
\begin{figure}[t]
		\centering
		\includegraphics[width=0.7\textwidth]{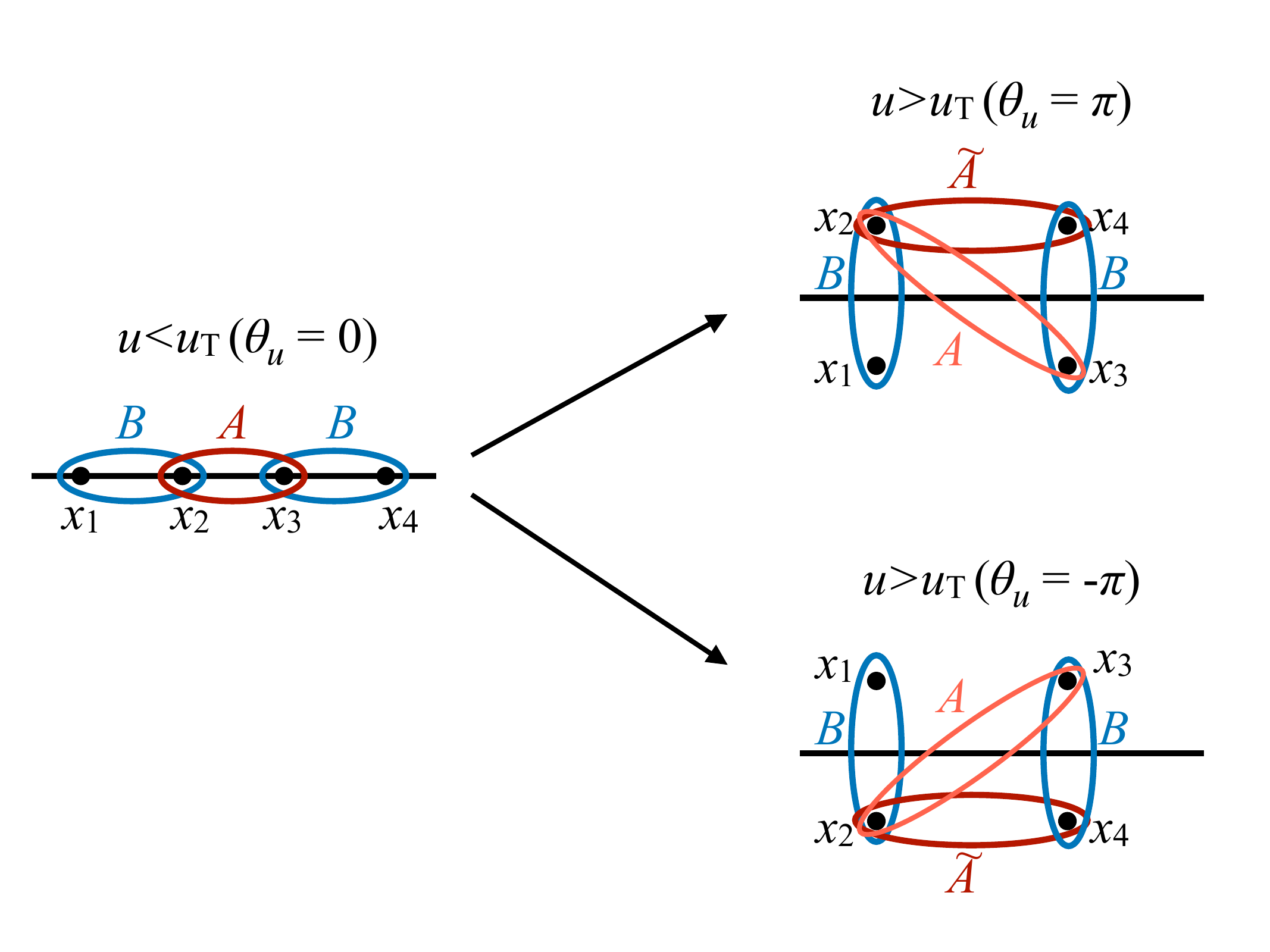}
		\caption{The transition of turning points and the cycles between $u<u_\mrmT$ and $u>u_\mrmT$ in the periodic potential, which is characterized by the orientation of the turning points and the geometry of the $A$ and $B$ WKB cycles.} \label{Figure: TurningPoints_Periodic}
	\end{figure}

	In this section, we apply the Airy and Weber-type EWKB methods to quantize one-dimensional periodic potential in all sectors and examine the connection between different sectors. Throughout this section, we consider the potential 
	\begin{equation}
		V_\mrmP(x) = \frac{1}{2}\sin^2 x\, ,  \label{Potential_Periodic}
	\end{equation} 
	and the exact quantization conditions are given by the Bloch-wave boundary condition: 
	\begin{equation}\label{BoundaryCondition_Peridoic}
		\psi(x+\pi) = e^{-i\t} \psi(x).
	\end{equation}
	The corresponding quantum mechanical system has two sectors, namely $u<u_\mrmT$ and $u>u_\mrmT$ for $u_\mrmT = \frac{1}{2}$. In the EWKB framework, the change between the two sectors is characterized by the orientation of the turning points and equivalently, the geometry of the $A$ and $B$ WKB cycles. We illustrate this change in Fig.~\ref{Figure: TurningPoints_Periodic}.

	The entire spectrum in both sectors are encoded by the actions $a(u)$ and $a^\mrmD(u)$. With the dictionary we provide in Section~\ref{Section: AiryWeber_Dictionary}, their explicit expressions are written as in the following:
	\begin{itemize}
		\item \underline{$u<u_\mrmT$ sector}:
		\begin{align}
			\Pi_A &= e^{-\frac{2\pi i}{g} a(u,g)} = e^{-2\pi i \mcalF\left(\frac{u}{g},g\right)} \, ,\label{AcycleBelow_Periodic}\\
			\Pi_B &= e^{-\frac{2\pi i}{g} a^\mrmD(u,g)} = e^{-\mcalG(\frac{u}{g},g)} \frac{2\pi \left(\frac{g}{\mcalC_\mathrm{P}}\right)^{-2 \mcalF\left(\frac{u}{g},g\right)}}{\left[\G\left(\frac{1}{2} + \mcalF\left(\frac{u}{g},g\right)\right)\right]^2} \, . \label{BcycleBelow_Periodic}
		\end{align} 
		\item \underline{$u>u_\mrmT$ sector}: 
		\begin{align}
			\Pi_{\tA} &= e^{-\frac{2\pi i}{g} a(u,g)} = e^{- \mcalG^\mrmD(i \frac{1-u}{g},ig)} \frac{2\pi \left(\frac{ g}{\mcalC_\mathrm{P}}\right)^{-2 \mcalF^\mrmD\left(i\frac{1-u}{g},ig\right)}}{\left[\G\left(\frac{1}{2} +   \mcalF^\mrmD \left(i\frac{1-u}{g},ig\right)\right)\right]^2}\, , \label{AcycleAbove_Periodic}\\ 
			\Pi_B &= e^{-\frac{2\pi i}{g} a^\mrmD(u,g)} = e^{-2\pi i \, \left( \mcalF^\mrmD\left(i\frac{1-u}{g},i g\right)\right)}\, . \label{BcycleAbove_Periodic}
		\end{align}
	\end{itemize}
	For both cases, we get the constant $\mcalC_\mrmP$ as 
	\begin{equation}
		\mcalC_\mrmP = 2\,\o \,\exp \int_0^{\pi} \mrmd x\, \left(\frac{1}{\sqrt{2 V(x)}} - \frac{1}{x} - \frac{1}{x - \pi} \right) = 8\o\, , \label{Weber_PrefactorConstant_Periodic}
	\end{equation}
	where $\o=1$ for $u<u_\mrmT$ case and $\o=i$ for $u>u_\mrmT$ case. Note also that initially the integral of the constant $\mcalC_\mrmP$ around the barrier top is written in a slightly different form in view of \eqref{PrefactorLog1}. But it is straightforward to bring it to the form of \eqref{Weber_PrefactorConstant_Periodic}. Finally, the functions $\mcalF$, $\mcalG$ and their dual $\mcalF^\mrmD$, $\mcalG^\mrmD$ in \eqref{AcycleBelow_Periodic}-\eqref{BcycleAbove_Periodic} can be expressed in series expansions of $u$ and $g$. We provide the first few orders at the end of this section, when we discuss the $S$-duality between $u\sim 0$ and $u\sim u_\mrmT$ regions.

	In the literature, the sector $u<u_\mrmT$ and its resurgence structure was well-discussed using various methods including the EWKB method \cite{Sueishi:2021xti}. 
	In particular, the trans-series of $u$ was uncovered and the cancellation of ambiguous neutral bion contributions against the Borel summation of the divergent perturbation series was shown in \cite{Zinn-Justin:2004vcw,Zinn-Justin:2004qzw, Dunne:2013ada, Dunne:2014bca,Sueishi:2021xti}. In addition to that an explicit constructive resurgence relation between perturbative and non-perturbative sector was established in \cite{Dunne:2013ada, Dunne:2014bca}. In our notation this P-NP relation for the periodic potential is written for $u<u_\mrmT$ as
	\begin{equation}\label{DunneUnsal_Periodic}
		\mcalG(\tu,g)= - 2\int \frac{\mrmd g}{g}\left[\, \frac{S_\mrmI}{g}\frac{\dee \tu(\mcalF,g)}{\dee \mcalF} + \mcalF(\tu,g)\right]\, ,
	\end{equation}
	where 
	\begin{equation}\label{InstantonAction_DW}
		S_\mrmI = \int_{0}^{\pi} \mrmd x\, \sqrt{2 V(x)} = 2 \, ,
	\end{equation}
	is the instanton action and it is directly related to the classical dual action and the function $\mcalG$ as
	\begin{equation}
		S_\mrmI = \lim_{g\rightarrow 0} \left[g\, \mcalG(\tu,g)\right]\, .
	\end{equation}
	Note that in \eqref{DunneUnsal_Periodic}, we used $\tu$ rather than the original spectral parameter $u$.
	
	The $u>u_\mrmT$ sector, on the other hand, is less studied \cite{Connor84,weinstein1985hill,weinstein1987asymptotic,Basar:2015xna}. Of particular importance for our discussion, in \cite{Basar:2015xna}, Ba\c{s}ar and Dunne carried out a resurgence analysis using all order WKB expansions for all values of $u$. They analyzed the connection between the two sectors using the all order WKB method. They showed the smooth transition of the trans-series of $u$ between the two sectors despite the singularity of the WKB action $a(u)$ at $u=u_\mrmT$ which should be taken care of via appropriate analytic continuations.  \\
	
	In the following, we analyze the transition between $u<u_\mrmT$ and $u>u_\mrmT$ sectors in the EWKB framework. In particular,
	
	\begin{itemize}
		\item We discuss the analytic continuation of the Stokes geometry between the two sectors based on our construction in Section \ref{Section: AiryTypeEWKB}. We show that the smooth transition of the spectrum shows itself in the median summations of the exact quantization conditions which stay intact upon the transition with appropriate redefinitions of the WKB cycles. In this way, we make the estimations in \cite{Basar:2015xna} manifestly exact. 
		\item We provide explicit series expansions of $\mcalF$, $\mcalG$ around $u\sim0$ and $\mcalF^\mrmD$, $\mcalG^\mrmD$ around $u\sim u_\mrmT$. In this way, we uncover the P-NP relation around the top of the barrier and show that it has the same form with \eqref{DunneUnsal_Periodic}. We elaborate that the P-NP for $u\sim u_\mrmT$, in fact, is the resurgence relation for the dual theory which happens to be the same in this particular case of the periodic potential.
	\end{itemize}

	\subsection{Transition between different sectors}
\begin{figure}[t]
		\centering
		\begin{subfigure}[h]{0.45\textwidth}
			\caption{\underline{Diagrams at $\t_\d=0^\mp$}}	\label{Figure: StokesDiagrams_PeriodicBelow}
			\includegraphics[width=\textwidth]{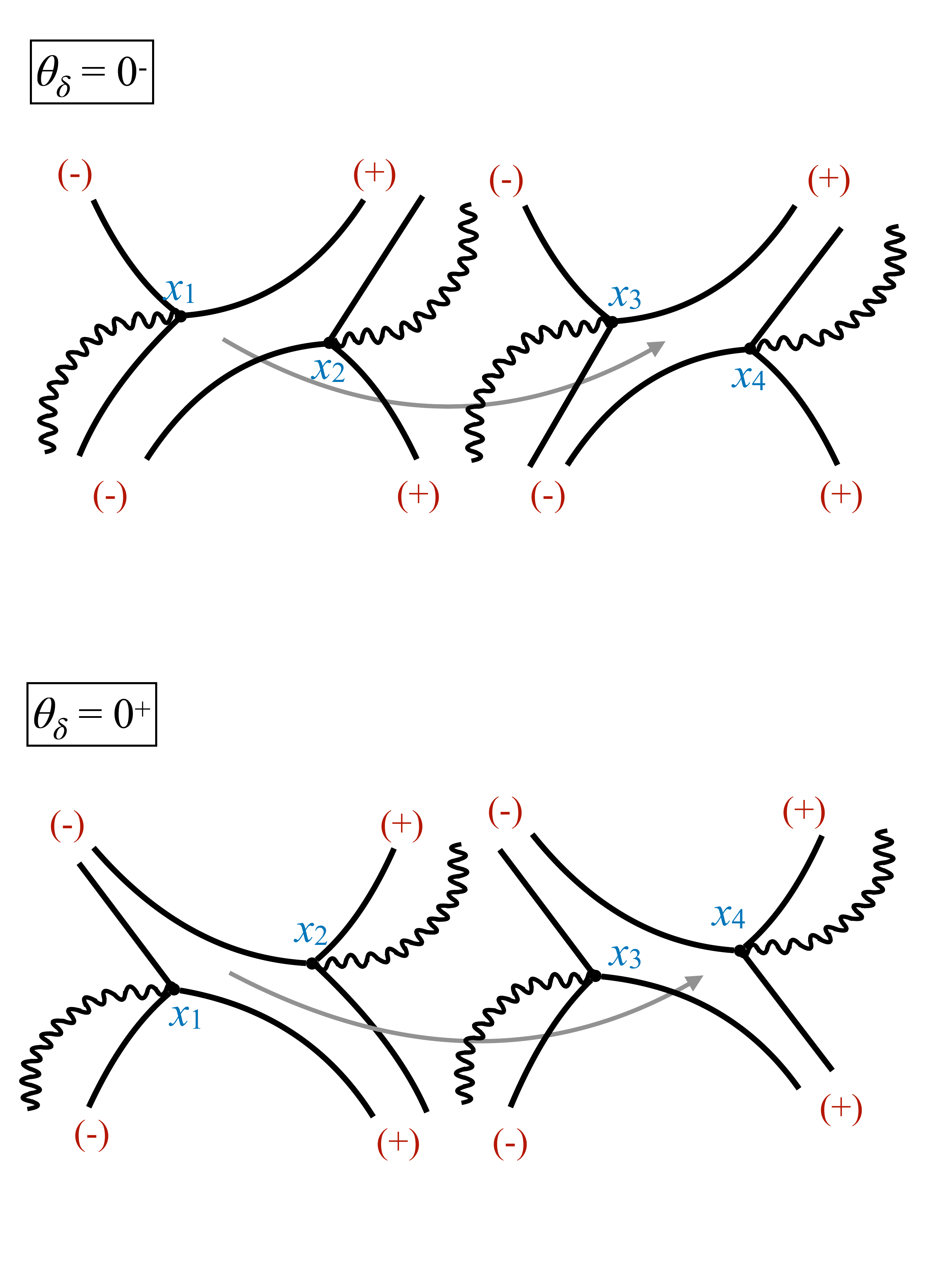}
		\end{subfigure}
		~\hfill 
		\begin{subfigure}[h]{0.45\textwidth}
			\caption{\underline{Diagrams at $\t_\d=\mp \pi^\pm$}}	\label{Figure: StokesDiagrams_PeriodicAbove}
			\includegraphics[width=\textwidth]{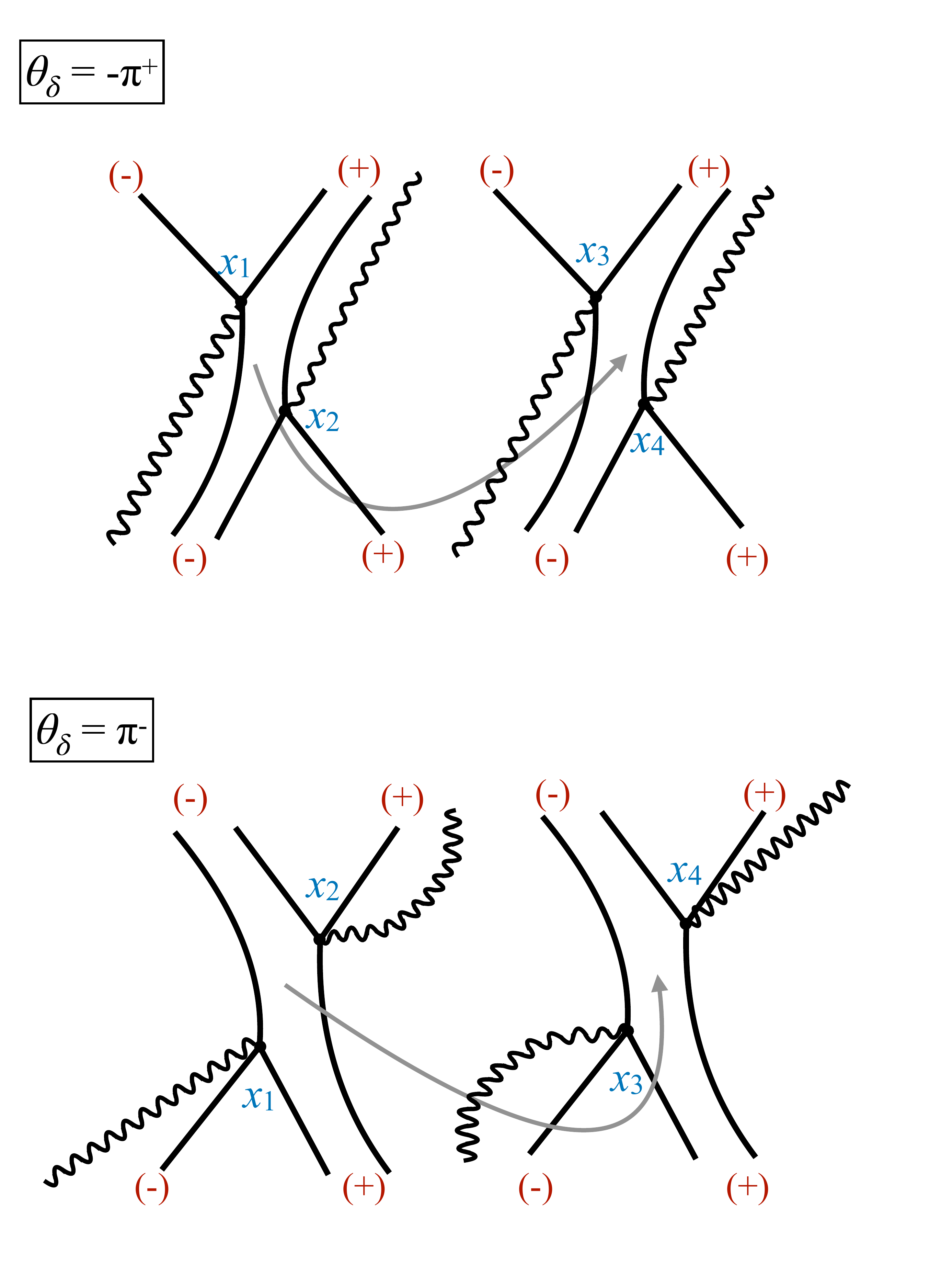}
		\end{subfigure}
		\caption{Stokes diagrams of the periodic potential. \textbf{(a)}~The diagrams in $u<u_\mrmT$ sector. The analytic continuations $\t_\d = 0^\mp$ are equivalent to $\arg g = 0^\pm$. \textbf{(b)}~The diagrams in $u>u_\mrmT$ sector. The diagrams at phases $\t_\d = \mp \pi^\pm$ are continuously connected to $\t_\d=0^\pm$.} \label{Figure: StokesDiagrams_Periodic}
	\end{figure}

	Let us start with the Airy-type EWKB. For $u\in\left(0,u_\mrmT\right)$, The Stokes geometry remains the same. Without an analytic continuation, as $u$ passes through $u=u_\mrmT$, the simple turning points first merge and then scatter off. As a result, we end up with a different Stokes geometry for the sector $u>u_\mrmT$. (See Fig.~\ref{Figure: StokesDiagrams_Periodic}).
	
	As we discussed in Section \ref{Section: AiryTypeEWKB}, we use the analytic continuation of $u$ to connect different sectors. For this reason, we parameterize $u$ as $u = u_\mrmT - \d$ and the Airy-type algebraic curve becomes 
	\begin{equation}
		P_\mrmA = 2\left(V_\mrmP(x) - u_\mrmT + |\d| e^{i\t_\d}\right), \label{Curve_Periodic}
	\end{equation}
	where we introduced $\t_\d$ as complexification parameter. In this way, we can carry out the analytic continuation of the Stokes diagrams in the same manner as the inverted harmonic oscillator. Note that this is provided by the fact that all turning points in the Airy-type analysis are simple and around a turning point $\s_\mrmA$ has the same form as \eqref{Airy_AroundTurningPoint}; therefore, the Stokes lines deform in the same way we discuss in Fig.~\ref{Figure: StokesDiagramRotation_All}. 
	
	In order to understand the transition between the two sectors, we consider $\d$ as a small parameter and impose the analytic continuation around the top of the barrier, where the potential $V(x)$ approximates to the inverted oscillator. Then, we observe that the transition of individual Stokes diagrams for $\t_\d:0 \rightarrow \pm \pi$ takes place in the same way as in Fig.~\ref{Figure: IHO_Transition}. Restricting ourselves to the region $x\in \left[-\pi,\pi\right]$, we sketched the Stokes diagrams for both sectors with the corresponding analytic continuations in Fig.~\ref{Figure: StokesDiagrams_Periodic}. 
	
	We first observe that for the sector $u<u_\mrmT$, the diagrams for $\t_\d = 0^\mp$ are equivalent to the ones with the analytic continuation of $g$ with $\Im \, g = 0^\pm$. (See e.g.~\cite{Sueishi:2021xti}.) As a result the exact quantization conditions for $\t_\d = 0^\mp$ is expected to recover the known results in \cite{Sueishi:2021xti}. In addition to that in Fig.~\ref{Figure: StokesDiagrams_Periodic}, it is clear that with the analytic continuations $\t_\d:0^\mp \rightarrow \mp \pi^\pm$  we introduced, the paths connecting $x=-\frac{\pi}{2}$ to $x=\frac{\pi}{2}$ pass through the same type of Stokes lines and branch points in the same order. Therefore, the transition matrices in both sectors are the same:
	\begin{align}
		T_{\t_\d=0^-} &= N_\mrmA^{3,4} M_\mrmA^- M_\mrmA^\mrmB N_\mrmA^{2,3} M_\mrmA^+ M_\mrmA^-  = T_{\t_\d=-\pi^+}\, , \label{TransitionMatrix_PeriodicBelow1}\\ \nonumber \\
		T_{\t_\d=0^+} &= N_\mrmA^{3,4} M_\mrmA^+ M_\mrmA^- M_\mrmA^\mrmB N_\mrmA^{2,3} M_\mrmA^+ = T_{\t_\d=\pi_-}  \, . \label{TransitionMatrix_PeriodicBelow2}
	\end{align}
	Then, imposing the periodic boundary conditions, we end up with the same exact quantization conditions for both sectors:
	\begin{align}
		D_{\t_\d = 0^\mp} = D_{\t_\d = \pm\pi^\pm} = 2\cos\t - \sqrt{\Pi_A^{-1} \Pi_B^{-1} } - \sqrt{\Pi_A \Pi_B^{-1}} - \sqrt{\Pi_A^{\mp 1} \Pi_B} = 0. \label{EQC_PeriodicBelow}
	\end{align}
	
	For $u<u_\mrmT$, it is known that this is equal to the ZJJ's exact quantization condition \cite{Zinn-Justin:2004vcw,Zinn-Justin:2004qzw} which becomes apparent when the Weber-type approach is incorporated. For completeness we write it using the explicit expressions in \eqref{AcycleBelow_Periodic} and \eqref{BcycleBelow_Periodic}:
	\begin{equation}\label{ExactQC_ZJJ}
		2\cos\t = \frac{\sqrt{2\pi}\, e^{\frac{\mcalG(\tu,g)}{2}}}{\Gamma\left(\frac{1}{2} - \mcalF(\tu,g)\right)} \left(\frac{g}{C_\mrmP}\right)^{\mcalF(\tu,g)} +  \frac{\sqrt{2\pi}\, e^{-\frac{\mcalG(\tu,g)}{2}}}{\Gamma\left(\frac{1}{2} + \mcalF(\tu,g)\right)}\left(-\frac{g}{C_\mrmP}\right)^{-\mcalF(\tu,g)}\, .
	\end{equation}  
	As pointed out in \cite{Dunne:2013ada, Dunne:2014bca}, this equation can also be turned into a more explicit P-NP resurgence relation, which we write in \eqref{DunneUnsal_Periodic}.
	
	The ZJJ quantization condition, however, is before the Borel summation procedure and it contains the ambiguous multi-instanton (bion) terms which, as we mentioned above, are canceled against the Borel summation of the perturbative expansion \cite{Zinn-Justin:2004vcw,Zinn-Justin:2004qzw, Dunne:2013ada, Dunne:2014bca,Sueishi:2021xti}. This leads to the reality of the spectrum. 
	
	In addition to that, in \cite{Basar:2015xna}, it was argued that the transition between the different sectors of this real spectrum is smooth. This is the case after the Borel summation and the cancellation of the ambiguities. In the EWKB framework, this procedure is equivalent to taking the median summation of the exact quantization condition \eqref{EQC_PeriodicBelow}. In this paper, we don't introduce the median summation in connection with the directional Borel summation procedure. We summarize basic facts in the following and refer reader to \cite{Kamata:2023opn} for a detailed introduction and pedagogical discussion:
	
	The main object we use in our discussion is the \textit{Stokes automorphism} operation, $\mfrS$, which encodes the discontinuity when a Stokes diagram is degenerate.~In that sense, $\mfrS$ relates two exact quantization conditions connected by an analytic continuation which breaks the degeneracy. For example, for $u<u_\mrmT$ sector, around $\t_\d=0$, the exact quantization conditions $D_{\t_\d=0^\pm}$ are related via $\mfrS$, which we elaborate later.
	
	The median summation, on the other hand, is equal to  performing the Borel summation after removing the discontinuity from the exact quantization condition via the action of half-Stokes automorphism $\mfrS^{\pm 1/2}$. At $\t_\d=0$, the resulting object is expressed by \[\mcalS_\pm \left[\mfrS^{\mp 1/2} D_{\t_\d=0^\pm}\right]\, ,\] where $\mcalS_\pm$ is the directional Borel summation. This procedure leads to an ambiguity free and real spectrum and it is equivalent to the cancellation of the ambiguities in a more direct but formal manner. 
	
	Note that although, it is desirable to obtain the Borel summed expressions for the spectrum, it is almost never a tractable process. In the EWKB framework, on the other hand, performing $\mfrS^{\pm 1/2}$ on the exact quantization conditions brings them to manifestly real forms\footnote{For explicit examples on the disappearance of the imaginary terms in the trans-series arising from the exact quantization conditions, we refer to \cite{Kamata:2021jrs}. For simple toy model examples, see also Sections III.A.1 and III.A.2 in \cite{Kamata:2023opn}.} which indicates the reality of the spectrum \cite{DDP2}. In our discussion, we define the exact quantization condition at a phase $\t_\d$ with discontinuities removed as $D^\med_{\t_\d}$ and call it ``\textit{median QC}''. In view of the discussions in \cite{Basar:2015xna}, we expect $D^\med$ stays intact upon transition between the two sectors.

	

	\paragraph{\underline{Smooth transition of $\bm {D^{\med}}$}:}  
	
	Now, we discuss how $D^\med$ stays intact in both sectors of the periodic potential or equivalently at phases $\t=0$ and $\t=\pm \pi$. Note that along with the Stokes automorphism, the preservation of the quantization condition is provided by an appropriate analytic continuation of the $A$-cycle upon transition barrier in a compatible way with the analytic properties of the corresponding Airy-type WKB action.\\
	
	\noindent{\textit{\underline{Below the Barrier Top}}:} Let us first focus on the $u<u_\mrmT$ sector.  For the diagrams in Fig.~\ref{Figure: StokesDiagrams_PeriodicBelow}, this relationship is established via to the discontinuity of $\Pi_A$. At $\t_\d=0$, it is expressed as
	\begin{equation}
		\Pi_{A_{0^-}} = \Pi_{A_{0^+}} \left(1 + \Pi_{B_{0^+}}\right)^{\left\<A,B\right\>} \, , \label{StokesAuto_Periodic}
	\end{equation}
	where $\<A,B\>=2$ is the intersection number of the cycles $A$ and $B$. Then, around $\t_\d = 0$, the quantization conditions $D_{\t_u = 0^\mp}$ are related to each other as
	\begin{equation}
		D_{\t_\d=0^+} = \mfrS_{0} D_{\t_\d = 0^-}, \label{StokesAuto_PeriodicBelow}
	\end{equation}
	where 
	\begin{equation}
		\mfrS_{0}: \Pi_{A_{0^-}} \mapsto \Pi_{A_{0^+}} \left(1+\Pi_{B_{0^+}}\right)^2\, .
	\end{equation}
	
	The median QCs in this perspective can be obtained by $\mfrS^{\pm 1/2} D_{\t_\d =0^\mp}$. 
	From \eqref{StokesAuto_PeriodicBelow}, it is clear that the median QCs are the same for both $D_{\t_\d = 0^\mp}$ and using \eqref{EQC_PeriodicBelow} and \eqref{StokesAuto_Periodic}, we get 
	\begin{align}
		D^{\med}_{u<u_\mrmT} &= \mfrS_{0}^{1/2} D_{\t_\d=0^-} = \mfrS_0^{-1/2}D_{\t_\d=0^+} \\
		& = 2\cos \t - \left(1 + \Pi_B\right)^{1/2} \left(\sqrt{\Pi_A^{-1} \Pi_B^{-1}} + \sqrt{\Pi_A \, \Pi_B^{-1}}\right). \label{EQC_MedianSum_PeriodicBelow}
	\end{align}
	Note that $D^\med_{u<u_\mrmT}$ is manifestly real. 
	To see this, let us denote $\mcalC$ as complex conjugation operator. Then, since $\mcalC\left[\Pi_A\right] = \Pi^{-1}_A$ and $\mcalC\left[\Pi_B\right] = \Pi_B$, it is clear that
	\begin{equation}
		\mcalC\left[D^\med_{u<u_\mrmT}\right] = D^\med_{u<u_\mrmT} \, . \label{ComplexConj_EQC_Periodic}
	\end{equation}
	
	The reality of $D^\med_{u<u_\mrmT}$ means that the trans-series of $u$ is also real and ambiguity free \cite{DDP2}. As we discussed above, this is one of the crucial aspects of the resurgence analysis of the periodic potential \cite{Zinn-Justin:1981qzi,Zinn-Justin:1983yiy,Zinn-Justin:2004vcw,Zinn-Justin:2004qzw, Dunne:2013ada, Dunne:2014bca}. The EWKB analysis makes it manifest to all order. We can now extend it to the sector $u>u_\mrmT$. \\

	\noindent \textit{\underline{Above the Barrier Top}}: The extension to the section $u>u_\mrmT$ is not straightforward. One indication of this is the analytic continuation in the geometry of the cycles $A$ and $B$ upon the transition in Fig.~\ref{Figure: TurningPoints_Periodic}. This is equivalent to the analytic continuation for the leading order actions $a_0(u)$ and $a_0^\mrmD$ we discuss in Section \ref{Section: ChebyshevTransitionDuality}: 
	\begin{equation}
		a_0(u) \rightarrow a_0(u) \pm a_0^\mrmD(u)\, , \qquad a^\mrmD_0(u) \rightarrow a^\mrmD_0(u)\, . \label{AC_periodic}
	\end{equation} 
	Note that the analytic continuation of $a_0(u)$ is provided by $u \rightarrow u \pm i \ve$ and it is equivalent to $\t_\d: 0^\pm \rightarrow \pm \pi^\mp$ in our discussion. Then, since we search for a smooth transition between the sectors $u<u_\mrmT$ and $u>u_\mrmT$, we need to adjust the cycles $A$ and $B$ accordingly. The analytic continuation in \eqref{AC_periodic} suggests that for $\t_\d: 0^\pm \rightarrow \pm \pi^\mp $, these transitions are incorporated into the exact quantization conditions via
	\begin{equation}\label{AC_ExponentialAction_Periodic}
		\Pi_{A_{\pm\pi}} \rightarrow \Pi_{\tA_{\pm\pi}} = \Pi_{A_{\pm\pi}} \Pi_{B_{\pm\pi}}^{\pm 1}, \qquad  \Pi_{B_{\pm\pi}} \rightarrow \Pi_{B_{\pm\pi}}.
	\end{equation}
	
	Note that the transformation of the $A$-cycle can also be inferred from the application of the Weber-type approach around the barrier top. Therefore, in the EWKB framework, these transformations can also be understood from a geometric point of view. This is clear for $\t_\d:0^+ \rightarrow +\pi^-$ as the cycle $\tA_\pi$ is simply given by the combination of $A_\pi$ and $B_\pi$ cycles, i.e. $\tA_\pi = A_\pi + B_\pi$. For $\t_\d:0^- \rightarrow - \pi^+$, on the other hand, this connection is not so clear: 
	The key point in this case is the difference in the branch cut choices\footnote{We elaborate on the difference branch cut choices in Appendix~\ref{Section: BranchCut_Appendix}.} of the corresponding Stokes diagrams at $\t_\d = \pm \pi^\mp$. This difference leads to the quantization conditions at $\t = \pm \pi$ being no longer related to each other by a Stokes automorphism, contrary to the case at $\t_\d=0$. In addition to that the geometry of the Stokes diagrams at $\t_\d= \pm \pi^\mp$ in Fig.~\ref{Figure: StokesDiagrams_PeriodicAbove} indicates that they are related to each other via $\Pi_{A_\pi} = \left(\Pi_{A_{-\pi}}\right)^{-1}$ as well as $\Pi_{\tA_\pi} = \left(\Pi_{\tA_{-\pi}}\right)^{-1}$. Then, at $\t_\d = -\pi$, the geometric relationship $\tA_{-\pi} = A_{-\pi}+B_{-\pi}$ should be interpreted as $\Pi_{\tA_{-\pi}}^{-1} = \Pi^{-1}_{A_{-\pi}} \Pi_{B_{-\pi}}$ as in \eqref{AC_ExponentialAction_Periodic}.
	
	Now we can compute the median QCs at $\t_\d=\pm \pi$: For convenience in the computations, we first invert the $B$-cycle and rewrite $D_{\t_\d=\pm \pi^\mp}$ in \eqref{EQC_PeriodicBelow} as
	\begin{equation}
		D_{\t_\d=\pm \pi^\mp} = 2\cos\t - \sqrt{\Pi^{-1}_{A_{\pm\pi}} \Pi_{B_{\pm\pi}^{-1}}} - \sqrt{\Pi_{A_{\pm\pi}} \Pi_{B_{\pm\pi}^{-1}}} -\sqrt{\Pi_{A_{\pm\pi}} \left(\Pi_{B_{\pm\pi}^{\mp 1}}\right)^{-1}}\, .
	\end{equation}
	With this inversion, we get the Stokes automorphism for the discontinuity of the $A$-cycles as
	\begin{equation}
		\Pi_{A_{\pm \pi^\pm} } = \Pi_{ A_{\pm \pi^\mp}} \left(1 + \Pi_{B^{-1}_{\pm \pi^\mp}}\right)^{\<A,B^{-1}\>}\, , \quad \<A,B^{-1} \> = -2 \, , \label{StokesAuto_Periodic2}
	\end{equation}
	which is slightly different than \eqref{StokesAuto_Periodic} due to the inversion of the $B$-cycle. Then, we find the median summations at $\t_\d = \pm \pi$ as
	\begin{align}
		D^{\med}_{\t_\d=\pm\pi} 
		& = \mfrS_{\pm \pi}^{\mp 1/2} D_{\t_\d = \pm \pi^\mp}  = 2\cos\t - \left(1 - \Pi_{B_{\pm \pi}}\right)^{1/2} \left(\sqrt{\Pi^{\pm 1}_{A_{\pm \pi}}} + \sqrt{\Pi^{\mp 1}_{A_{\pm \pi}} \Pi^{-2}_{B_{\pm \pi}}} \,  \right)\,,
	\end{align}
	where we invert $B^{-1}$-cycle back to the $B$-cycle. 
	
	Finally, using \eqref{AC_ExponentialAction_Periodic}, we obtain the analytic continuation of the exact quantization conditions \eqref{EQC_MedianSum_PeriodicBelow} for $u>u_\mrmT$ as
	\begin{align}
		D^\med_{u>u_\mrmT} & = 2\cos\t - \left(1 + \Pi_{B}\right)^{1/2} \left(\sqrt{\Pi_{\tA}^{-1} \Pi_B^{-1}} - \sqrt{\Pi_{\tA} \Pi_B^{-1}}\right)\, . \label{EQC_MedianSum_PeriodicAbove}	
	\end{align}
	Note that we dropped the subscripts $\pm\pi$, since $\Pi_{\tA} = \Pi_{\tA_\pi} = \Pi_{\tA_{-\pi}}$ and  $\Pi_{B} = \Pi_{B_\pi} = \Pi_{B_{-\pi}}$ 
	as a result of the transformations in \eqref{AC_ExponentialAction_Periodic}.  
	
	Then, the equality of \eqref{EQC_MedianSum_PeriodicBelow} and \eqref{EQC_MedianSum_PeriodicAbove} shows that the median QCs, so that median summations, of the exact quantization conditions in the sectors $u<u_\mrmT$ and $u>u_\mrmT$ are linked to each other by continuous analytic continuations and appropriate transformations.  Note that for $u>u_\mrmT$, $D^{\med}_{u>u_\mrmT}$ is manifestly real as it is for $u<u_\mrmT$. This presents the smooth connection of the resurgent trans-series of $u$ throughout the spectrum. 
	
	\paragraph{\underline{Remark}:} Obtaining the trans-series structure for the spectrum around the barrier top is much more cumbersome than it is for the bottom of wells. The main reason is around $u=u_\mrmT$, both $\tA$ and $B$ cycles are order of 1. Therefore, contrary to $u=0$ region, there is no clear perturbative quantization prescription, e.g. Bohr-Sommerfeld condition, to begin with. On the other hand, with the smooth connection between different sectors, it is possible to use the trans-series around $u=0$ to predict the spectrum around $u=u_\mrmT$. For first few non-perturbative orders in the trans-series, this continuous transition was previously discussed in \cite{Basar:2015xna} using all order WKB expansion. With the EWKB method, we make this conjecture manifestly exact. Further analysis in this line can be performed in the future.
	
	\subsection{The duality}
	Finally, let us discuss the duality between $u\sim 0$ and $u\sim u_\mrmT$ regions in the particular case of the periodic potential. As we stated in Section \ref{Section: ChebyshevTransitionDuality}, the dual theory is characterized by the potential $V_\mrmD = u_\mrmT - V$. Then, for $V(x) = \frac{1}{2}\sin^2 x$, the dual potential becomes
	\[V_\mrmD = \frac{1}{2}\cos^2 x = \frac{1}{2}\sin^2 \left(x - \frac{\pi}{2}\right)\,,\] 
	and the resulting dual theory is equal to the original one: It is given by a simple translation of the original potential. 
	
	The equivalence is also apparent in explicit computations of the Weber-type EWKB. For example, the constants $\mcalC_\mrmP$ around $u\sim 0$ and $u\sim u_\mrmT$ are the same. More importantly, the functions $\mcalF$ and $\mcalG$ have the same series expansions around $u\sim 0$ and $u\sim u_\mrmT$ up to an imaginary factor at each order: For both sectors, we express first few orders of the corresponding series expansions as
	\begin{align}
		\mcalF^{u\sim0}(\tu,g) &= u+g \left(\frac{1}{16}+\frac{u^2}{4}\right)+g^2 \left(\frac{5 u}{64}+\frac{3 u^3}{16}\right) +g^3 \left(\frac{17}{2048}+\frac{35 u^2}{256}+\frac{25 u^4}{128}\right) +\dots \, ,  \label{WeberExpansionPeriodic_PerturbativeBelow}\\ \nonumber \\
		\mcalG^{u\sim0}(\tu,g) &= \frac{4}{g}+g \left(\frac{3}{16}+\frac{3 u^2}{4}\right)+g^2 \left(\frac{23 u}{64}+\frac{11 u^3}{16}\right)+g^3 \left(\frac{215}{4096}+\frac{341 u^2}{512}+\frac{199
			u^4}{256}\right) + \dots ,  \label{WeberExpansionPeriodic_NonPerturbativeBelow}
	\end{align}
	and 
	\begin{align}
		\mcalF^{u\sim u_\mrmT}(\tu,g) &= i u+g \left(\frac{i}{16}-\frac{i u^2}{4}\right)+g^2 \left(-\frac{1}{64} (5 i u)+\frac{3 i u^3}{16}\right)\nonumber \\
		&\;\; +g^3 \left(-\frac{17 i}{2048}+\frac{35 i u^2}{256}-\frac{25 i
			u^4}{128}\right) +\dots , \label{WeberExpansionPeriodic_PerturbativeAbove}  \\ \nonumber \\
		\mcalG^{u\sim u_\mrmT}(\tu,g) &= -\frac{4 i}{g}+g \left(\frac{3 i}{16}-\frac{3 i u^2}{4}\right)+g^2 \left(-\frac{1}{64} (23 i u)+\frac{11 i u^3}{16}\right) \nonumber \\
		& \;\; +g^3 \left(-\frac{215 i}{4096}+\frac{341 i
			u^2}{512}-\frac{199 i u^4}{256}\right) +\dots , \label{WeberExpansionPeriodic_NonPerturbativeAbove}
	\end{align}
	$\mcalF^{u\sim0}$ and $\mcalG^{u\sim0}$ obey the P-NP relation in \eqref{DunneUnsal_Periodic}. Then, we get the Pu-NP relation around $u\sim u_\mrmT$ as 
	\begin{equation}\label{DunneUnsal_Above_Periodic}
		\mcalG^{u\sim u_\mrmT}(\tu,g)= - 2\int \frac{\mrmd g}{g}\, \left[\, \frac{S^{u\sim u_\mrmT}_\mrmI}{i\, g}\,  \frac{\dee \tu(\mcalF^{u\sim u_\mrmT},g)}{\dee \mcalF^{u\sim u_\mrmT}} + \mcalF^{u\sim u_\mrmT}(\tu,g)\right]\,,
	\end{equation}
	where \begin{equation}
		S^{u\sim u_\mrmT}_\mrmI  = \int_{0}^\pi \mrmd x\, \sqrt{2\left(V(x) - u_\mrmT\right)} = 2 \, i  \, ,
	\end{equation}
	and it is also given by
	\[S^{u\sim u_\mrmT}_\mrmI =\lim_{g\rightarrow 0} \left[g\, \mcalG^{u\sim u_\mrmT}(\tu,g)\right] \, .\]
	To obtain the P-NP relation for the dual theory, we get the series expansions using
	\begin{equation}
		\mcalF^\mrmD(i \tu,i g) = \mcalF^{u\sim u_\mrmT}(\tu,g)  \, , \qquad  \mcalG^\mrmD(i \tu,i g) = \mcalG^{u\sim u_\mrmT}(\tu,g) \, , \label{WeberExpansionPeriodic_Dual}
	\end{equation} 
	and observe that 
	\begin{equation}
		\mcalF(\tu,g) = \mcalF^\mrmD(\tu,g) \, ,\qquad \mcalG(\tu,g) = \mcalG^\mrmD(\tu,g).
	\end{equation}
	As a result, the P-NP relation for the dual theory becomes the same with \eqref{DunneUnsal_Periodic}:
	\begin{equation}\label{DunneUnsal_Dual_Periodic}
		\mcalG^\mrmD(\tu,g)= - 2\int \frac{\mrmd g}{g}\, \left[\frac{S^\mrmD_\mrmI}{g}\frac{\dee \tu(\mcalF^\mrmD,g)}{\dee \mcalF^\mrmD} + \mcalF^\mrmD(\tu,g)\right]\,,
	\end{equation}
	where 
	\[S_\mrmI^\mrmD = \int_{-\frac{\pi}{2}}^{\frac{\pi}{2}} \mrmd x \,  \sqrt{2 V_\mrmD} = S_\mrmI \,,\] 
	is the instanton action of the dual theory. Note that this is in total agreement with the transformation \eqref{DU_parameterDuality} as we get $\k_\mrmP = 1$ from \eqref{WeberExpansionPeriodic_Dual}.

	
	\section{Symmetric double-well potential}\label{Section: DoubleWell}

	In this section, we focus on the symmetric double-well potential of form 
	\begin{equation}\label{Potential_DW}
		V_\mathrm{DW}(x) = \frac{1}{16}T^2_2(x) = \frac{1}{16}\left(1-2x^2\right)^2 \,  ,
	\end{equation}
	which admits two sectors, i.e.~$u<u_\mrmT$ and $u>u_\mrmT$ for $u_\mrmT = \frac{1}{16}$ and implementing the boundary conditions 
	\begin{equation}\label{BoundaryCondition_DW}
		\lim_{x\rightarrow \pm \infty} \psi(x)= 0\,,
	\end{equation}
	for both sectors, we perform an exact quantization in a parallel manner to the periodic potential. Note that we introduce the simple rescaling by $\frac{1}{16}$ to keep the behaviour around a saddle point $x_0$ as $\pm \frac{1}{2}\left(x - x_0\right)$. This doesn't alter arguments on the analytic continuation of the classical action in Section \ref{Section: ChebyshevTransitionDuality} which we continue to use as guidelines.
	
	\begin{figure}[t]
		\centering
		\includegraphics[width=0.7\textwidth]{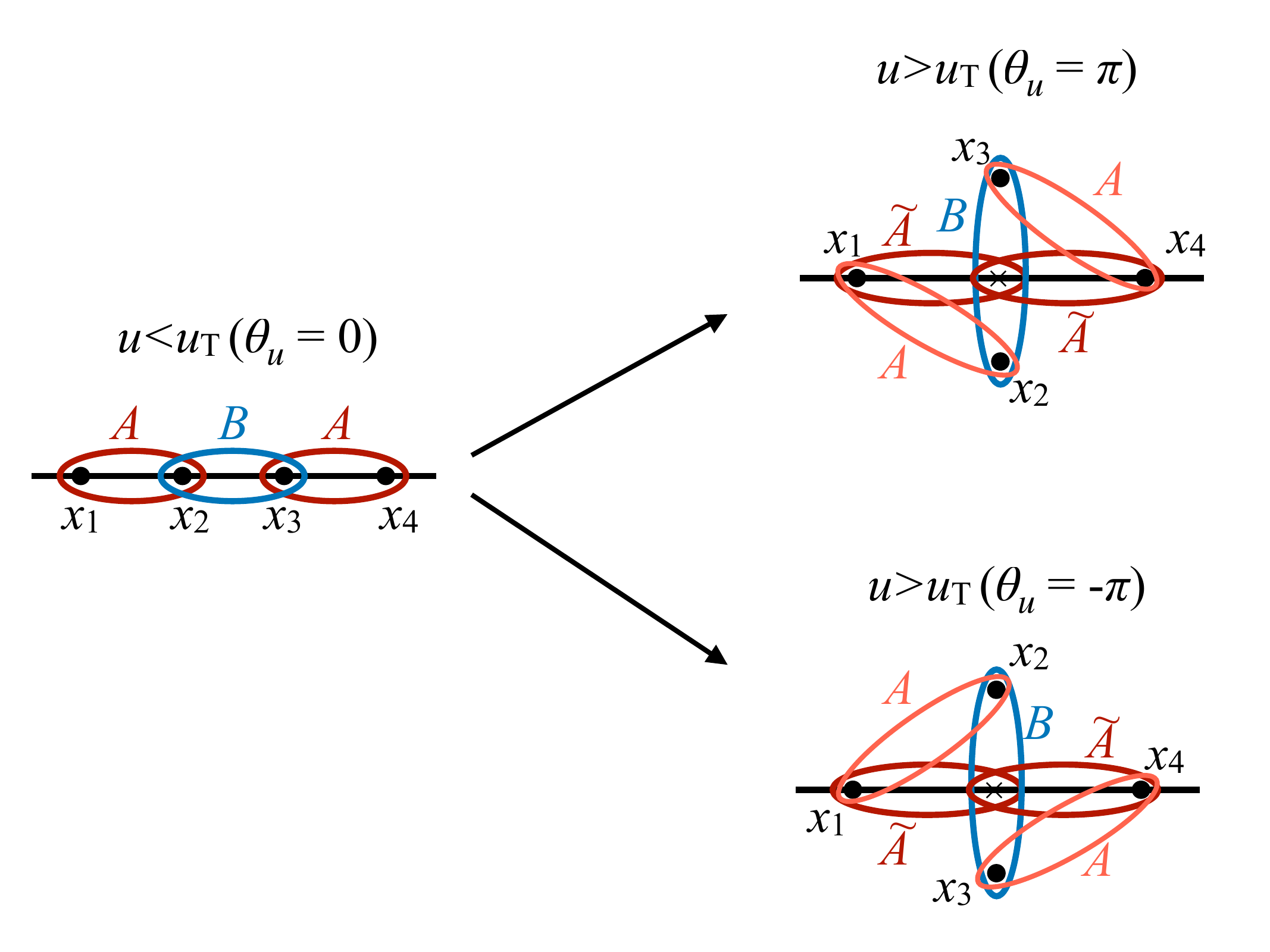}
		\caption{The transition of turning points and the cycles between $u<u_\mrmT$ and $u>u_\mrmT$ in the symmetric double-well potential, which is characterized by the orientation of the turning points and the geometry of the $A$ and $B$ WKB cycles.} \label{Figure: TurningPoints_DW}
	\end{figure}
	
	We illustrate the turning points and the geometry of WKB cycles in Fig.~\ref{Figure: TurningPoints_DW}. Note we picked the cycles in $u>u_\mrmT$ accordingly to the Weber-type construction. It is also possible to consider the combination of two $\tA$ cycles which would make the geometry of $u>u_\mrmT$ sector equivalent to the (stable) quartic potential. Below, we elaborate on this connection when we discuss the exact quantization of this sector.
	
	The explicit expressions for the actions for the cycles in Fig.~\ref{Figure: TurningPoints_DW} are written as in the following:
	\begin{itemize}
		\item \underline{$u<u_\mrmT$ sector}:
		\begin{align}
			\Pi_A &= e^{-\frac{2\pi i}{g} a(u,g)} = e^{-2\pi i \mcalF\left(\frac{u}{g},g\right)}\,, \label{AcycleBelow_DW}\\
			\Pi_B &= e^{-\frac{2\pi i}{g} a^\mrmD(u,g)} =	e^{-\mcalG(\frac{u}{g},g)} \frac{2\pi \left(\frac{g}{\mcalC_\mathrm{DW}}\right)^{-2 \mcalF\left(\frac{u}{g},g\right)}}{\left[\G\left(\frac{1}{2} + \mcalF\left(\frac{u}{g},g\right)\right)\right]^2} \,, \\  \label{BcycleBelow_DW}
		\end{align} 
		\item \underline{$u>u_\mrmT$ sector}:
		\begin{align}
			\Pi_{\tA} &= e^{-\frac{2\pi i}{g} a(u,g)} = e^{- \mcalG^\mrmD(i \frac{1-u}{g},ig)} \frac{\sqrt{2\pi} \left(\frac{ g}{\mcalC_4}\right)^{- \mcalF^\mrmD\left(i\frac{1-u}{g},i g\right)}}{\G\left(\frac{1}{2} +   \mcalF^\mrmD \left(i\frac{1-u}{g},ig\right)\right)}\,, \label{AcycleAbove_DW}\\ 
			\Pi_B &= e^{-\frac{2\pi i}{g} a^\mrmD(u,g)} = e^{-2\pi i \, \left( \mcalF^\mrmD\left(i\frac{1-u}{g},i g\right)\right)} \,,\label{BcycleAbove_DW}
		\end{align}
	\end{itemize}
	where 
	\begin{equation}\label{Prefactor_DW}
		\mcalC_\mathrm{DW} = 2\, \left(\sqrt{2}\right)^2 \exp \left\{\int_{-\frac{1}{\sqrt{2}}}^{\frac{1}{\sqrt{2}}}\, \mrmd x\, \left[\frac{1}{\sqrt{2 V(x)}} - \frac{1}{x + \frac{1}{\sqrt{2}}} + \frac{1}{\frac{1}{\sqrt{2}} - x} \right]\right\} = 4  \, 
	\end{equation}
	and
	\begin{equation}\label{Prefactor_Quartic}
		\mcalC_4 = \frac{2 i}{\sqrt{2}} \exp \left\{2\int_{0}^{1} \mrmd x \, \left[\frac{1}{\sqrt{4 V_\mrmD}} +\frac{1}{0-x}  \right] \right\} = 4\sqrt{2}i\, .
	\end{equation}
	The dual potential
	\[V_\mrmD(x) =u_\mrmT - V_\mathrm{DW}(x) = \frac{1}{4}\left(x^2 - x^4\right)\,\] 
	and it represents an unstable quartic oscillator.  
	
	Like periodic potential, when $u<u_\mrmT$ , the symmetric double-well and its resurgence structure is well-studied via several methods 
	\cite{Zinn-Justin:2004vcw,Dunne:2013ada,Dunne:2014bca,Behtash:2015loa} as well as the EWKB method \cite{DDP2,DP1,Sueishi:2020rug}. 
	The P-NP relation is also well-known in this sector \cite{Alvarez3,Dunne:2013ada, Dunne:2014bca}. For our particular choice in \eqref{Potential_DW}, it is written as
	\begin{equation}\label{DunneUnsal_DW}
		\mcalG(\tu,g) = -2\int\frac{\mrmd g}{g}\left[\frac{S_\mrmI}{ g}\frac{\dee \tu\left(\mcalF,g\right)}{\dee \mcalF} +  \mcalF \right]\, ,
	\end{equation} 
	where 
	\begin{equation}\label{InstantonAction_DW}
		S_\mrmI = \int_{-\frac{1}{\sqrt{2}}}^{\frac{1}{\sqrt{2}}} \mrmd x\, \sqrt{2 V(x)} = \frac{1}{3}\,
	\end{equation}
	is the instanton action of the symmetric double-well potential and it is directly related to the function $\mcalG$ as
	\begin{equation}
		S_\mrmI =  \lim_{g\rightarrow 0} \left[g\, \mcalG(\tu,g)\right]\, .
	\end{equation}
	
	The $u>u_\mrmT$ sector, on the other hand, was studied in \cite{DDP2,DP1}, where the computations for $\mcalF^\mrmD$, $\mcalG^\mrmD$ and $\mcalC_4$ was performed to get the expressions in \eqref{AcycleAbove_DW}. However, the explicit relationship of \eqref{AcycleAbove_DW} with the unstable quartic oscillator wasn't discussed. This also left out the possibility of observing the $S$-duality in the EWKB framework.

	Finally, note that contrary to the case for the periodic potential, the dual theory of the double-well potential is not the same. Therefore, it is expected that the P-NP of the dual theory would be different than the P-NP relation in \eqref{DunneUnsal_DW}. In this particular case, a similar discussion was carried out in \cite{Codesido:2017dns} where the duality transformation was performed for the free energy (prepotential) of the quantum system and the P-NP relation for the quartic oscillator was indicated in the dual theory. In our discussion, on the other hand,  we perform a transformation for $\mcalF$ and $\mcalG$ which leads to the dual P-NP relation to \eqref{DunneUnsal_DW} directly. 
	
	In the rest of the section, following our arguments in Section \ref{Section: PeriodicPotential}, 
	\begin{itemize}
		\item We discuss the analytic continuation of the exact quantization conditions and show the continuity of the spectrum via their median summations and the proper redefinitions of the WKB cycles.
		\item Obtaining the explicit expressions for $\mcalF$ and $\mcalG$, we uncover the P-NP relation around $u\sim u_\mrmT$. Then, we show that the P-NP relation for the dual theory is recovered by the transformation in \eqref{DU_parameterDuality}, which supports our hypothesis. 
	\end{itemize}
	Since we follow the same path in Section \ref{Section: PeriodicPotential}, we don't discuss the same arguments which exists there; instead, we emphasize different points.

	\subsection{Transition between different sectors}
\begin{figure}
		\centering
		\begin{subfigure}[h]{0.48\textwidth}
			\caption{\underline{Diagrams at $\t_\d=0^\mp$}}	\label{Figure: StokesDiagrams_DoubleWellBelow}
			\includegraphics[width=\textwidth]{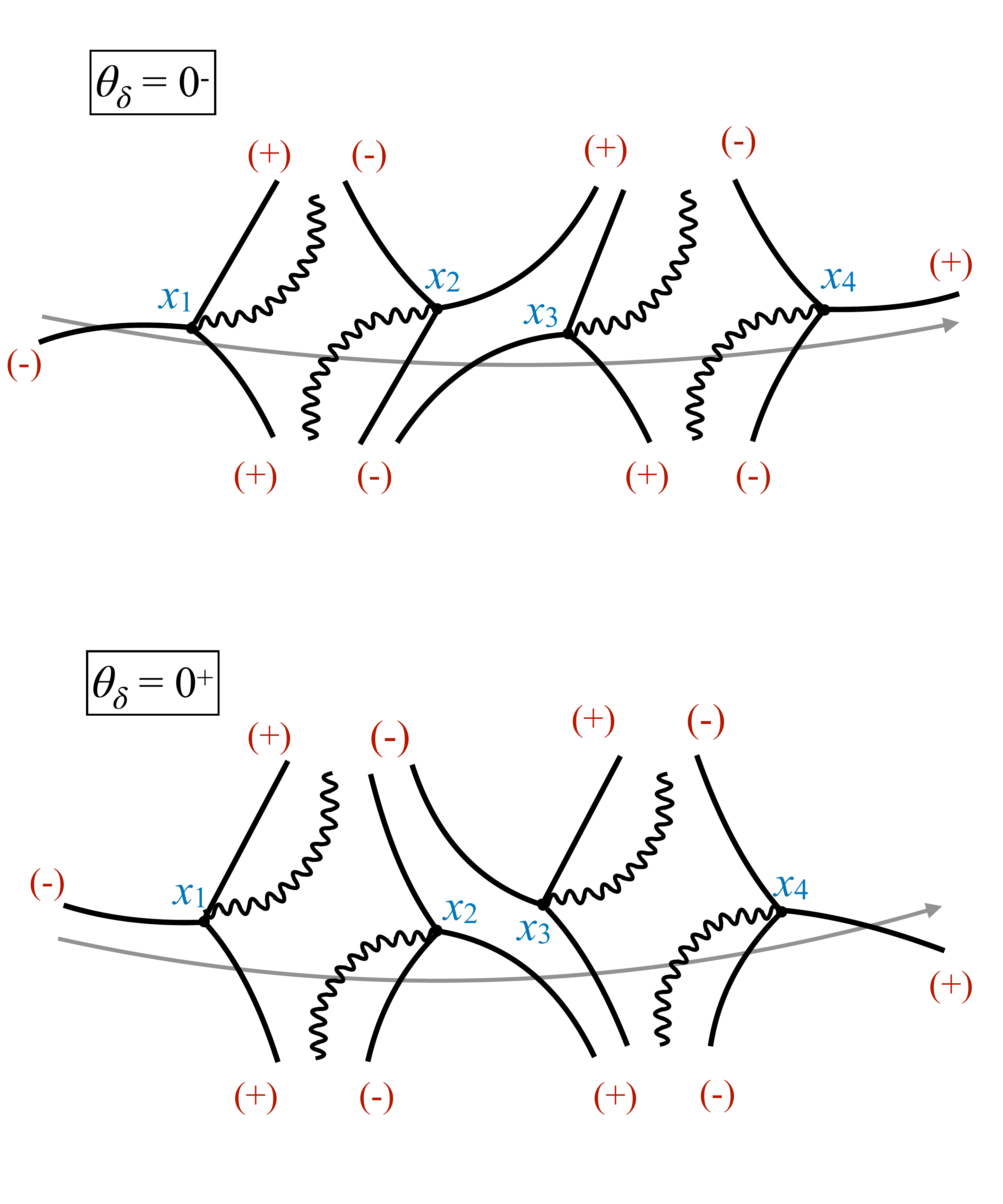}
		\end{subfigure}
		~\hfill 
		\begin{subfigure}[h]{0.48\textwidth}
			\caption{\underline{Diagrams at $\t_\d=\mp \pi^\pm$}}	\label{Figure: StokesDiagrams_DoubleWellAbove}
			\includegraphics[width=\textwidth]{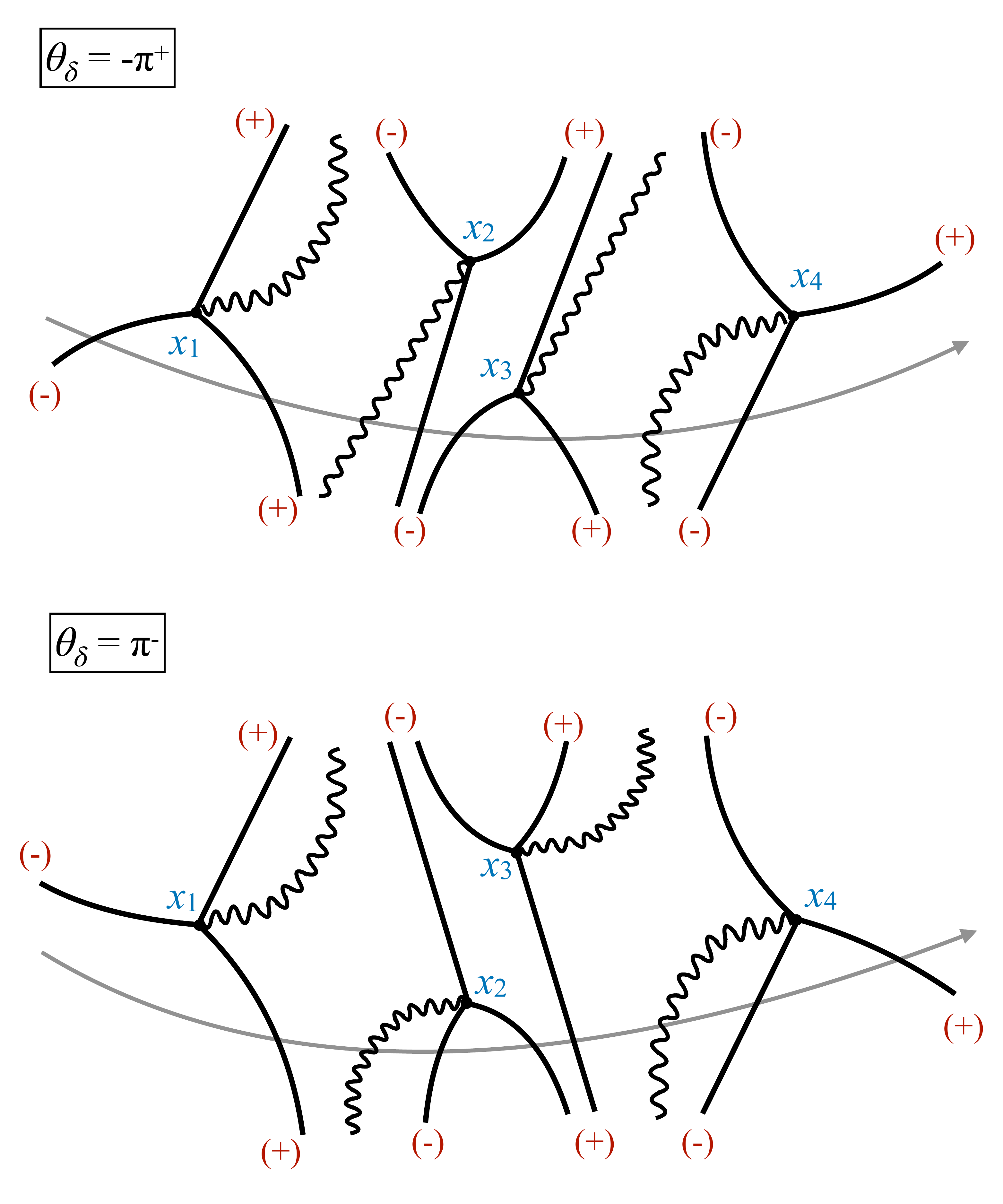}
		\end{subfigure}
		\caption{Stokes diagrams of the symmetric double-well potential. \textbf{(a)}~The diagrams in $u<u_\mrmT$ sector. The analytic continuations $\t_\d = 0^\mp$ are equivalent to $\arg g = 0^\pm$. \textbf{(b)}~The diagrams in $u>u_\mrmT$ sector. The diagrams at phases $\t_\d = \mp \pi^\pm$ are continuously connected to $\t_\d=0^\mp$.} \label{Figure: StokesDiagrams_DoubleWell}
	\end{figure}

	To be able to perform the analytic continuation between the sectors $u<u_\mrmT$ and $u>u_\mrmT$, we parameterize $u = u_\mrmT - \d$ and express $P_\mrmA$ as 
	\begin{equation}
		P_\mrmA = 2\left(V_\mathrm{DW} - u_\mrmT + |\d| e^{i\t_\d}\right)\, . \label{Curve_DW}
	\end{equation}
	Then, $\t_\d=0^\pm$ are the two analytic continuations which break the degeneracy in $u<u_\mrmT$ sector and the analytic continuations $\t_\d: 0^\pm \rightarrow \pm \pi^\mp$ connect the Stokes geometries of the two sectors continuously. Note that in this case, only two of the Stokes diagrams emerging from simple turning points are subject to the analytic continuation as crossing the level $u=u_\mrmT$. Two possible analytic continuations of these diagrams are illustrated in Fig.~\ref{Figure: StokesDiagrams_DoubleWell}. 
	
	Despite the significant change in the Stokes geometry, Fig.~\ref{Figure: StokesDiagrams_DoubleWell} shows that the transition matrices for $\t_\d=0^\pm$ are equal to $\t_\d = \pm \pi^\mp$ since the path connecting the infinities passes through the Stokes lines and branch cuts in the same order. The corresponding transition matrices are given as
	\begin{align}
		T_{\t_\d=0^-} &=  M_\mrmA^- M_\mrmA^\mrmB N_\mrmA^{3,4} M_\mrmA^+ M_\mrmA^- N_\mrmA^{2,3} M_\mrmA^- M_\mrmA^\mrmB N_\mrmA^{1,2} M_\mrmA^+ M_\mrmA^- = T_{\t_\d=-\pi^+}\, , \label{TransitionMatrix_DoubleWellBelow1}\\ \nonumber \\
		T_{\t_\d=0^+} &= M_\mrmA^+ M_\mrmA^- M_\mrmA^\mrmB N_\mrmA^{3,4} M_\mrmA^+ N_\mrmA^{2,3} M_\mrmA^+  M_\mrmA^- M_\mrmA^\mrmB N_\mrmA^{1,2} M_\mrmA^+ = T_{\t_\d=\pi^-}  \, . \label{TransitionMatrix_DoubleWellBelow2}
	\end{align}
	Then, imposing the boundary condition \eqref{BoundaryCondition_DW}, we get 
	\begin{align}
		D_{\t_\d = 0^-} &= D_{\t_\d = -\pi^+} = \sqrt{\Pi_A^{-2} \Pi_B^{-1}} \left[\left(1 + \Pi_A\right)^2 + \Pi_B \right] = 0    \, , \label{EQC_DoubleWell_1}\\
		D_{\t_\d = 0^+} &= D_{\t_\d = +\pi^-} =  \sqrt{\Pi_A^{-2} \Pi_B^{-1}} \left[\left(1 + \Pi_A\right)^2 + \Pi_A^2 \Pi_B  \right] = 0  \, , \label{EQC_DoubleWell_2}
	\end{align}
	as exact quantization conditions.
	
	Note that the square root prefactors don't play any role in the physical interpretation of the exact quantization procedure and to obtain the quantized spectrum, the expressions in the brackets are set to zero. However, we observe that $\sqrt{\Pi_A^{-2} \Pi_B^{-1}}$ factors are important in order to capture the precise relation between $D_{\t_\d=0^\pm}$ via a Stokes automorphism and in the computation of the median summations at $\t_\d=0$ as well as at $\t_\d=\pm \pi$. Therefore, we choose to leave the prefactor in the quantization conditions.
	
	For $u<u_\mrmT$ sector, \eqref{EQC_DoubleWell_1} and \eqref{EQC_DoubleWell_2} equal to ZJJ exact quantization conditions 
	\cite{Zinn-Justin:2004vcw,Dunne:2013ada,Dunne:2014bca,Sueishi:2020rug} which becomes apparent when \eqref{AcycleBelow_DW} and \eqref{BcycleBelow_DW} are used to write them as
	\begin{equation}
		\pm i = \frac{e^{-\mcalG(\tu,g)}\, \Gamma\left(\frac{1}{2} - \mcalF\left(\tu,g\right)\right)}{\sqrt{2\pi}}\, \left(-\frac{g}{\mcalC_{\mathrm{DW}}}\right)^{-\mcalF(\tu,g)} \, .
	\end{equation}
	This is the quantization condition prior to the Borel summation procedure and contains all the ambiguous terms to be canceled, which is carried out by the acting Stokes automorphism on the quantization conditions in the EWKB framework. 
	
\begin{figure}[t]
		\centering
		\includegraphics[width=0.85\textwidth]{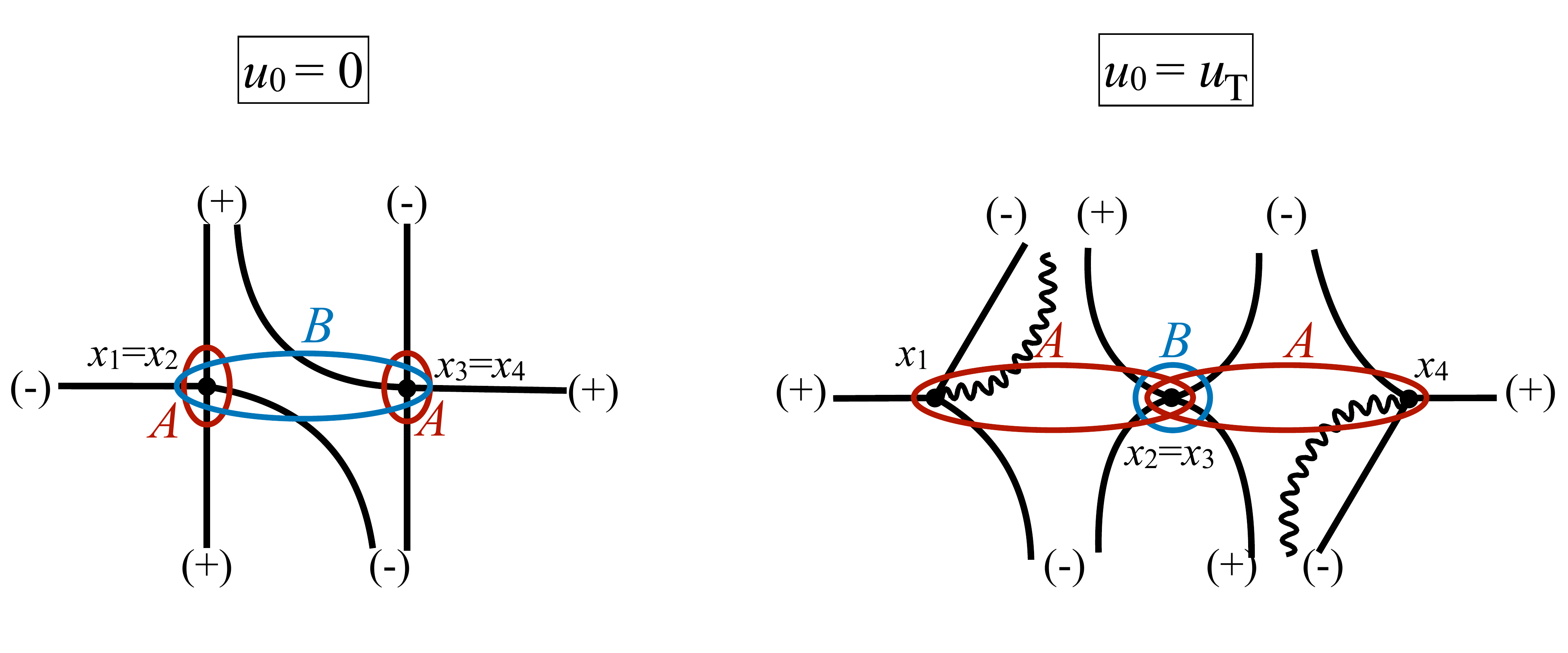}
		\caption{Weber type stokes diagram for the symmetric double-well potential.} \label{Figure: StokesDiagrams_Weber_DW}
	\end{figure}

	As we mentioned above, the geometry of the $u>u_\mrmT$ sector is the same as the stable quartic oscillator \cite{Bucciotti:2023trp,Kamata:2023opn}. Indeed, it is possible to interpret \eqref{EQC_DoubleWell_1} and \eqref{EQC_DoubleWell_2} as the exact quantization conditions for the quartic oscillator\footnote{There are differences between \eqref{EQC_DoubleWell_1} and \eqref{EQC_DoubleWell_2} and the quantization conditions in \cite{Bucciotti:2023trp,Kamata:2023opn}. This is simply due to the branch cut choices in the Stokes diagrams which leads to different interpretations of $\Pi_A$ and $\Pi_B$ in different settings. (See Appendix~\ref{Section: BranchCut_Appendix}.)}. The main difference with the quartic oscillator and $u>u_\mrmT$ sector of the symmetric double-well, on the other hand, becomes apparent when the Weber-type approach is incorporated: The associated Stokes diagram in the double-well potential for $u = u_\mrmT + g\tu$ is given in Fig.\ref{Figure: StokesDiagrams_Weber_DW} and this is quite different than the Stokes diagram for the quartic oscillator at $u=0 + g\tu$. (See Fig.~6 in \cite{Kamata:2023opn}.) 
	
	The reason behind this is the difference in the merging turning points at the singular points of the corresponding cases. In the double-well potential, $x_2$ and $x_3$ in Fig.\ref{Figure: StokesDiagrams_DoubleWellBelow} merge at $u=u_\mrmT$ and scatter to the imaginary axis. As a result, the perturbative cycle for $u>u_\mrmT$ connects these turning points on the imaginary axis 
	as shown in Fig.\ref{Figure: StokesDiagrams_DoubleWellAbove}. For the quartic oscillator, on the other hand, the turning points at $x_1$ and $x_4$ merge when $u=0$. Therefore, the perturbative cycle would connect $x_1$ and $x_4$ as in \cite{Kamata:2023opn,Kamata:2024tyb} and the non-perturbative cycles would lie on the imaginary axis.

	\paragraph{\underline{Smooth transition of $\bm {D^{\med}}$}:}
	Now we turn our attention to the median summations of the exact quantization conditions in \eqref{EQC_DoubleWell_1} and \eqref{EQC_DoubleWell_2}, and the smooth transition of the spectrum across $u=u_\mrmT$. As in the periodic potential case, we present the preservation of the median QC across the barrier top provided by appropriate analytic continuations.\\

	\noindent{\textit{\underline{Below the Barrier Top}}:} In the $u<u_\mrmT$ sector, the quantization conditions at $\t_\d = 0^\pm$ are related to each other by
	\begin{equation}
		D_{\t_\d=0^+} = \mfrS_{0} D_{\t_\d = 0^-}, \label{StokesAuto_DoubleWellBelow}
	\end{equation}
	where the Stokes automorphism associated to the discontinuity of the $A$-cycle at $\t_\d=0$ is given by
	\begin{equation}
		\mfrS_{0}: \Pi_{A_{0^-}} \mapsto \Pi_{A_{0^+}} \left(1+\Pi_{B_{0^+}}\right)\, . \label{StokesAuto_DW}
	\end{equation}
	Then, we get the median QC as 
	\begin{align}
		D^{\med}_{u<u_\mrmT} &= \mfrS_{0}^{1/2} D_{\t_\d=0^-} = \mfrS_0^{-1/2}D_{\t_\d=0^+} \\
		& =   \sqrt{\Pi_A^{-2} \Pi_B^{-1}} \left[ \left(1 + \Pi_A^2\right)\sqrt{1+\Pi_B} + 2 \Pi_A \right]\, , \label{EQC_MedianSum_DoubleWellBelow}
	\end{align}
	which is manifestly real, i.e. $\mcalC\left[D^\med_{u<u_\mrmT}\right] = D^\med_{u<u_\mrmT}$, so that ambiguity free as expected. Therefore, \eqref{EQC_MedianSum_DoubleWellBelow} is in total agreement with the cancellation of ambiguous terms upon Borel summation procedure in 
	\cite{Zinn-Justin:2004vcw,Zinn-Justin:2004qzw, Dunne:2013ada,Dunne:2014bca,DDP2,Zinn-Justin:1981qzi,Zinn-Justin:1983yiy}. \\
	
	\noindent \textit{\underline{Above the Barrier Top}}: As in the periodic potential case, the transition to $u>u_\mrmT$ needs an incorporation of the analytic continuation of $a(u,g)$ and $a^\mrmD(u,g)$. Recalling the discussion in Section~\ref{Section: ChebyshevTransitionDuality}, for the symmetric double-well potential, the leading order analytic continuations are given as
	\begin{equation}
		a_0(u) \rightarrow a_0(u) \pm \frac{1}{2} a_0^\mrmD(u)\, ,  \qquad  a^\mrmD_0(u) \rightarrow a_0^\mrmD(u)\, . 
	\end{equation} 
	This suggests the following transformations of $\Pi_A$ and $\Pi_B$ for the analytic continuations $\t_\d : 0^\mp \rightarrow \pm \pi^\mp$:
	\begin{equation}
		\Pi_A \rightarrow \Pi_{\tA} = \Pi_A \Pi_B^{\pm \frac{1}{2}}, \qquad  \Pi_B \rightarrow \Pi_B. \label{AC_ExponentialAction_DW}
	\end{equation}
	For $\t_\d: 0^+ \rightarrow + \pi^-$, this is compatible with the geometry of the WKB cycles above the barrier top which indicates $\tA = A+\frac{1}{2}B$. For $\t_\d: 0^- \rightarrow -\pi^+$, on the other hand, as in the periodic potential case, we find that $A = A_\pi = A_{-\pi}^{-1}$ due to the differences in the branch cut conventions for $\t_\d = \pm \pi^\mp$. Then, at the phase $\t_\d = -\pi$, $\tA = A + \frac{1}{2}B$ should be interpreted as $ \Pi_{\tA}^{-1} = \Pi^{-1}_A \Pi_B^{1/2}$ and in this way, we get the transition in \eqref{AC_ExponentialAction_DW}. (See Appendix~\ref{Section: BranchCut_Appendix} for further discussion.) 
	
	Now, we compute the median QCs at $\t_\d = \pm \pi$: As in Section~\ref{Section: PeriodicPotential}, we start with inverting $B$-cycle and rewrite
	\begin{align}
		D_{\t_\d = +\pi^-} & = \sqrt{\Pi_A^{-2} \Pi_{B^{-1}}}\, \left[\left(1 + \Pi_A\right)^2 + \Pi_A^2 \Pi_{B^{-1}}^{-1} \right] , \\
		D_{\t_\d = -\pi^+} & = \sqrt{\Pi_A^{-2} \Pi_{B^{-1}}}\,\Pi_{B^{-1}}^{-1} \left[1 + \left(1 + \Pi_A\right)^2 \Pi_{B^{-1}}  \right]\, .
	\end{align}
	Then, the Stokes automorphisms at $\t_\d = \pm \pi$ becomes
	\begin{equation}
		\Pi_{A_{\pm \pi^\pm} } = \Pi_{ A_{\pm \pi^\mp}} \left(1 + \Pi_{B^{-1}_{\pm \pi^\mp}}\right)^{\<A,B^{-1}\>}\, , \quad \<A,B^{-1} \> = -1 \, , \label{StokesAuto_DoubleWell2}
	\end{equation}
	and we get the median QC as
	\begin{align}
		D^{\med}_{\t_\d = \pm\pi} & = \mfrS_{\pm \pi}^{\mp 1/2} D_{\t_\d = \pm \pi^\mp} \nonumber \\
		& = \sqrt{\Pi_A^{-2} \Pi_B^{-1}} \,  \Pi_B^{\mp 1} \left[ \sqrt{1+\Pi_B} + 2\Pi_A \sqrt{\Pi_B^{\pm1} } + \Pi_A^2 \Pi_B^{\pm 1} \sqrt{1 + \Pi_B}\,  \right]\, , \label{MedianSumDW_BeforeCycleRedefinitions} 
	\end{align}
	where we invert $B^{-1}$ cycle back to $B$-cycle. Finally, using \eqref{AC_ExponentialAction_DW}, for both cases, we get
	\begin{equation}
		D^\med_{u>u_\mrmT}  = \sqrt{\Pi_{\tA}^{-2} \Pi_B^{-1}} \, \left[ \left(1 + \Pi_{\tA}^2\right) \sqrt{1 + \Pi_B} + 2\Pi_{\tA} \right] \, . \label{EQC_MedianSum_DoubleWellAbove}
	\end{equation}
	Again, we emphasize the role of the square-root prefactors in the computations leading the same results for both analytic continuations although they don't have any physical implications. For example, when imposing \eqref{AC_ExponentialAction_DW} in \eqref{MedianSumDW_BeforeCycleRedefinitions}, they led to the cancellations of $\Pi^{\mp}_B$ terms which appear differently at $\t_\d = +\pi$ and $\t_\d = -\pi$. Therefore, we observe that there is a single median QC, in the $u>u_\mrmT$ region and it has exactly the same form as \eqref{EQC_MedianSum_DoubleWellBelow}, the median QC in $u<u_\mrmT$ sector. 
	
	In addition to the equality of the median QCs \eqref{EQC_MedianSum_DoubleWellBelow} and \eqref{EQC_MedianSum_DoubleWellAbove}, they are also manifestly real, which indicates the reality of the spectrum for all $u$. As a result, we conclude that there is a smooth continuous transition of the spectrum between the two sectors and the resurgence structure stays intact for the entire spectrum.
	
	\paragraph{\underline{Remark}:} As in the periodic potential discussion, since both $\tA$ and $B$ cycles around $u=u_\mrmT$ have equivalent contributions, the quantization around $u=u_\mrmT$ is not as easy as it is for $u=0$. Note that this also shows that the spectrum around $u=u_\mrmT$ behaves different than the spectrum of the quartic anharmonic oscillator although their Stokes geometry are the same. A further analysis on this difference would be interesting.

	
	\subsection{The duality}
	Let us now discuss the duality between the bottom of wells and the barrier top of the symmetric double-well potential. As we stated above, the dual theory is the unstable quartic oscillator. The perturbative and non-perturbative quantities for these theories are not the same and the transformations between them are non-trivial contrary to the periodic potential case. For example, we get the series expansions of $\mcalF$ and $\mcalG$ around $u\sim 0$ and $u \sim u_\mrmT$ first few orders in $g$ as
	\begin{align}
		\mcalF^{u\sim0}(\tu,g) &= \tu+g \left(\frac{1}{8}+\frac{3 \tu^2}{2}\right)+g^2 \left(\frac{25 \tu}{16}+\frac{35 \tu^3}{4}\right)+g^3 \left(\frac{175}{256}+\frac{735	\tu^2}{32}+\frac{1155 \tu^4}{16}\right) \nonumber \\
		& \;\; +g^4 \left(\frac{31185 \tu}{1024}+\frac{45045 \tu^3}{128}+\frac{45045 \tu^5}{64}\right) + \dots , \label{WeberExpansionDW_PerturbativeBelow}\\ \nonumber \\
		\mcalG^{u\sim0}(\tu,g) &= \frac{2}{3 g}+g \left(\frac{19}{24}+\frac{17 \tu^2}{2}\right)+g^2 \left(\frac{187 \tu}{16}+\frac{227 \tu^3}{4}\right)+g^3 \left(\frac{28829}{4608}+\frac{34121 \tu^2}{192}+\frac{47431 \tu^4}{96}\right) \nonumber \\
		& \;\;+g^4 \left(\frac{842909 \tu}{3072}+\frac{264725 \tu^3}{96}+\frac{317629 \tu^5}{64}\right)+\dots , \label{WeberExpansionDW_NonPerturbativeBelow}
	\end{align} 
	and 
	\begin{align}
		\mcalF^{u\sim u_\mrmT}(\tu,g) &= i \sqrt{2} \tu+g \left(\frac{3 i}{8 \sqrt{2}}-\frac{3 i  \tu^2}{\sqrt{2}}\right)+g^2 \left(-\frac{85 i \tu}{16 \sqrt{2}}+\frac{35 i	\tu^3}{2 \sqrt{2}}\right) \nonumber \\ & \;\; +g^3 \left(-\frac{1995 i}{512 \sqrt{2}}+\frac{2625 i \tu^2}{32 \sqrt{2}}-\frac{1155 i \tu^4}{8\sqrt{2}}\right) \nonumber \\
		&\;\; +g^4
		\left(\frac{400785 i \tu}{2048 \sqrt{2}}-\frac{165165 i \tu^3}{128 \sqrt{2}}+\frac{45045 i \tu^5}{32 \sqrt{2}}\right) + \dots , \label{WeberExpansionDW_PerturbativeAbove}\\ \nonumber \\
		\mcalG^{u\sim u_\mrmT}(\tu,g) &= -\frac{i \sqrt{2}}{3 g}+g \left(\frac{67 i}{48 \sqrt{2}}-\frac{17 i \tu^2}{2 \sqrt{2}}\right)+g^2 \left(-\frac{671 i \tu}{32 \sqrt{2}}+\frac{227i \tu^3}{4 \sqrt{2}}\right)\nonumber \\ 
		& \;\; +g^3 \left(-\frac{372101 i}{18432 \sqrt{2}}+\frac{125333 i \tu^2}{384 \sqrt{2}}-\frac{47431 i \tu^4}{96\sqrt{2}}\right) \nonumber \\
		& \;\; +g^4 \left(\frac{3839943 i \tu}{4096 \sqrt{2}}-\frac{82315 i \tu^3}{16 \sqrt{2}}+\frac{317629 i \tu^5}{64 \sqrt{2}}\right) +\dots \; .  \label{WeberExpansionDW_NonPerturbativeAbove}
	\end{align}
	This recovers the expressions in \cite{DP1} and adds a couple of more orders. Note that for our purpose, the series of $\mcalF$ and $\mcalG$ are important due to their relationship via the P-NP relation and establishing the $S$-transformation for the latter. In that sense, we observe that while $\mcalF^{u\sim0}$ and $\mcalG^{u\sim0}$ obey the P-NP relation around $u\sim0$ which is given in \eqref{DunneUnsal_DW}, $\mcalF^{u\sim u_\mrmT}$ and $\mcalG^{u\sim u_\mrmT}$ satisfy a slightly different P-NP relation: 
	\begin{equation}
		\mcalG^{u\sim u_\mrmT}(\tu,g)= -2 \int \frac{\mrmd g}{g}\, \left[\sqrt{2}\, \frac{S_\mrmI^{u\sim u_\mrmT}}{i g} \frac{\dee \tu\left(\mcalF^{u\sim u_\mrmT},g\right)}{\dee \mcalF^{u\sim u_\mrmT}} + \frac{\mcalF^{u\sim u_\mrmT}}{\sqrt{2}} \right] \, ,\label{DunneUnsal_DWTop}
	\end{equation}
	where \begin{equation}
		S_\mrmI^{u\sim u_\mrmT}  = \int_{0}^1 \mrmd x\, \sqrt{2\left(V(x) - u_\mrmT\right)} = \frac{i}{3\sqrt{2}} \, .
	\end{equation}
	It corresponds to the classical WKB action at $u=u_\mrmT$:
	\[S_\mrmI^{u\sim u_\mrmT} = \frac{a_0(u_\mrmT)}{2}=\lim_{g\rightarrow 0} \left[g\, \mcalG^{u\sim u_\mrmT}(\tu,g)\right] \, .\] Thus, it can be seen as the $a_0^\mrmD(u=0)$ for the dual theory as expected.
	
	This further indicates the duality between P-NP relations of the symmetric double-well potential and the unstable quartic oscillator. To make the connection explicit, we write the series expansions for the dual theory using 
	\begin{equation}
		\mcalF^\mrmD(\tu, g) = \mcalF^{u\sim u_\mrmT}(-i\tu,-ig)  \, , \qquad  \mcalG^\mrmD( \tu, g) = \mcalG^{u\sim u_\mrmT}(-i\tu,-i g) \, , \label{WeberExpansionDW_Dual}
	\end{equation}
	and get the associated P-NP relation as
	\begin{equation}\label{DunneUnsal_DW_Dual}
		\mcalG^\mrmD(\tu,g) = -2\int \frac{\mrmd g}{g}\, \left[\sqrt{2} \frac{S_\mrmI^\mrmD}{ g} \frac{\dee \tu\left(\mcalF^\mrmD,g\right)}{\dee \mcalF^\mrmD} + \frac{\mcalF^\mrmD}{\sqrt{2}} \right]  \,,
	\end{equation}
	where 
	\begin{equation}
		S_\mrmI^\mrmD  = \int_{0}^1 \mrmd x\, \sqrt{2V_\mrmD(x)} = \frac{1}{3\sqrt{2}} = \frac{S_\mrmI}{\sqrt{2}}\,,
	\end{equation}
	is equal to the half bounce action of the unstable quartic oscillator. The P-NP relation in \eqref{DunneUnsal_DW_Dual} is for the unstable quartic oscillator and it is indeed related to \eqref{DunneUnsal_DW} via the transformations given in \eqref{DU_parameterDuality} for $\k_2 = 2$, i.e.
	\begin{equation}
		\mcalF \rightarrow \frac{\mcalF^\mrmD}{\sqrt{2}}\, , \qquad \mcalG \rightarrow \sqrt{2} \mcalG^\mrmD\, ,\qquad  S_\mrmI \rightarrow \sqrt{2} S_\mrmI^\mrmD\, .
	\end{equation}
	Note that as we discussed in Section \ref{Section: ChebyshevTransitionDuality}, while $\k_2 = 2$ is the ratio of the classical (dual) actions of both theories but it is not the case for higher order terms. Equivalently, in the Weber-type EWKB framework, it can also be deduced from the relations
	\begin{equation}
		\mcalF_0^\mrmD\Big(\frac{u}{g}\Big) = \sqrt{2}\mcalF_0\Big(\frac{u}{g}\Big)\, , \qquad \mcalG_0^\mrmD\Big(\frac{u}{g}\Big) = \frac{\mcalG_0\Big(\frac{u}{g}\Big)}{\sqrt{2}}\, .
	\end{equation}
	However, its effect in the resurgence structure exceeds the leading order to cover all orders in $g$ as we observe in the duality transformation of the P-NP relations.

	
	\section{Discussion and outlook}
	
	The analysis presented in this paper advances our understanding of the EWKB method, especially in its applications to systems with degenerate locally harmonic saddles. We introduced several innovative approaches, including the complexification of the energy parameter $u$, and performed a detailed study of the Airy and Weber-type EWKB methods. By doing so, we have refined the EWKB methods to provide more accurate solutions in quantum systems.
	Let us summarize what we have done in this work:
	
	\begin{itemize}
		\item
		{\it \underline{Generalization of the EWKB Method}}:
		In Section \ref{Section: Setup_AiryWeber}, we discussed the generalization of the EWKB method to incorporate all sectors and saddle points for locally harmonic potentials. We focused on two distinct complementary approaches: the Airy-type and Weber-type EWKB methods.
		
		In Section \ref{Section: AiryTypeEWKB}, introducing the complexification of the parameter $u$, we extended the application realm of the Airy-type approach, which was historically limited to below-barrier sectors of potentials, to include regions above the barrier. This extension provides a systematic way to track the quantization conditions through analytic continuation across different sectors for generic locally harmonic potentials. 
		
		
		In Section \ref{Section: WeberTypeEWKB}, we examined the Weber-type EWKB, which provides exact estimations of the quantum actions around saddle points. Unlike the Airy-type method, which is more qualitative, the Weber-type method allows for precise quantitative estimates. Along with identifying local and non-local components of quantum actions, we extended their analysis to all saddle points which presents a generalization of Zinn-Justin and Jentschura's results \cite{Zinn-Justin:2004qzw,Zinn-Justin:2004vcw,Jentschura:2010zza,Jentschura:2011zza}. This generalization is particularly relevant for understanding the role of non-perturbative effects in complex quantum systems. The exact estimates of quantum actions we obtained could be applied to a variety of systems, ranging from simple harmonic oscillators to more exotic potentials with rich symmetry structures. Notably, we also identified an exchange between perturbative and non-perturbative behaviors near minima and maxima, which is controlled by $S$-duality --a powerful symmetry that links weak and strong coupling regimes in quantum systems.

		
		\item
		{\it \underline{Chebyshev Potentials, Transition, and Duality}}: Section \ref{Section: ChebyshevTransitionDuality} discusses the generic features of Chebyshev potentials \cite{Basar:2017hpr} which present ideal guidelines for the discussion of the transition between different sectors and the duality transformation between minima and maxima. We utilized the analytic properties of the hypergeometric function and demostrate how exact quantization conditions, as well as resurgence structures, can be carried between different sectors continuously via the complexification $u$ parameter. This provides us a significant insight in quantization of generic potentials in different sectors.
		
		The transition between different potential sectors revealed the role of $S$-duality in maintaining consistency between perturbative and non-perturbative contributions. This duality effectively swaps the behavior around minima and maxima, offering insights into the system's symmetry structure. Importantly, we found that the duality can be understood in terms of the Weber-type EWKB elements, namely the functions $\mcalF$ and $\mcalG$, further extending its utility in analyzing complex quantum mechanical systems.
		
		\item
		{\it \underline{Periodic and symmetric double-well Potentials}}:
		In Sections \ref{Section: PeriodicPotential} and \ref{Section: DoubleWell}, we examined periodic and symmetric double-well potentials, two archetypes of systems with multiple saddle points. They present ideal settings for studying the connection between different quantum sectors. Here we utilized both the Airy and Weber type approaches. The first shows that how the exact quantization conditions and their median summations remain unchanged upon the transition across the barrier top. In this way, we presented the smooth transition of the spectrum between different sectors in a manifest and exact manner. The latter, on the other hand, provides explicit expressions for both perturbative and non-perturbative contributions around all saddle points which is crucial to understand the underlying resurgence structures, such as P-NP relations, in all sectors.
		
		
		
		\item
		{\it \underline{Resurgence and $S$-Duality}}:~	Another central theme in our study is the resurgence theory which exhibits the deep connection between perturbative and non-perturbative sectors. Through carefully tracking of quantum actions across different sectors of potentials and using the Stokes automorphism and the mehdian resummation along with appropriate analytic continuations, we showed that the resurgence structure is preserved in all sectors. 
		
		The role of $S$-duality was also highlighted throughout our analysis. This duality allows the exchange between strong and weak coupling regimes, a feature that was evident in our analysis of quantum actions around minima and maxima. The exact-WKB method, particularly in its Weber-type formulation, provides a natural framework for studying such dualities, as it is able to capture both perturbative and non-perturbative behaviors consistently. One of the highlights in our discussion is the conjecture on the mapping of the P-NP relation to the dual theory in Eqs.~(\ref{DunneUnsal_GeneralDual})(\ref{DU_parameterDuality}).
		
	\end{itemize}
	
	In conclusion, our study has significantly extended the exact-WKB method, making it a powerful tool for analyzing quantum systems with degenerate saddle points, and complex potentials. By introducing the complexification of the energy parameter $u$, we have enabled seamless transitions between different sectors of a potential, preserving the exact quantization condition and resurgence structure. The Weber-type EWKB method, in particular, offers precise estimates of quantum actions, shedding new light on the role of $S$-duality and the interplay between perturbative and non-perturbative behaviors in quantum mechanics.
	
	This paper lays the groundwork for future research into the exact-WKB method and its applications. Future research could build on these findings to explore even more complex quantum systems and their corresponding physical phenomena. One straightforward extension would be the analysis of the connection of different sectors for the potentials with non-degenerate saddle points, which possess a slightly different resurgence structure \cite{Cavusoglu:2023bai,Cavusoglu:2024usn}. Another promising direction would be the exploration of higher-dimensional quantum systems, where the methods developed here could be applied to analyze more complex potentials via the intricate connection between EWKB method and phase space path integrals \cite{Sueishi:2020rug,Sueishi:2021xti,Kamata:2021jrs,Ture:2024nbi}. 
	
	Another avenue would be the detailed study of multi-instanton effects, which play a crucial role in the non-perturbative dynamics of quantum systems around multiple saddles (both minima and maxima). In this sense, the presented extension of the exact-WKB method 
	also opens up new possibilities for studying the relationship between classical and quantum solutions in integrable systems. Future work could explore the implications of our complexification approach for other quantum mechanical systems, including those with chaotic behavior or those in which quantum tunneling plays a dominant role.

	\section*{Acknowledgements}
	This work of T. M. is supported by the Japan Society for the Promotion of Science (JSPS) Grant-in-Aid for Scientific Research (KAKENHI) Grant Numbers 23K03425 and 22H05118. C.P. thanks for the hospitality of Kindai University where the initial steps of the project took place.
	The beginning of this work is also owed to the discussion in the workshop ‘Invitation to Recursion, Resurgence and Combinatorics’ at Okinawa Institute of Science and Technology Graduate University (OIST). We also thank Alireza Behtash for his comments on the first manuscript.

	\appendix
	\section{The relationship between different branch cut conventions}\label{Section: BranchCut_Appendix}
	
	In the EWKB framework, the solutions to the Schrödinger equation are encoded in the Stokes geometry which is mainly determined by a given potential $V(x)$ and deforms continuously as variables such as the spectral parameter $u$ or coupling $g$ changes. There is also a freedom of choice remaining for any Stokes diagram: The placement of the branch cuts. From the physical point of view, any choice of branch cut convention leads to the same physical results. However, it is also interesting to see the relationships between different cut conventions. \
	
	In addition to this, in Sections~\ref{Section: PeriodicPotential} and \ref{Section: DoubleWell}, we observe that as the parameter $u$ is analytically continued from $u<u_\mrmT$ to $u>u_\mrmT$ sector, the WKB actions are also needed to be transformed appropriately. More specifically, we showed that for $\t_\d: 0^\pm \rightarrow \pm \pi^\mp$, the actions corresponding to the new cycle $\tA$ are defined as \[\Pi_{\tA} = \Pi_A \Pi_B^{\pm m}\, , \] where $m=1$ for the periodic potential and $m=\frac{1}{2}$ for the symmetric double-well potential. $(\pm)$ signs are related to the analytic continuation $u\rightarrow u \pm i\ve$ which is imposed in $u>u_\mrmT$ sector for the action $a(u)$. In the EWKB framework, on the other hand, the sign difference is associated to the differences in the branch cut conventions of the corresponding diagrams in Figures~\ref{Figure: StokesDiagrams_PeriodicAbove} and \ref{Figure: StokesDiagrams_DoubleWellAbove}.  
	
	In this Appendix, we discuss the relationship between different branch cut conventions and their effect on the WKB cycles. We first start with the simple harmonic oscillator and investigate the entire transition matrix for different branch cut choices. Then, we present the periodic and double-well potentials using branch cut conventions  different than the ones Sections~\ref{Section: PeriodicPotential} and \ref{Section: DoubleWell}. This helps us to illuminate the relationship between the resulting Stokes diagrams in $u>u_\mrmT$ sector in those sections as well as the necessary transformation of $\Pi_A$ which keeps the median summations intact throughout the spectrum.

	\subsection*{\underline{The simple harmonic oscillator}:}
\begin{figure}[t]
		\centering
		\begin{subfigure}[h]{0.85\textwidth}
			\caption{{\bf(Left)} The branch cut convention we use in the main text. $\Im \sqrt{V_2(x) - u} >0$ in the principal branch. {\bf (Right)} An equivalent branch cut convention} \hrulefill 	\label{Figure: HarmonicChoice1}
			
			\includegraphics[width=\textwidth]{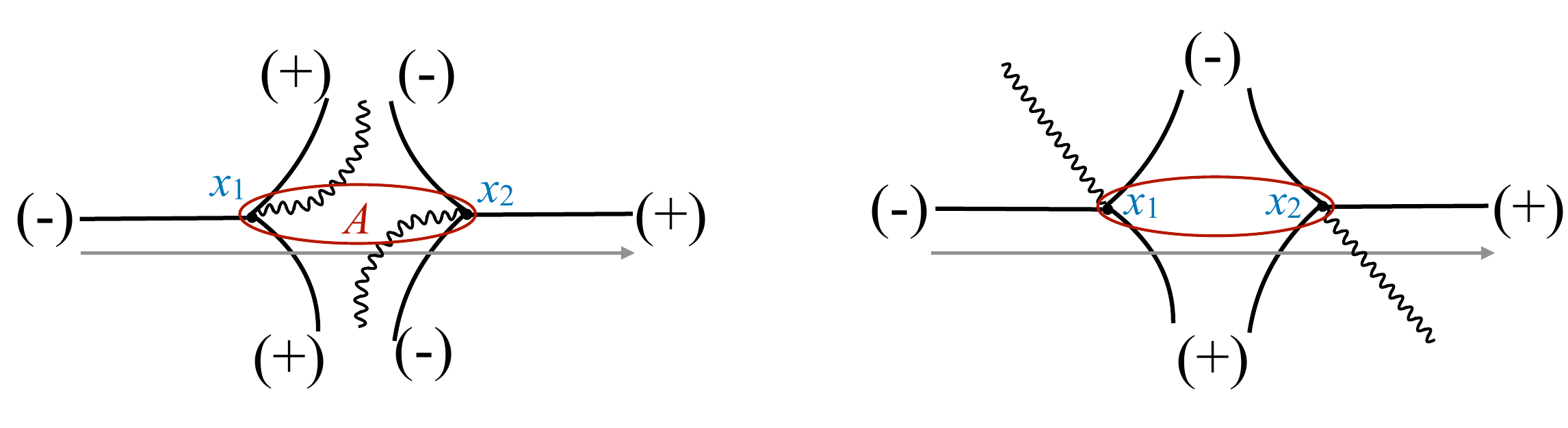}
		\end{subfigure}
		~\hfill \vfill \vspace{1cm}
		\begin{subfigure}[h]{0.85\textwidth}
			\caption{{\bf (Left)} Branch cut convention which set $\Im \sqrt{V_2(x) - u}<0$ in the principal branch. {\bf (Right)} An equivalent branch cut convention.} \hrulefill	\label{Figure: HarmonicChoice2}
			\includegraphics[width=\textwidth]{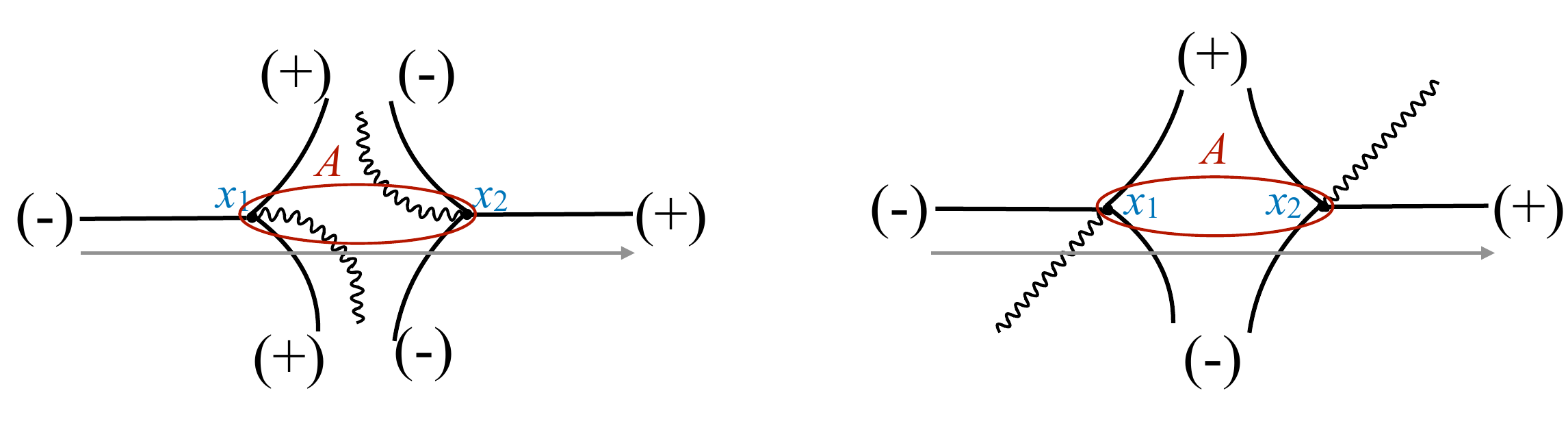}
		\end{subfigure}
		\caption{Different branch cut choices for the Stokes diagrams of simple harmonic oscillator. The WKB cycles in (a) and (b) are related to each other via an inversion which also relates the transition matrices in \eqref{ConnectionMatrix_Harmonic1} and \eqref{ConnectionMatrix_Harmonic2}.} \label{Figure: HarmonicDiagram}
	\end{figure}

	Let us begin with the simple harmonic oscillator: 
	\begin{equation}
		V_2(x) = x^2\, .
	\end{equation}
	We illustrate 4 different branch cut choices in Fig.~\ref{Figure: HarmonicDiagram}. The diagrams in Fig.~\ref{Figure: HarmonicChoice1} are exactly equivalent to each other. This stems from the following equality
	\begin{equation}\label{MonodromyMatrix_OrderChange}
		M_\mrmA^B M_\mrmA^\pm = M_\mrmA^\mp M_\mrmA^B.
	\end{equation}  
	Then, the corresponding transition matrix connecting $x=-\infty$ and $x=+\infty$ in these cases are the same:
	\begin{equation}\label{ConnectionMatrix_Harmonic1}
		T^{(+)}   = \begin{pmatrix}
			0 & -i \Pi_{A_+}^{-1/2} \\[1em]
			-i \Pi_{A_+}^{1/2} & \Pi_{A_+}^{-1/2} + \Pi_{A_+}^{1/2}
		\end{pmatrix} \, .
	\end{equation}
	Similarly, the diagrams in Fig.~\ref{Figure: HarmonicChoice2} are equal to each other and the corresponding transition matrix connecting $x=-\infty$ and $x=+\infty$ is written as
	\begin{equation}\label{ConnectionMatrix_Harmonic2}
		T^{(-)}  = \begin{pmatrix}
			0 & -i \Pi_{A_-}^{1/2} \\[1em]
			-i \Pi_{A_-}^{-1/2} & \Pi_{A_-}^{-1/2} + \Pi_{A_-}^{1/2}
		\end{pmatrix} \, .
	\end{equation}
	In both cases, 
	\begin{equation}\label{ExponentiatedAction}
		\Pi_{A_i} = \exp\left\{\oint_{A_i} \mrmd x\, \ts(x,u)\right\} = \exp\left\{2\int_{x_1}^{x_2} \mrmd x\, \ts(x,u)\right\}
	\end{equation} 
	is the exponentiated action of the $A_i$-cycle for the corresponding diagram and the quantization conditions are obtained by setting $T_{2,2}^\pm = 0$, which leads to 
	\begin{equation}
		\Pi_{A_\pm}^{-1/2} + \Pi_{A_\pm}^{1/2} = 0 \, . 
	\end{equation}
	Then, the quantization conditions for all the choices in Fig.~\ref{Figure: HarmonicDiagram} have the same form and it is tempting to conclude that $\Pi_{A_\pm}$ are exactly the same. 
	
	The connection matrices in \eqref{ConnectionMatrix_Harmonic1} and \eqref{ConnectionMatrix_Harmonic2}, however, are in slightly different forms. More precisely, the non-diagonal elements, $T^{\pm}_{1,2} = - i \Pi_{A_\pm}^{\mp1/2}$ and $T^{\pm}_{2,1} = - i \Pi_{A_\pm}^{\pm1/2}$ are not the same but they are related to the each other by an inversion of one of the corresponding cycles (or equivalently the action). This suggests that the relation 
	\begin{equation}\label{Inversion_Acycle}
		\Pi_{A_+} = \Pi_{A_-}^{-1}
	\end{equation} 
	should hold so that all the diagrams represent \underline{exactly} the same physical system\footnote{Note that in the search for an exact quantization condition, only $T^\pm_{2,2}$ elements of \eqref{ConnectionMatrix_Harmonic1} and \eqref{ConnectionMatrix_Harmonic2} are relevant. But in different settings, such as scattering problems, other elements could possess the relevant physical information \cite{Taya:2020dco,Enomoto:2020xlf,Enomoto:2022mti}.}. 
	
	To understand this property better let us compare two diagrams on in Fig.~\ref{Figure: HarmonicDiagram}(Left). The difference in the resulting $\Pi_{A_\pm}$ terms stems from the orientations of the branch cuts which present itself in the transition matrix via
	\begin{equation}\label{MonodromyMatrix_AcycleInversion}
		M_\mrmA^B N_\mrmA^{x_1,x_2} = \left(N_\mrmA^{x_1,x_2}\right)^{-1} M_\mrmA^B\, .
	\end{equation}
	The compensation of this change arises from the sign difference of $\Im \sqrt{V_2(x) - u}$ in the principal branches of the two sheeted Riemann surfaces. 
	In particular \cite{chew1999waves}, when $V_2(x)<u$, for the choice in Fig.~\ref{Figure: HarmonicChoice1}, we have
	\begin{equation}
		\Im \sqrt{V_2(x) -u} >0\, ,
	\end{equation} 
	while for the choice in Fig.~\ref{Figure: HarmonicChoice2}, we have
	\begin{equation}
		\Im \sqrt{V_2(x) - u}<0\, ,
	\end{equation}
	in respective principal branches. Note that this sign difference affects all orders in the WKB expansion\footnote{In fact, for the simple harmonic oscillator, upon integration over the closed $A_\pm$ cycles, only the leading order term contributes to $\Pi_{A_\pm}$ but here we emphasize the presence sign difference at all orders since the integration of the WKB expansion in more general cases is not trivial as in the simple harmonic oscillator.} and as a result, we get two complex conjugate expressions in \eqref{ExponentiatedAction}:
	\begin{equation}\label{TwoActions_HO}
		\Pi_{A_\pm} = \exp\left\{\mp\frac{i}{g} \oint_{A_\pm} \mrmd x\, p(x,u) \right\}  = e^{\mp \frac{2\pi i}{g} a(u)}\, ,
	\end{equation}
	where we used the definition of $p(x,u)$ in Section~\ref{Section: AiryWeber_Dictionary}: $\ts(x,u) = \mp i g p(x,u)$. Then, these two expressions are indeed inverse of each other as indicated in \eqref{Inversion_Acycle} and the transition matrices \eqref{ConnectionMatrix_Harmonic1} and \eqref{ConnectionMatrix_Harmonic2} are equal to each other; thus, they represent the same physical system.
	
	\subsection*{\underline{Generic definitions for different branch cut conventions}:}
	The harmonic oscillator Stokes diagrams in Fig.~\ref{Figure: HarmonicDiagram} forms basic building blocks for more generic potentials. In the main text, we originally use in Fig.~\ref{Figure: HarmonicChoice1}(Left) as the main branch-cut convention. Then, the exponentiated WKB actions for $A$ and $B$ cycles are defined accordingly to \eqref{TwoActions_HO} as
	\begin{equation}\label{ActionWKB_Convention}
		\Pi_A = e^{-\frac{2\pi i}{g}a(u)}\, , \quad \Pi_B = e^{-\frac{2\pi i}{g}\aD(u)}\, .
	\end{equation} 
	This, as we observe here, is consistent with our convention. 
	
	For the alternative branch cut conventions, we introduce the corresponding actions as
	\begin{equation}\label{ActionWKB_inverse}
		\tPi_A = \left(\Pi_A\right)^{-1} = e^{+\frac{2\pi i}{g} a(u)} \, , \quad \tPi_B  = \left(\Pi_B\right)^{-1} = e^{+\frac{2\pi i}{g} a^\mrmD(u)} \,  . 
	\end{equation}
	Note that the inversion of $\Pi_B$ has no practical effect as it is a real quantity. More precisely, in the associated principal Riemann sheets for the cases in Fig.~\ref{Figure: AlternativePeriodic_Below1} and Fig.~\ref{Figure: AlternativePeriodic_Below2}, the actions of $A$ and $B$ cycles are related as $\aD(u) = -i a(u_\mrmT-u)$ and $\aD(u) = + i a(u_\mrmT-u)$, respectively. Then, in both settings, $\tPi_{B}$ an exponentially suppressed expression, but expressing them in terms of $a^\mrmD(u)$ induces a sign change.
	
	\subsection*{\underline{Periodic potential}:}
\begin{figure}[t]
		\centering
		\begin{subfigure}[h]{0.48\textwidth}
			\caption{\underline{$\t_\d=0^-$}}	\label{Figure: AlternativePeriodic_Below1}
			\includegraphics[width=\textwidth]{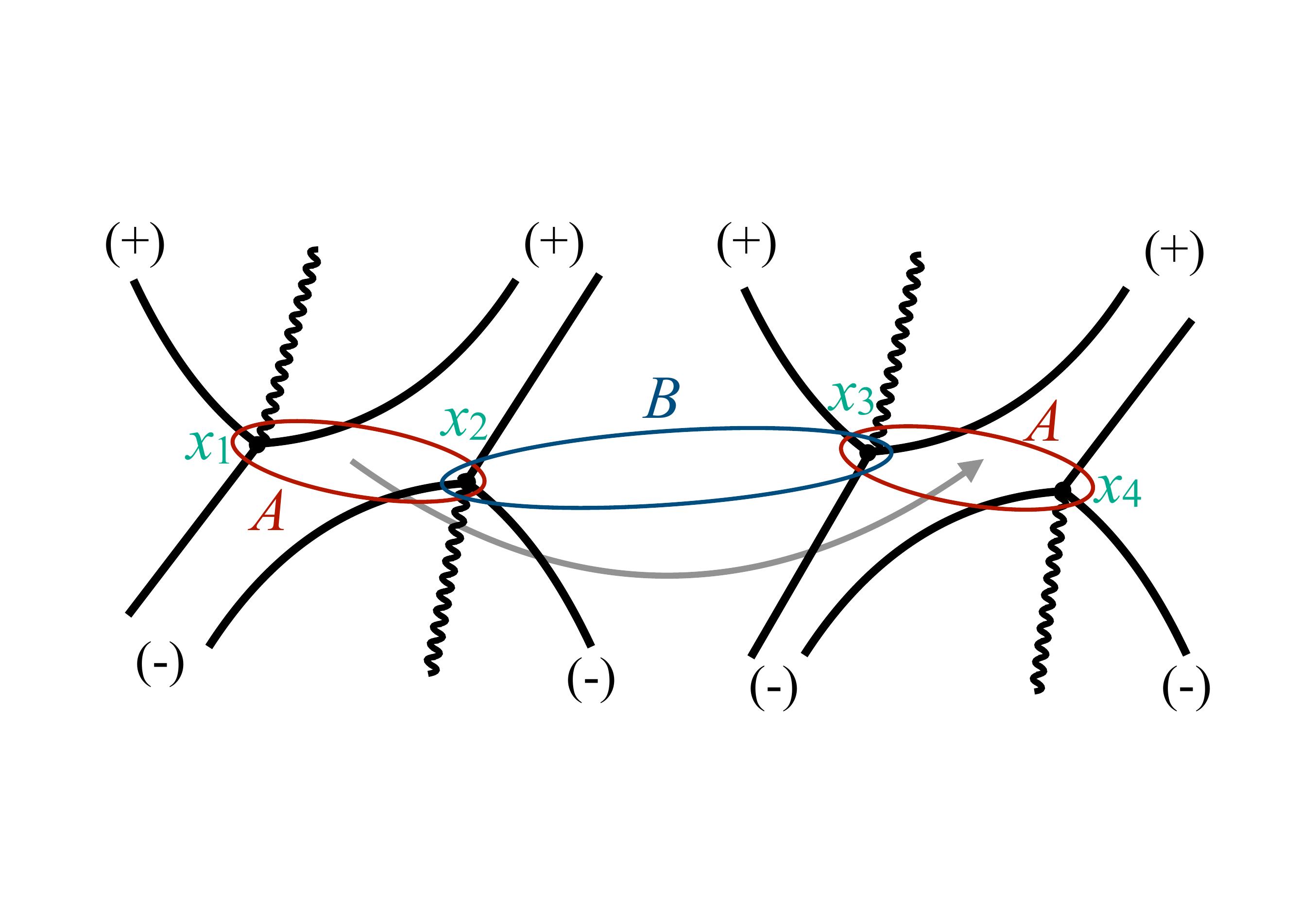}
		\end{subfigure}
		~\hfill 
		\begin{subfigure}[h]{0.48\textwidth}
			\caption{\underline{ $\t_\d=0^+$}}	\label{Figure: AlternativePeriodic_Below2}
			\includegraphics[width=\textwidth]{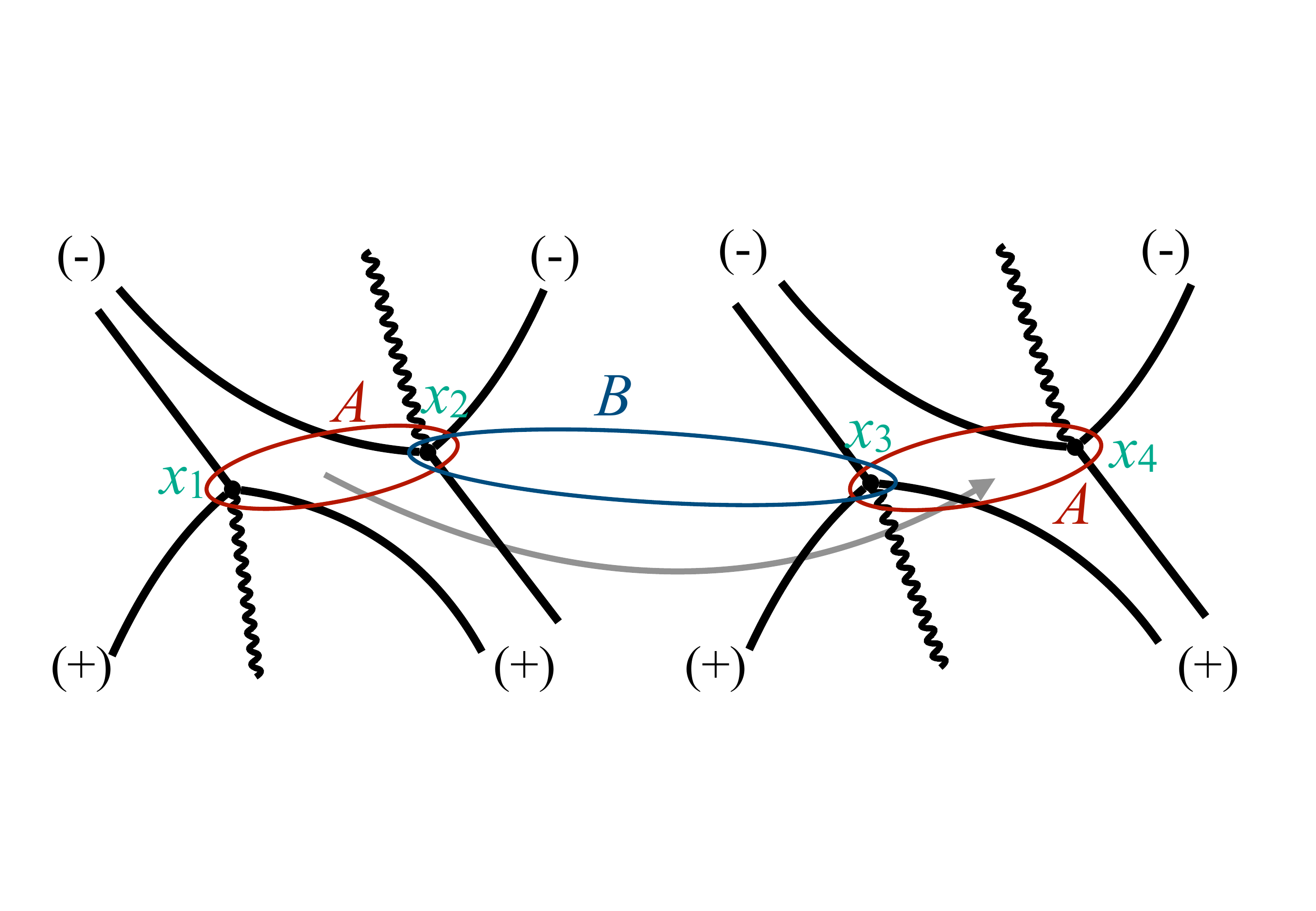}
		\end{subfigure}
		\vspace{-0.2cm}
		\caption{Stokes diagrams for periodic potential with alternative branch cut conventions. {\bf{(a)}} The choice of the cuts is equivalent to the original choice in Section~\ref{Section: PeriodicPotential}. {\bf(b)} A different branch cut convention which is related to (a) via an inversion of $A$-cycle.} \label{Figure: AlternativeDiagramsPeriodic_Below}
	\end{figure}

	Let us now consider the branch cut conventions in the case of periodic potential different than we discuss in Section~\ref{Section: PeriodicPotential}. In the main text, our choice was equivalent to Fig.~\ref{Figure: HarmonicChoice1}(Left). In Fig.~\ref{Figure: AlternativeDiagramsPeriodic_Below}, we illustrate the Stokes diagrams, for $u<u_\mrmT$ sector with two branch cut conventions of Fig.~\ref{Figure: HarmonicDiagram}(Right).
	
	As we discussed in the simple harmonic oscillator example, Fig.~\ref{Figure: AlternativePeriodic_Below2} is equivalent to our original choice in Section~\ref{Section: PeriodicPotential} while Fig.~\ref{Figure: AlternativePeriodic_Below2} represent a diagram in a different Riemann sheet and as a result the action associated to the $A$-cycle is inverted. The effects of these relations are apparent in the transition matrices connecting $x= \pm \frac{\pi}{2}$ in Fig.~\ref{Figure: AlternativeDiagramsPeriodic_Below}:
	\begin{align}
		T_{\t_\d=0^-} &= N_\mrmA^{3,4} M_\mrmA^- {\color{red} N_\mrmA^{2,3}  M_\mrmA^- M_\mrmA^\mrmB}  M_\mrmA^- = N_\mrmA^{3,4} M_\mrmA^- {\color{red} M_\mrmA^\mrmB \left(N_\mrmA^{2,3}\right)^{-1}  M_\mrmA^+ }  M_\mrmA^- \label{TransitionMatrix_PeriodicAlternative1}\, ,  \\ \nonumber \\
		T_{\t_\d=0^+} &= N_\mrmA^{3,4} M_\mrmA^+ {\color{blue} M_\mrmA^\mrmB M_\mrmA^+} N_\mrmA^{2,3} M_\mrmA^+ = N_\mrmA^{3,4} M_\mrmA^+ {\color{blue} M_\mrmA^- M_\mrmA^\mrmB} N_\mrmA^{2,3} M_\mrmA^+ \, ,  \label{TransitionMatrix_PeriodicAlternative2} 
	\end{align}
	where we used \eqref{MonodromyMatrix_OrderChange} and \eqref{MonodromyMatrix_AcycleInversion} to rearrange the monodromy matrices. With the rearrangements, \eqref{TransitionMatrix_PeriodicAlternative2} is in the same form with \eqref{TransitionMatrix_PeriodicBelow2}, while in \eqref{TransitionMatrix_PeriodicAlternative1} the Voros matrix, $N_\mrmA^{2,3}$ is inverted with respect to \eqref{TransitionMatrix_PeriodicBelow1} as we expect.  
	
	The quantization condition for $\t_\d = 0^-$ which is in a different form than \eqref{EQC_PeriodicBelow}:
	\begin{equation}\label{EQC_PeriodicBelow_Alternative}
		D_{\t_\d = 0^-} = 2\cos\t - \sqrt{\tPi_A^{-1} \tPi_B^{-1} } - \sqrt{\tPi_A \tPi_B^{-1}} - \sqrt{\tPi_A \tPi_B} = 0\, .
	\end{equation}
	For $\t_\d =0^+$, on the other hand, we recover the quantization condition in \eqref{EQC_PeriodicBelow}
	\begin{equation}\label{EQC_PeriodicBelow_Same}
		D_{\t_\d = 0^+} = 2\cos\t - \sqrt{\Pi_A^{-1} \Pi_B^{-1} } - \sqrt{\Pi_A \Pi_B^{-1}} - \sqrt{\Pi_A \Pi_B} = 0\, ,
	\end{equation}
	
	The differences in different branch cut conventions are totally compatible with each other and in all cases, the median summations lead to the same manifestly real quantization condition. To understand the last point, we observe that the change in branch cut conventions alters the orientation of the $A$ and $B$ cycles as well. Then, the median summations of \eqref{EQC_PeriodicBelow_Same} and \eqref{EQC_PeriodicBelow_Alternative} are given by
	\begin{align}
		\mfrS_0^{-1/2}D_{\t_\d=0^-} &= 2\cos \t - \left(1 + \tPi_B\right)^{1/2} \left(\sqrt{\tPi_A^{-1} \tPi_B^{-1}} + \sqrt{\tPi_A \, \tPi_B^{-1}}\right)\, ,\label{EQC_MedianSum_PeriodicAlternative1}\\
		\mfrS_0^{-1/2}D_{\t_\d=0^+} &= 2\cos \t - \Big(1 + \Pi_B\Big)^{1/2} \left(\sqrt{\Pi_A^{-1} \Pi_B^{-1}} + \sqrt{\Pi_A \, \Pi_B^{-1}}\right)\label{EQC_MedianSum_PeriodicAlternative2} \, .
	\end{align}
	Both recover the median summation \eqref{EQC_MedianSum_PeriodicBelow}. At this point, we emphasize that the expressions in \eqref{EQC_MedianSum_PeriodicAlternative1} and \eqref{EQC_MedianSum_PeriodicAlternative2} are not exactly the same due to the relationship $\tPi_A = \left(\Pi_A\right)^{-1}$ and the equivalence is due to the reality of the exact quantization condition.  
	
	\begin{figure}[t]
		\centering
		\begin{subfigure}[h]{0.48\textwidth}
			\caption{\underline{$\t_\d=-\pi^+$}}	\label{Figure: AlternativePeriodic_Above1}
			\includegraphics[width=\textwidth]{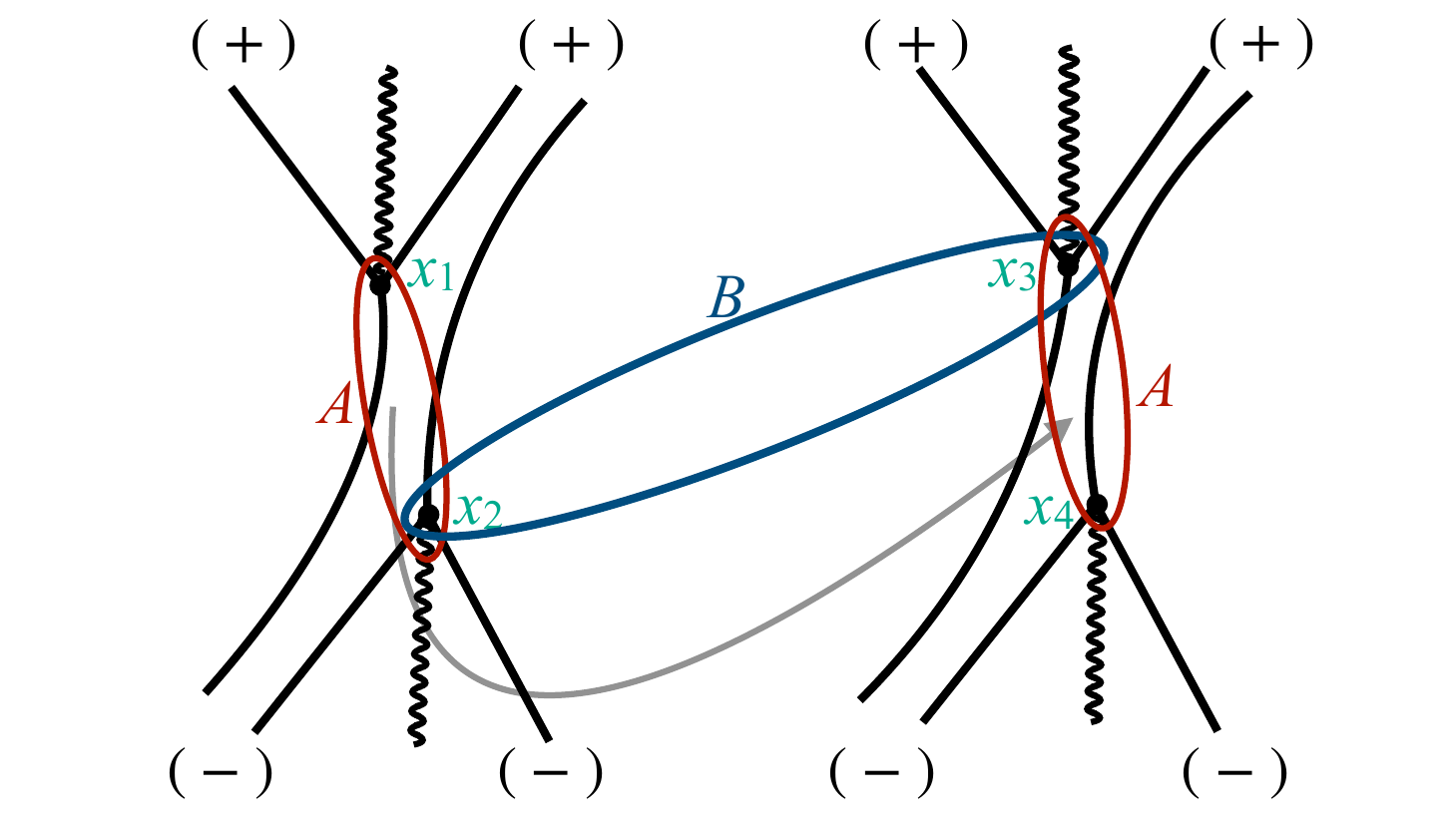}
		\end{subfigure}
		~\hfill 
		\begin{subfigure}[h]{0.48\textwidth}
			\caption{\underline{$\t_\d=\pi^-$}}	\label{Figure: AlternativePeriodic_Above2}
			\includegraphics[width=\textwidth]{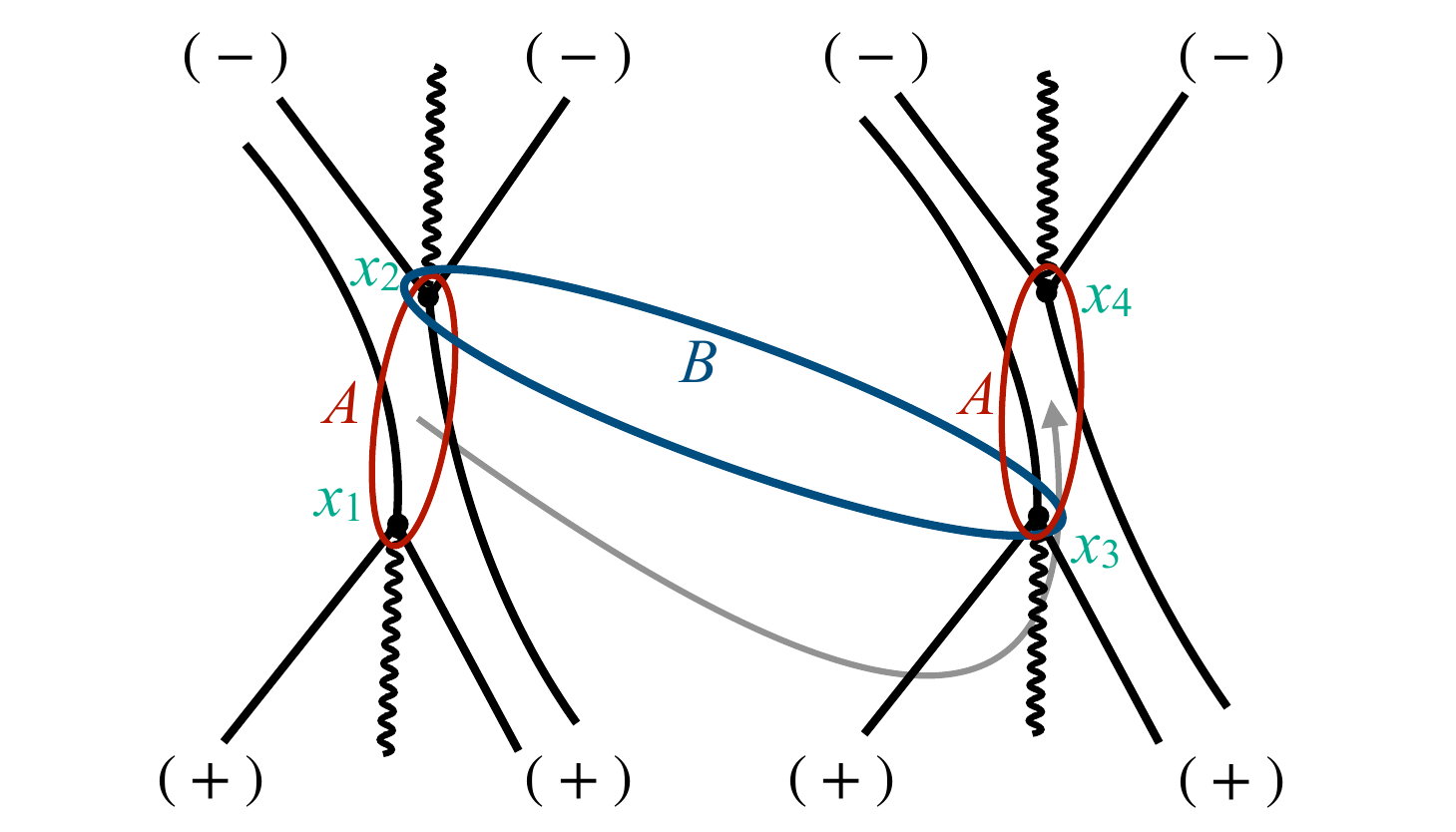}
		\end{subfigure}
		\vspace{5pt}
		\caption{Analytic continuations of the diagrams in Fig.~\ref{Figure: AlternativeDiagramsPeriodic_Below}.} \label{Figure: AlternativeDiagramsPeriodic_Above}
	\end{figure}
	
	Now, we turn our focus to $u>u_\mrmT$ sector. In Fig.~\ref{Figure: AlternativeDiagramsPeriodic_Above}, we illustrate the Stokes diagrams at $\t_\d = \mp\pi^\pm$. As we discuss in Section~\ref{Section: PeriodicPotential}, for $\t_\d=\pi^-$ and $\t_\d = -\pi^+$, the resulting exact quantization conditions are the same with \eqref{EQC_PeriodicBelow_Same} and \eqref{EQC_PeriodicBelow_Alternative}, respectively. To get their median summations, we follow the derivation in Section~\ref{Section: PeriodicPotential} as well. Then, we get the median summation  at $\t_\d = -\pi$ as
	\begin{align}
		D^{\med}_{\t_\d=-\pi} & =  2\cos\t - \left(1 - \tPi_{B_{-\pi}}\right)^{1/2} \left(\sqrt{\tPi_{A_{ -\pi}}} + \sqrt{\tPi^{-1}_{A_{-\pi}} \Pi^{-2}_{B_{\pi}}} \,  \right) \nonumber \\ 
		& = 2\cos\t - \left(1 + \tPi_{B_{-\pi}}\right)^{1/2} \left(\sqrt{\tPi_{\tA_{-\pi}}^{-1} \tPi_{B_{-\pi}}^{-1}} - \sqrt{\tPi_{\tA_{-\pi}} \tPi_{B_\pi}^{-1}}\right) \, , \label{EQC_MedianSum_PeriodicAlternateve_Above2}
	\end{align}
	and  at $\t_\d = \pi$ as
	\begin{align}
		D^{\med}_{\t_\d=\pi} & =  2\cos\t - \left(1 - \Pi_{B_{\pi}}\right)^{1/2} \left(\sqrt{\Pi_{A_{ \pi}}} + \sqrt{\Pi^{-1}_{A_{\pi}} \Pi^{-2}_{B_{\pi}}} \,  \right) \nonumber \\ 
		& = 2\cos\t - \left(1 + \Pi_{B_\pi}\right)^{1/2} \left(\sqrt{\Pi_{\tA_\pi}^{-1} \Pi_{B_\pi}^{-1}} - \sqrt{\Pi_{\tA_\pi} \Pi_{B_\pi}^{-1}}\right) \, , \label{EQC_MedianSum_PeriodicAlternateve_Above1}
	\end{align}
	In both cases, to get the last equations, we used
	\begin{equation}\label{AC_ExponentiatedAction_PeriodicAlternative}
		\Pi_{\tA_\pi} = \Pi_{A_\pi} \Pi_{B_\pi}\, , \quad \tPi_{\tA_{-\pi}} = \tPi_{A_{-\pi}} \tPi_{B_{-\pi}}\, , 
	\end{equation} which are equivalent to $a(u)\rightarrow a(u) + a^\mrmD(u)$ in the terms of WKB actions. Note that since $\aD(u) = \mp i a(u_\mrmT-u)$ for the phases $\t_\l = \pm \pi$, more precise analytic continuations are written as $a(u) \rightarrow a(u) \mp i a(u_\mrmT - u)$.  
	
	In light of this, we can now compare the Stokes diagrams at the phase $\t_\d = -\pi^+$ which we illustrate in Fig.~\ref{Figure: AlternativePeriodic_Above1} and in Fig.~\ref{Figure: StokesDiagrams_PeriodicAbove}. Since they originated from different type of branch cut conventions, the exponetiated actions corresponding to $A$-cycles in these cases are complex conjugate to each other. This fact leads to the differences in the interpretations of the $\tA$ cycle in $u>u_\mrmT$ sector: While, for Fig.~\ref{Figure: AlternativePeriodic_Above1}, $\tPi_{\tA_{-\pi}}$ is defined as in \eqref{AC_ExponentialAction_Periodic}, for Fig.~\ref{Figure: StokesDiagrams_PeriodicAbove}, the same analytic continuation yields $\Pi_{\tA_{-\pi}} = \Pi_{A_{-\pi}} \Pi^{-1}_{B_{-\pi}}$ as we in Section~\ref{Section: PeriodicPotential}.
	
	
	Finally, let us comment on the straightforward comparison of the cases in 
	for Fig.~\ref{Figure: AlternativePeriodic_Above1} and Fig.~\ref{Figure: AlternativePeriodic_Above2} without referring to the analytic continuations they are originated: In the Stokes geometry, only differences appear in the signs of individual Stokes lines and at all steps, all the expressions are in the same form. The latter can be understood by observing that their Stokes geometries can be considered as the deformations of the parameter $g$ as $\Im g = 0^-$ and $\Im g  = 0^+$, respectively. This compensates the sign differences of the Stokes lines. As a result, the only possible differences between their exact quantization conditions can arise from the interpretations (or more precisely the definitions) of the WKB actions which can be fixed by the connection to the original Stokes geometry at $\t_\l=0$. 
	
	\begin{figure}[t]
		\centering
		\begin{subfigure}[h]{0.49\textwidth}
			\caption{\underline{$\t_\d=0^-$}}	\label{Figure: AlternativeDW_Below1}
			\includegraphics[width=\textwidth]{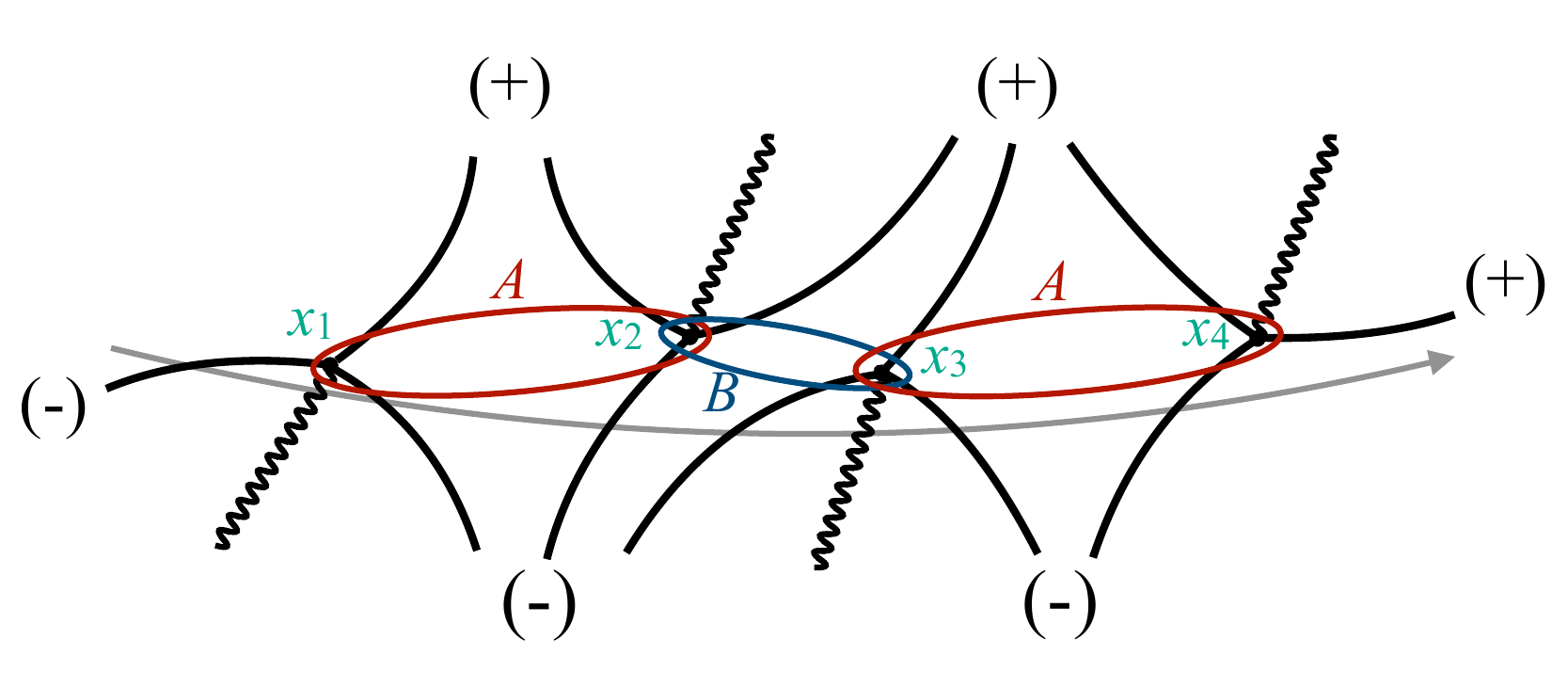}
		\end{subfigure}
		~\hfill 
		\begin{subfigure}[h]{0.49\textwidth}
			\caption{\underline{$\t_\d=0^+$}}	\label{Figure: AlternativeDW_Below2}
			\includegraphics[width=\textwidth]{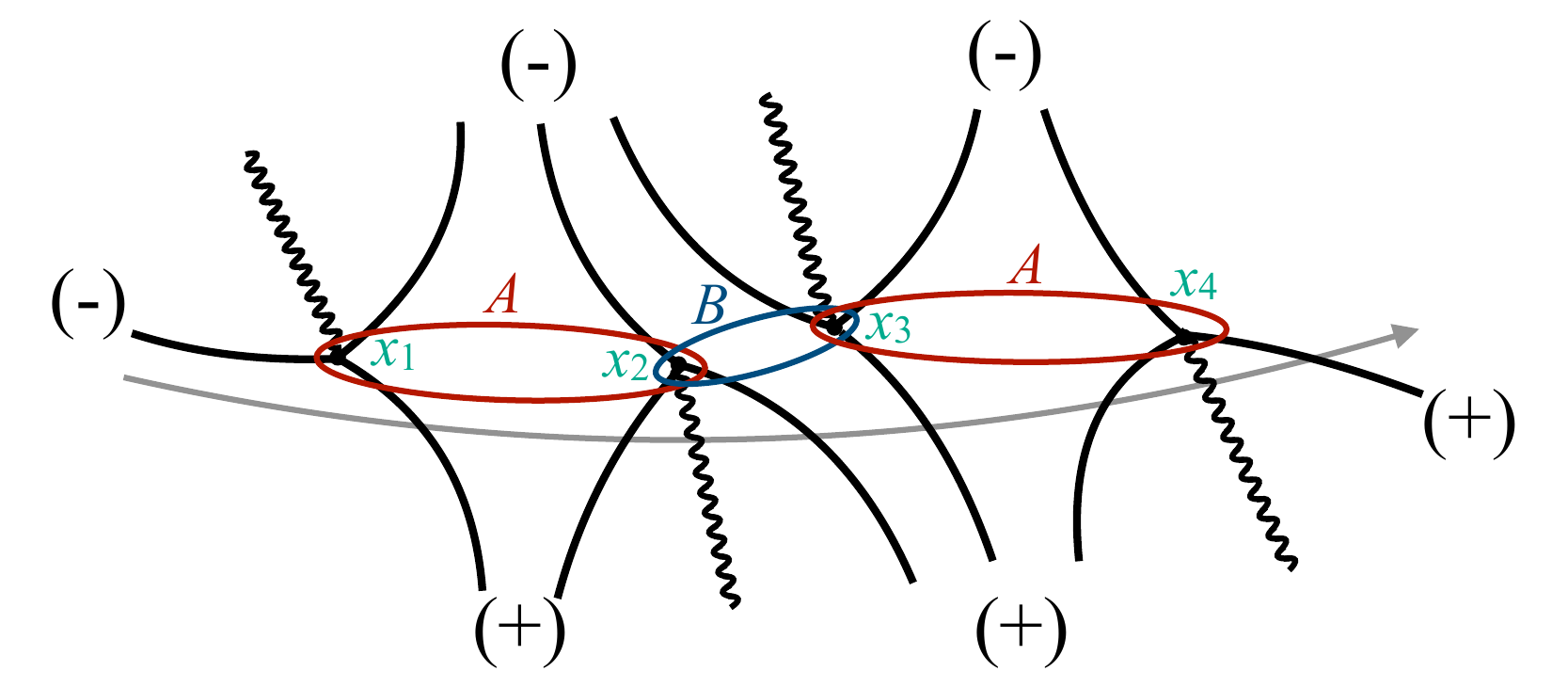}
		\end{subfigure}
		\vspace{10pt}
		\caption{Stokes diagrams for symmetric double-well potential with alternative branch cut conventions. {\bf{(a)}} The choice of the cuts is equivalent to the original choice in Section~\ref{Section: DoubleWell}. {\bf(b)} A different branch cut convention which is related to (a) via an inversion of $A$-cycle.} \label{Figure: AlternativeDiagramsDW_Below}
	\end{figure}
	
	\subsection*{\underline{Symmetric double-well potential}:}
	Finally, we briefly discuss the different branch cut conventions for the symmetric double-well potential. For the sector, $u<u_\mrmT$, we illustrate two branch cut conventions in Fig.~\ref{Figure: AlternativeDiagramsDW_Below}, which stem from the diagrams in Fig.~\ref{Figure: HarmonicDiagram}(Right). The associated transition matrices connecting $x = \pm \infty$ are 
	\begin{align}
		T_{\t_\d = 0^-} & = M_\mrmA^- {\color{red} N_\mrmA^{3,4} M_\mrmA^{-} M_\mrmA^\mrmB} M_\mrmA^- N_\mrmA^{2,3} M_\mrmA^- {\color{red}N_\mrmA^{1,2} M_\mrmA^- M_\mrmA^\mrmB} M_\mrmA^- \nonumber \\ 
		& =	M_\mrmA^- {\color{red}M_\mrmA^\mrmB \left(N_\mrmA^{3,4}\right)^{-1} M_\mrmA^+} M_\mrmA^- N_\mrmA^{2,3} M_\mrmA^- {\color{red}M_\mrmA^\mrmB \left(N_\mrmA^{1,2}\right)^{-1} M_\mrmA^+} M_\mrmA^- \, , \label{EQC_DoubleWellBelow_Alternative}\\ \nonumber \\
		T_{\t_\d = 0^+} & = M_\mrmA^+ {\color{blue}M^\mrmB_\mrmA M_\mrmA^+}   N_\mrmA^{3,4} M_\mrmA^+ N_\mrmA^{1,2} M_\mrmA^+ {\color{blue}M_\mrmA^\mrmB M_\mrmA^-} N_\mrmA^{1,2} M_\mrmA^+ \nonumber \\
		& = M_\mrmA^+ {\color{blue}M_\mrmA^- M_\mrmA^\mrmB }N_\mrmA^{3,4} M_\mrmA^+ N_\mrmA^{2,3} M_\mrmA^+  {\color{blue}M_\mrmA^- M_\mrmA^\mrmB} N_\mrmA^{1,2} M_\mrmA^+\, , \label{EQC_DoubleWellBelow_Same}
	\end{align}
	where we used \eqref{MonodromyMatrix_OrderChange} and \eqref{MonodromyMatrix_AcycleInversion} to obtain final expressions. For $\t_\d = 0^+$, the matrix is in the same form with \eqref{TransitionMatrix_DoubleWellBelow1}, while for $\t_\d = 0^-$, the Voros matrices associated to $A$-cycles are inverted. The effect of this inversion appears also in the quantization conditions:
	\begin{align}
		D_{\t_\d = 0^-} & = \sqrt{\tPi_A^{-2} \tPi_B^{-1}} \left[\left(1 + \tPi_A\right)^2 + \tPi_A^2 \tPi_B  \right] = 0\, , \\
		D_{\t_\d = 0^+} & = \sqrt{\Pi_A^{-2} \Pi_B^{-1}} \left[\left(1 + \Pi_A\right)^2 + \Pi_A^2 \Pi_B  \right] = 0\, ,
	\end{align}
	where we use the definitions in \eqref{ActionWKB_inverse}. Note that $D_{\t_\d=0^+}$ is in the same form with \eqref{EQC_DoubleWell_2}, while $D_{\t_\d = 0^-}$ has a different form than \eqref{EQC_DoubleWell_1} and they are related to each other by inversion of $\tPi_A$ terms.

	Despite their differences, the branch cut conventions do not affect the properties of the physical system and both Stokes diagram choices in Fig.\ref{Figure: AlternativeDiagramsDW_Below} leads to equivalent median summations:
	\begin{align}
		\mfrS_0^{-1/2}D_{\t_\d=0^-} =  \sqrt{\tPi_A^{-2} \tPi_B^{-1}} \left[ \left(1 + \tPi_A^2\right)\sqrt{1+\tPi_B} + 2 \tPi_A \right]\, , \label{EQC_MedianSum_DW_Alternative1}\\
		\mfrS_0^{-1/2}D_{\t_\d=0^+} = \sqrt{\Pi_A^{-2} \Pi_B^{-1}} \left[ \left(1 + \Pi_A^2\right)\sqrt{1+\Pi_B} + 2 \Pi_A \right]\, . \label{EQC_MedianSum_DW_Alternative2} 
	\end{align}
	Note that the orientation of $A$ and $B$ cycles in Fig.~\ref{Figure: AlternativeDW_Below1} is different than Fig.~\ref{Figure: AlternativeDW_Below2} which introduces the negative power to the Stokes automorphism in \eqref{EQC_MedianSum_DW_Alternative1}. 
	
	\begin{figure}[t]
		\centering
		\begin{subfigure}[h]{0.48\textwidth}
			\caption{\underline{$\t_\d=-\pi^+$}}	\label{Figure: AlternativeDW_Above1}
			\includegraphics[width=\textwidth]{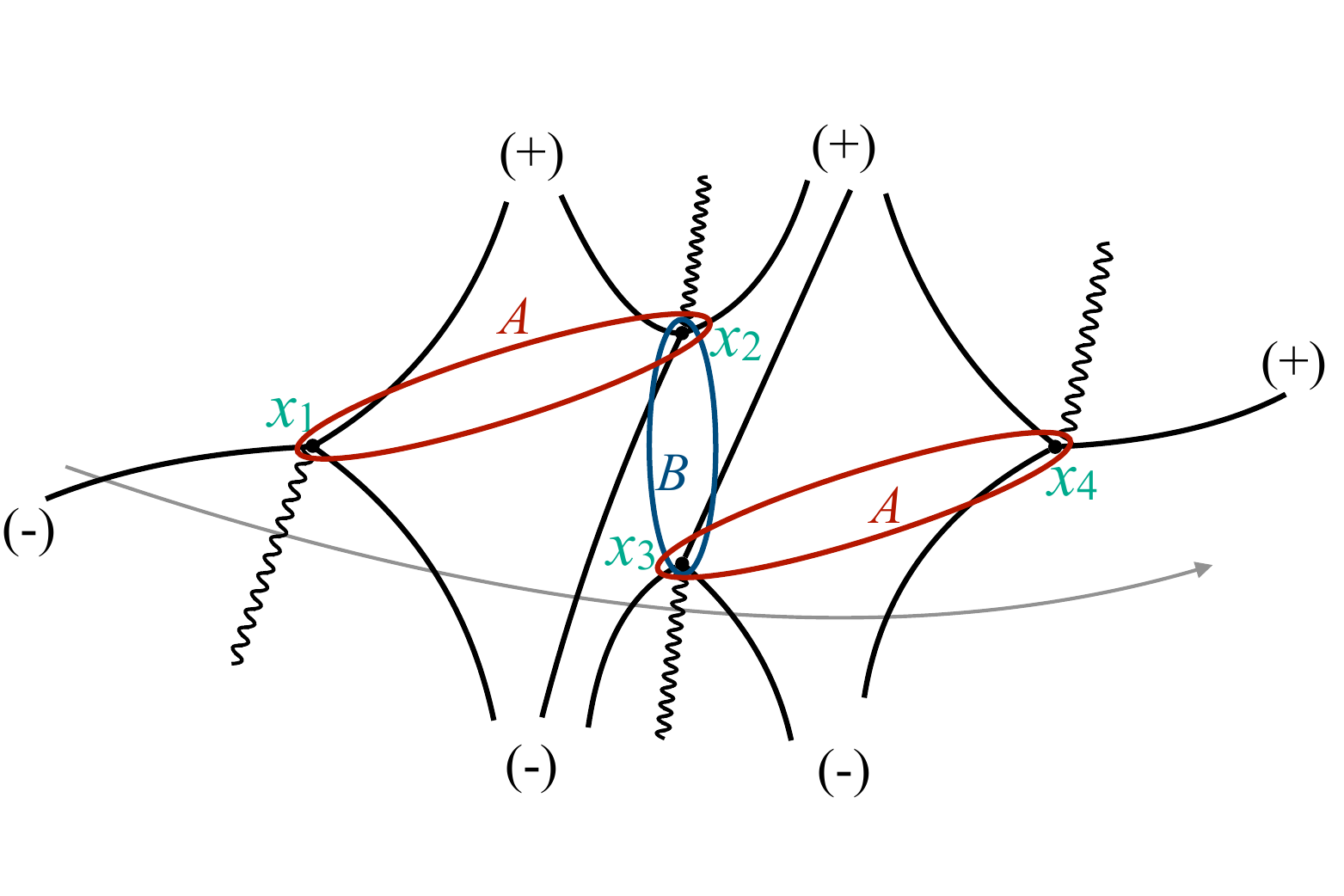}
		\end{subfigure}
		~\hfill 
		\begin{subfigure}[h]{0.48\textwidth}
			\caption{\underline{$\t_\d=\pi^-$}}	\label{Figure: AlternativeDW_Above2}
			\includegraphics[width=\textwidth]{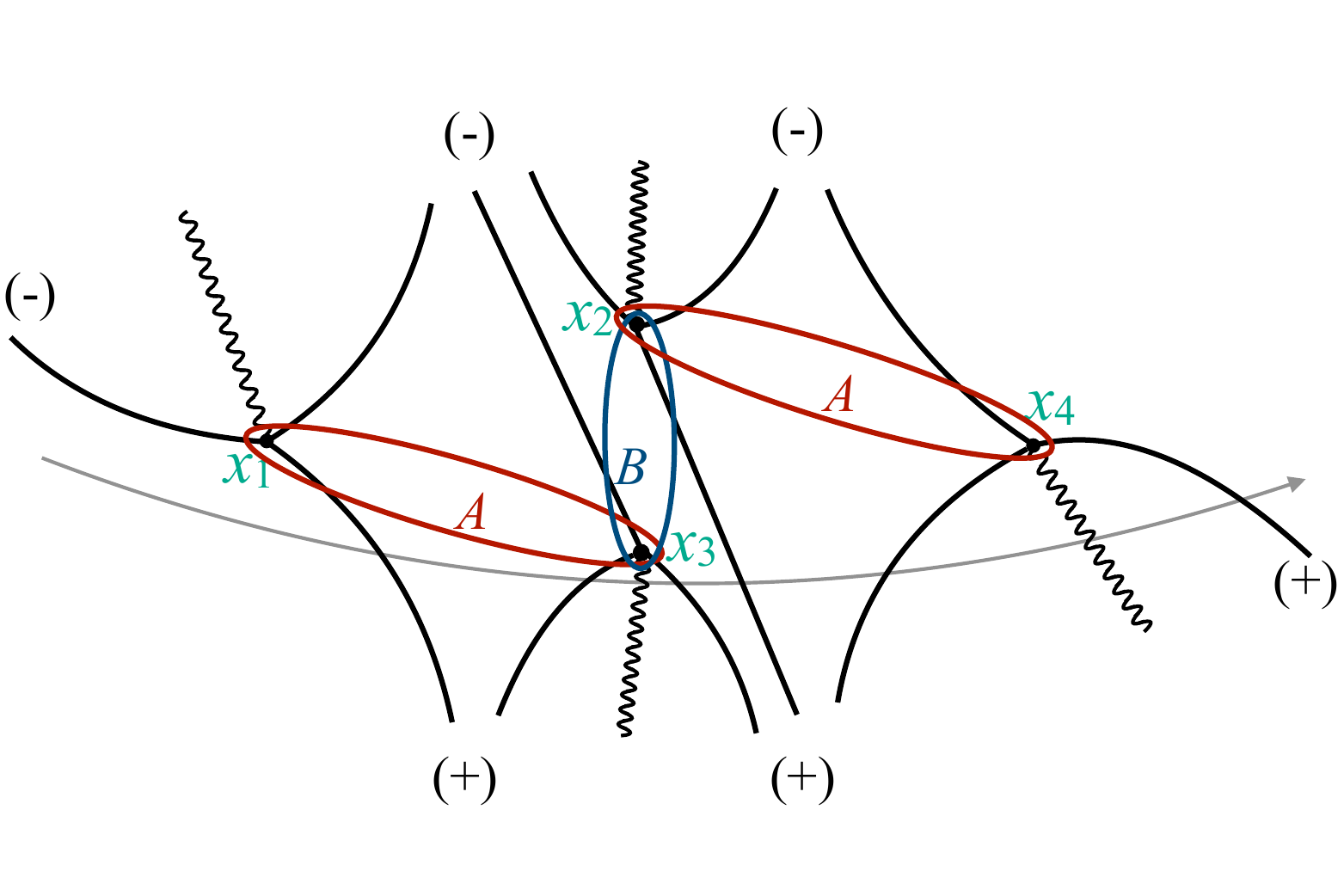}
		\end{subfigure}
		\caption{Analytic continuations of the diagrams in Fig.~\ref{Figure: AlternativeDiagramsDW_Below}.} \label{Figure: AlternativeDiagramsDW_Above}
	\end{figure}
	
	The analytic continuations of the diagrams in Fig.~\ref{Figure: AlternativeDiagramsDW_Below} to $u>u_\mrmT$ sector are illustrated in Fig.~\ref{Figure: AlternativeDiagramsDW_Above}. As always, the exact quantization conditions in \eqref{EQC_DoubleWellBelow_Alternative} and \eqref{EQC_DoubleWellBelow_Same} stays intact in both transitions. To obtain their median summations, we follow the discussion in Section~\ref{Section: DoubleWell}. Then, we get at $\t_\d = -\pi$
	\begin{align}
		D^\med_{\t_\d = -\pi} & =  \sqrt{\Pi_{\tA_{-\pi}}^{-2} \Pi_{B_{-\pi}}^{-1}} \, \left[ \left(1 + \Pi_{\tA_{-\pi}}^2\right) \sqrt{1 + \Pi_{B_{-\pi}}} + 2\Pi_{\tA_{-\pi}} \right]\, ,
	\end{align}
	and at $\t_\d=\pi$
	\begin{align}
		D^\med_{\t_\d = \pi} & = \sqrt{\tPi_{\tA_\pi}^{-2} \Pi_{B_\pi}^{-1}} \, \left[ \left(1 + \tPi_{\tA_\pi}^2\right) \sqrt{1 + \tPi_{B_\pi}} + 2\tPi_{\tA_\pi} \right] \, ,
	\end{align}
	where we define
	\begin{equation}\label{AC_ExponentiatedAction_DoubleWellAlternative}
		\Pi_{\tA_\pi} = \Pi_{A_\pi} \Pi_{B_\pi}^{\frac{1}{2}}\, , \quad \tPi_{\tA_{-\pi}} = \tPi_{A_{-\pi}} \tPi_{B_{-\pi}}^{\frac{1}{2}}\, .
	\end{equation}
	As in the periodic potential case, these transformations are associated to the analytic continuations of the WKB actions in respective cases as $a(u)\rightarrow a(u) \pm \frac{1}{2}a^\mrmD(u)$. In the latter transformation in \eqref{AC_ExponentiatedAction_DoubleWellAlternative}, the sign change is compensated by using different branch cut convention in Fig.~\ref{Figure: AlternativeDW_Above2} and it is related to the analytic continuation in \eqref{AC_ExponentialAction_DW} via $\Pi_A = \left(\tPi_A\right)^{-1}$ and $\Pi_{\tA} = \left(\tPi_{\tA}\right)^{-1}$.

	\section{Notation and other definitions}
	A list of variables which we use in the main text to clarify the notation and definitions.
	\begin{table}[H]
		\begin{center} 
			\begin{tblr}{|c || c |}	
				\hline
				\SetCell[c=2]{c}{\bf{General}} &  \\ 	
				\hline
				$P$: Generic algebraic curve & $V$: Classical potential  \\
				$\Pi_X$: Exponentiated action of a cycle $X$ & $p(x,g)$: All order quantum momentum \\ 
				$a(u,g)$ : Quantum Action &$a^\mrmD (u,g)$: Dual Quantum Action \\
				$D_{\t} = 0$: Quantization condition at phase $\t$& $\mathfrak{S}_\t$: Stokes Automorphism at phase $\t$\\
				$D^\med=0$: Median quantization condition & $\tau$: Modular Parameter \\
				$\mcalF(\tu,g)$: Perturbative data in P-NP relation & $\mcalG(\tu,g)$: Nonperturbative data in P-NP relation\\
				\hline
			\end{tblr}
		\end{center}
	\end{table}
	\begin{table}[H]
		\begin{center} 
			\begin{tblr}{|c|| c | c|}
				\hline 
				{\bf Airy-type approach} & \SetCell[c=2]{c} {\bf Weber-type approach} \\ 
				\hline
				&	\textbf{In local coordinates} & \textbf{In global coordinates} \\ 
				\hline
				$x$: Spatial variable & $y$: Spatial variable,  & $x$: Spatial variable, \\
				\hline 
				$P_\mrmA$: Airy-type algebraic curve &$Q$: Algebraic curve& $P_\mrmW$: Weber-type algebraic curve \\
				\hline 
				$s$: Riccati function &  $\O$: Riccati function & $S$: Riccati function\\
				\hline 
				$\ts$: Odd terms & $\ttO$: Odd terms & $\tS$: Odd terms  \\
				\hline 
				$\psi$: Wave function&  $\Psi_\mrmL$: Local wave function& $\Psi_\mrmG$: Global wave function\\ 
				\hline 
			\end{tblr}	
		\end{center}	
	\end{table}

	
	\bibliographystyle{JHEP}
	\bibliography{EWKB.bib}

\end{document}